\documentclass[manuscript]{aastex}
\usepackage{graphicx}
\usepackage[lofdepth,lotdepth,caption=false]{subfig}
\begin{document}

\title{Near-Nucleus Photometry of Outbursting Comet 17P/Holmes}

\author{Rachel Stevenson$^{1,2}$ 
and 
David Jewitt$^{1,3}$} 
 
\affil{$^1$Department of Earth and Space Sciences, UCLA, 
595 Charles Young Drive East, \\
Los Angeles, CA 90095-1567} 
 
\affil{$^2$Now at: Jet Propulsion Laboratory, California Institute of Technology, 4800 Oak Grove Drive,
Pasadena, CA 91109}

\affil{$^3$Department of Physics and Astronomy, UCLA, 
430 Portola Plaza, Box 951547, \\
Los Angeles, CA 90095-1547}
\email{Rachel.A.Stevenson@jpl.nasa.gov, jewitt@ucla.edu}

\begin{abstract}

Comet 17P/Holmes underwent the largest cometary outburst in recorded history on UT 2007 Oct.\ 23, releasing massive quantities of dust and gas.  We used the Canada-France-Hawaii telescope to obtain wide-field images of 17P/Holmes on 15 dates over a period of 3 months following the outburst and employ them here to examine the subsequent activity of the nucleus and the nature of the ejecta closest to the nucleus.  Through aperture photometry we observed the inner coma (within 2500 km of the nucleus) to fade from an apparent magnitude of 11.7 mag to 17.6 mag, corresponding to absolute magnitudes of 8.1 mag and 12.4 mag, between UT 2007 Nov.\ 6 and 2008 Feb.\ 12.  A second much smaller outburst occurred on UT 2007 Nov. 12, three weeks after the original outburst, suggesting that the nucleus remained unstable.  The surface brightness profile of the inner coma was consistently shallow relative to the expected steady-state profile, and showed a persistent brightness enhancement within $\sim$ 5000 km of the nucleus.  We propose that sublimating ice grains created an ice grain halo around the nucleus, while fragmenting grains were responsible for the shallow surface brightness profile.  

\end{abstract}


\section{Introduction}

Comet 17P/Holmes underwent the largest cometary outburst in recorded history on UT 2007 Oct.\ 23.7 $\pm$ 0.2 \citep{2007IAUC.8886....1B}, brightening from $\sim$ 17$^{th}$ magnitude to $\sim$ 2$^{nd}$ in 42 hours \citep{2008ApJ...677L..63M,2011ApJ...728...31L}.  The Jupiter family comet is known to have experienced at least two prior outbursts: the first led to its discovery on UT 1892 Nov.\ 6 \citep{1892Obs....15..441H} while the second occurred in January 1893 \citep{1893Obs....16..142D}.  \cite{1984Icar...60..522W} speculated that the unusual outburst activity might have been triggered by a grazing encounter with a natural satellite.  In this scenario, the second outburst would have been caused by the final impact of the satellite with the nucleus.  A third major outburst practically eliminates the likelihood of this scenario, suggesting that an alternative mechanism causes these unusual periods of activity in 17P/Holmes.  The two largest outbursts - those that occurred in 1892 Nov.\ and 2007 Oct.\ - both took place approximately five months after 17P/Holmes passed through perihelion at 2.1 AU, suggesting that the activity is thermally-triggered, perhaps with a lag due to the time needed for heat to reach subsurface ice deposits \citep{2008ICQ....30....3S}.

The mass of material ejected has been estimated to be between 10$^{10}$ kg and 10$^{12}$ kg \citep{2008ICQ....30....3S,2010Icar..208..276R,2011ApJ...728...31L}, corresponding to $\sim$ 0.05\% to 5\% of the nucleus mass.  This ejected mass was dominated by the contribution from the dust coma that was observed to expand around the nucleus.  The peak dust production rate approached 10$^{6}$ kg s$^{-1}$ on UT 2007 Oct.\ 24.5 $\pm$ 0.01 \citep{2011ApJ...728...31L}, about 10$^{3}$ times the dust production rates typical of Jupiter Family comets when at similar heliocentric distances.

In addition to massive quantities of dust, 17P/Holmes released significant quantities of volatiles with H$_{2}$O production rates exceeding 4.5 $\times$ 10$^{29}$ s$^{-1}$ \citep{2008ApJ...680..793D,2009AJ....138.1062S}.  Results from large-aperture (radii $>$ 10,000 km) narrowband photometry in the weeks following the outburst led \cite{2009AJ....138.1062S} to conclude that enhanced water production rates were primarily due to prolonged nucleus activity, with a possible contribution from sublimating icy grains in the inner coma.

In this work, we use a sequence of high-resolution wide-field continuum images to examine the evolution of 17P/Holmes during the four months following the outburst.  By utilizing a single instrument and observing strategy we minimize the effect of systematic errors and are able to use a uniform procedure for obtaining, reducing, and analyzing the data.  This results in a consistent, high-quality data set.  In contrast to an earlier study of the integrated light from the comet (Li et al.~2011), we here study material near the nucleus, which is the most-recently ejected and the most pristine of the observable ejecta.  

\section{Observations and Data Reduction}

We used the 3.6 m Canada-France-Hawaii Telescope (CFHT) atop Mauna Kea with the wide-field imager MegaCam \citep{2003SPIE.4841...72B} to obtain high-quality, time-series imaging data.  MegaCam has a 0.96${^\circ}$ $\times$ 0.96${^\circ}$ field of view provided by an array of 36 charge-coupled devices (CCDs).  Each CCD has 2048 $\times$ 4068 pixels, with a pixel scale of $0^{\prime\prime}.185$ pixel$^{-1}$.  We obtained at least one set of five dithered images in the SDSS r$^{\prime}$ filter ($\lambda_{c} =$ 6250 \AA, bandwidth = 1210 \AA) on each of 15 nights between UT 2007 Nov.\ 6 and UT 2008 Feb.\ 12 UT.  Throughout 2007 Nov.\ and on UT 2007 Dec.\ 11 we obtained two sets of exposures with differing exposure times to allow us to probe the bright central region while still being sensitive to the faint outer coma.  In this work we use the set of exposures with the longest exposure time but no saturation of the central region.  We have analyzed all sets of the exposures and find the agreement between sets obtained on the same night to be very good, with differences in calculated magnitudes to be less than 0.02 mag.

The observations were taken in queue-scheduled mode.  In this mode, non-sidereal tracking was unavailable and the comet trailed across each image.  Exposure times varied between 5 s and 200 s, and were chosen to balance the need for sensitivity while minimizing trailing.   Table~(\ref{table:obs}) provides a journal of observations.  

\begin{figure}
\centering
\subfloat{\includegraphics[width=0.2\textwidth]{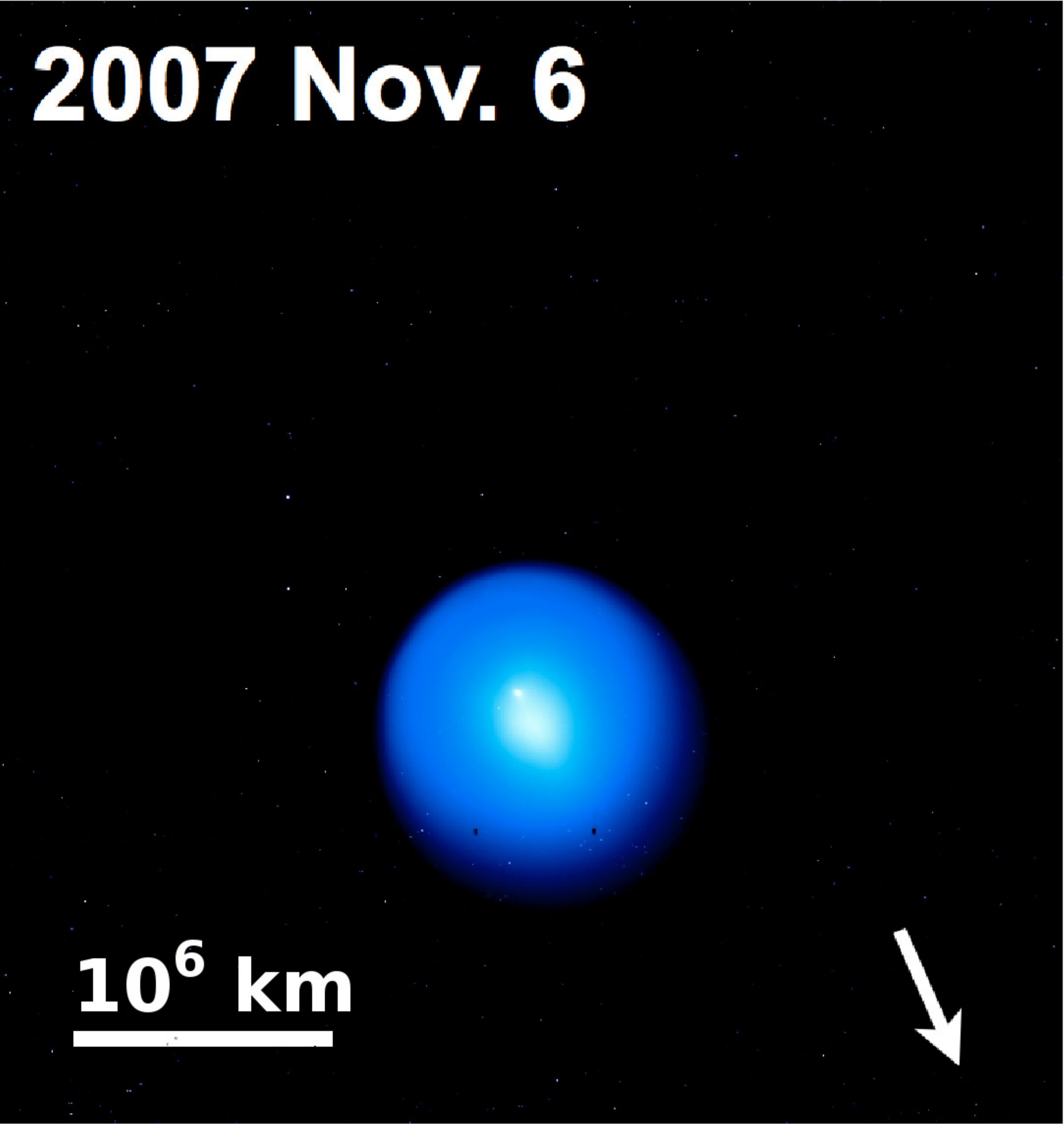}}
\hspace{0.5bp}
\subfloat{\includegraphics[width=0.2\textwidth]{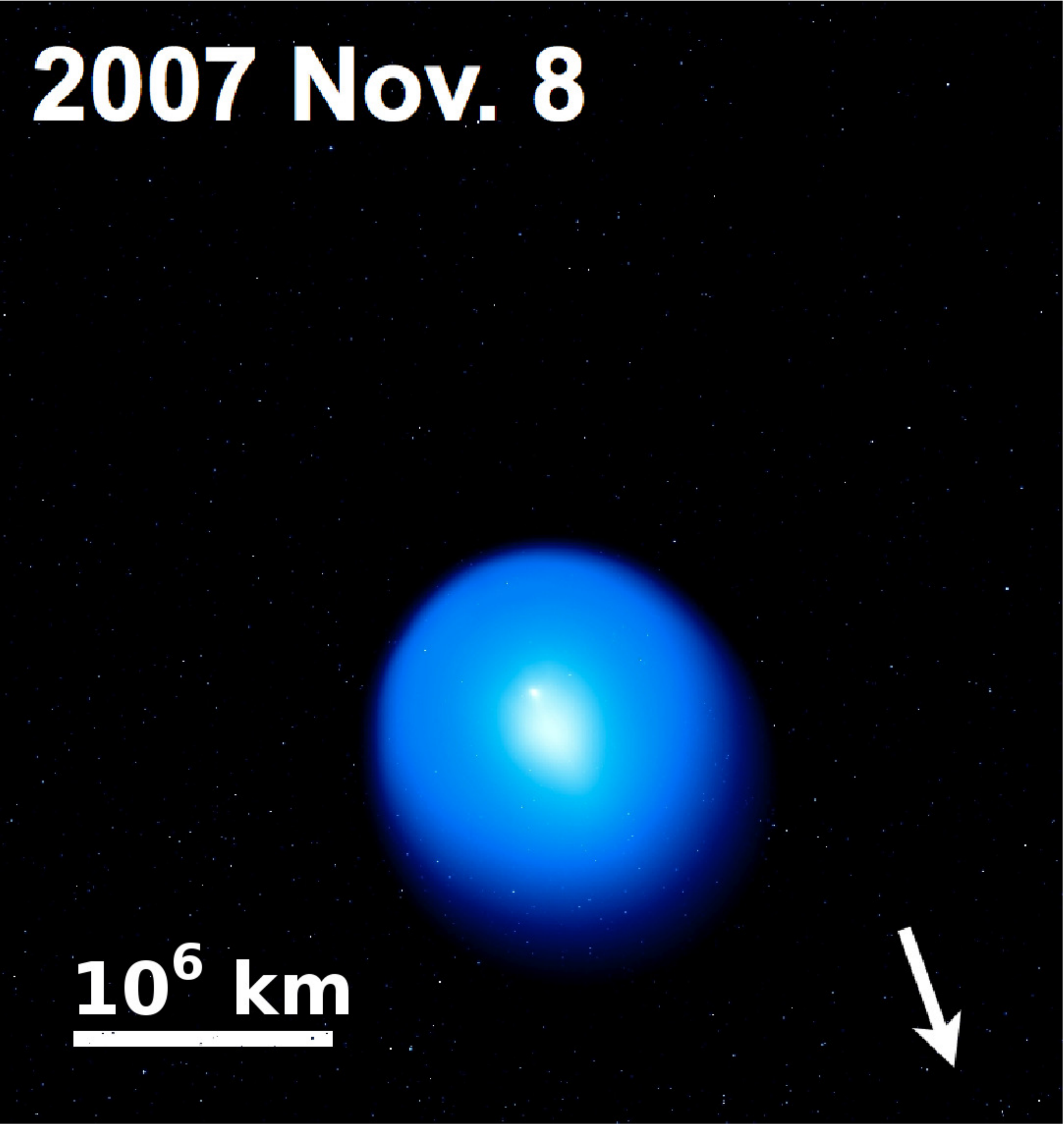}}
\hspace{0.5bp}
\subfloat{\includegraphics[width=0.2\textwidth]{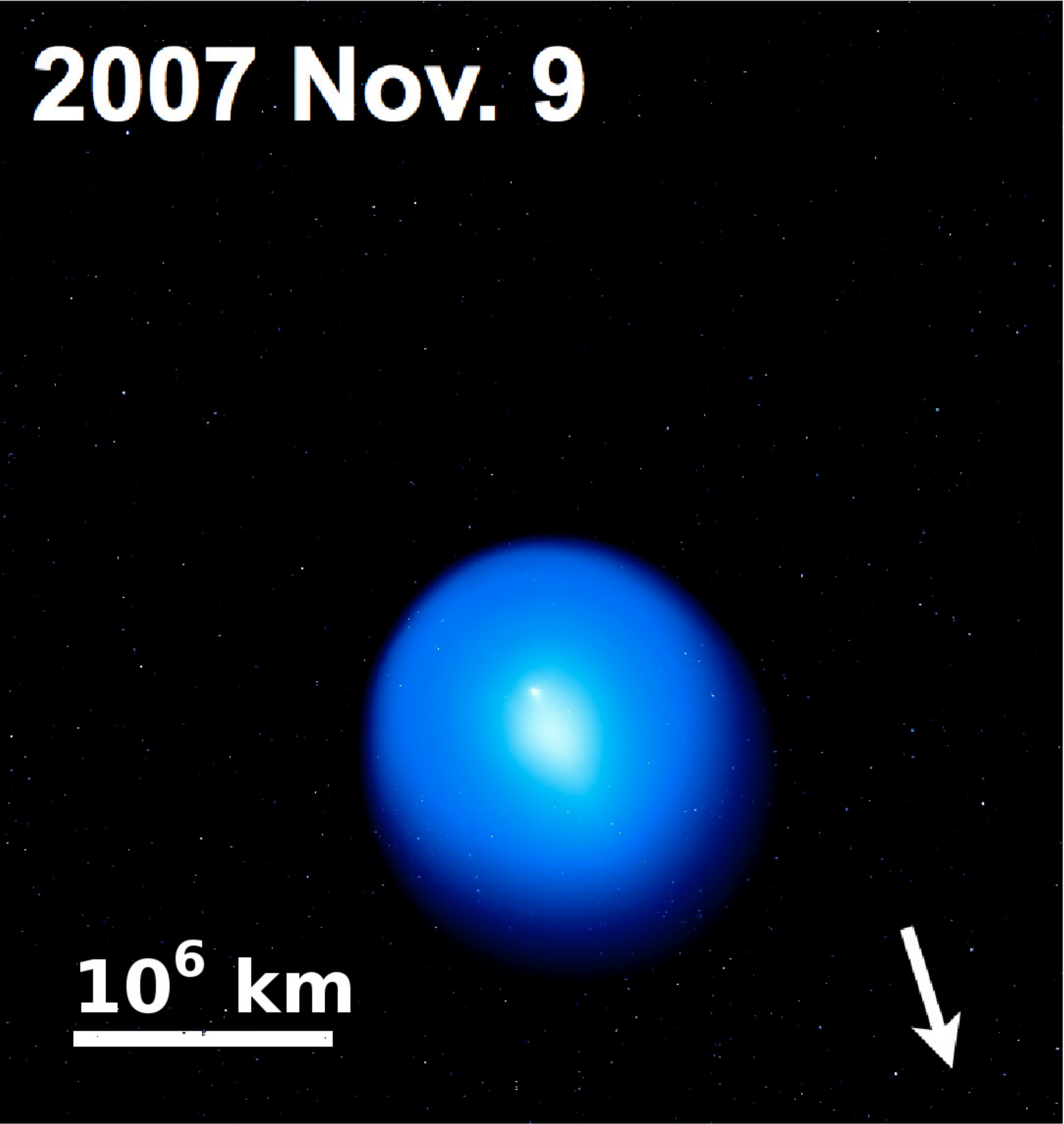}}
\hspace{0.5bp}
\subfloat{\includegraphics[width=0.2\textwidth]{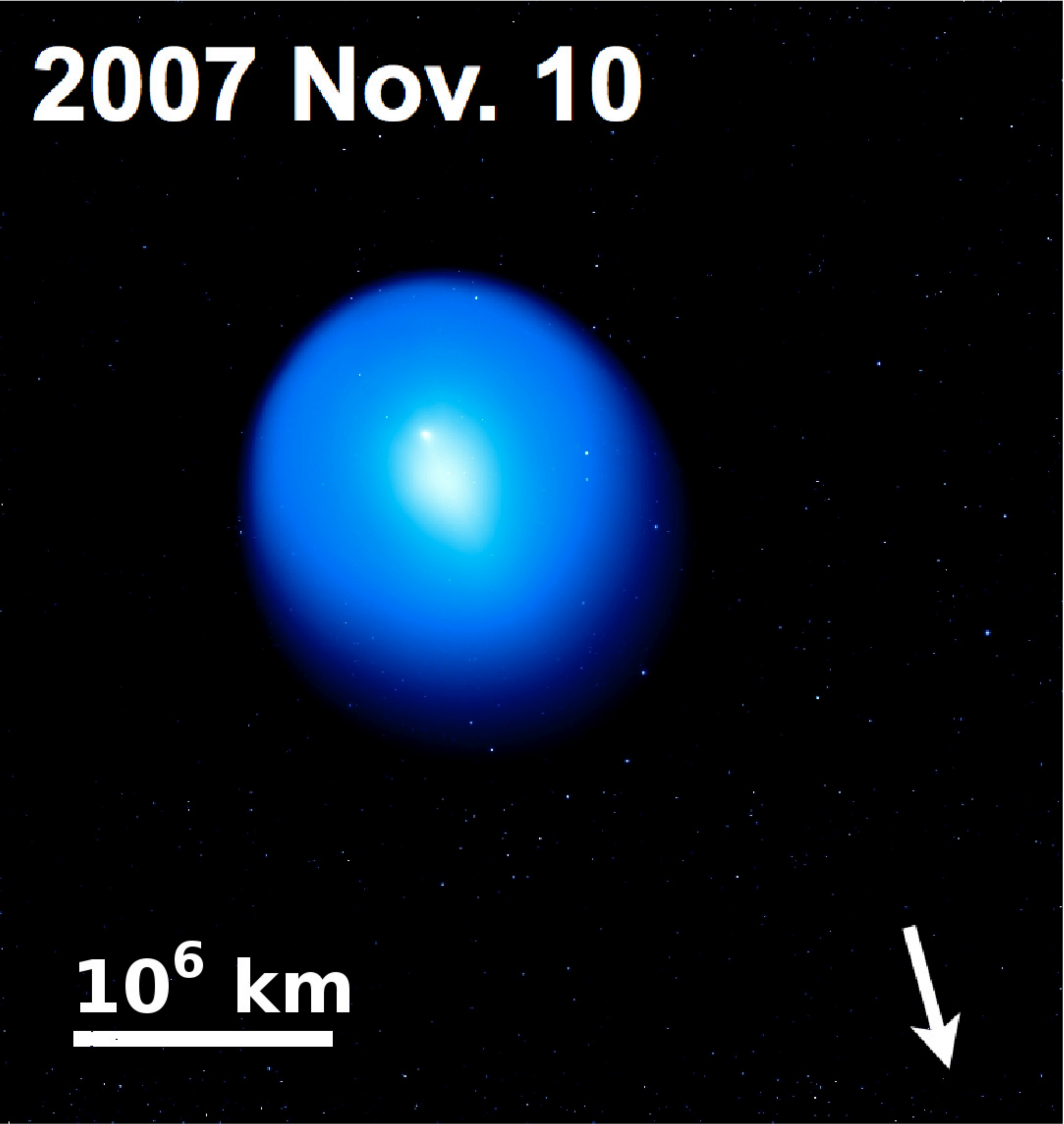}}
\vspace{2bp}
\subfloat{\includegraphics[width=0.2\textwidth]{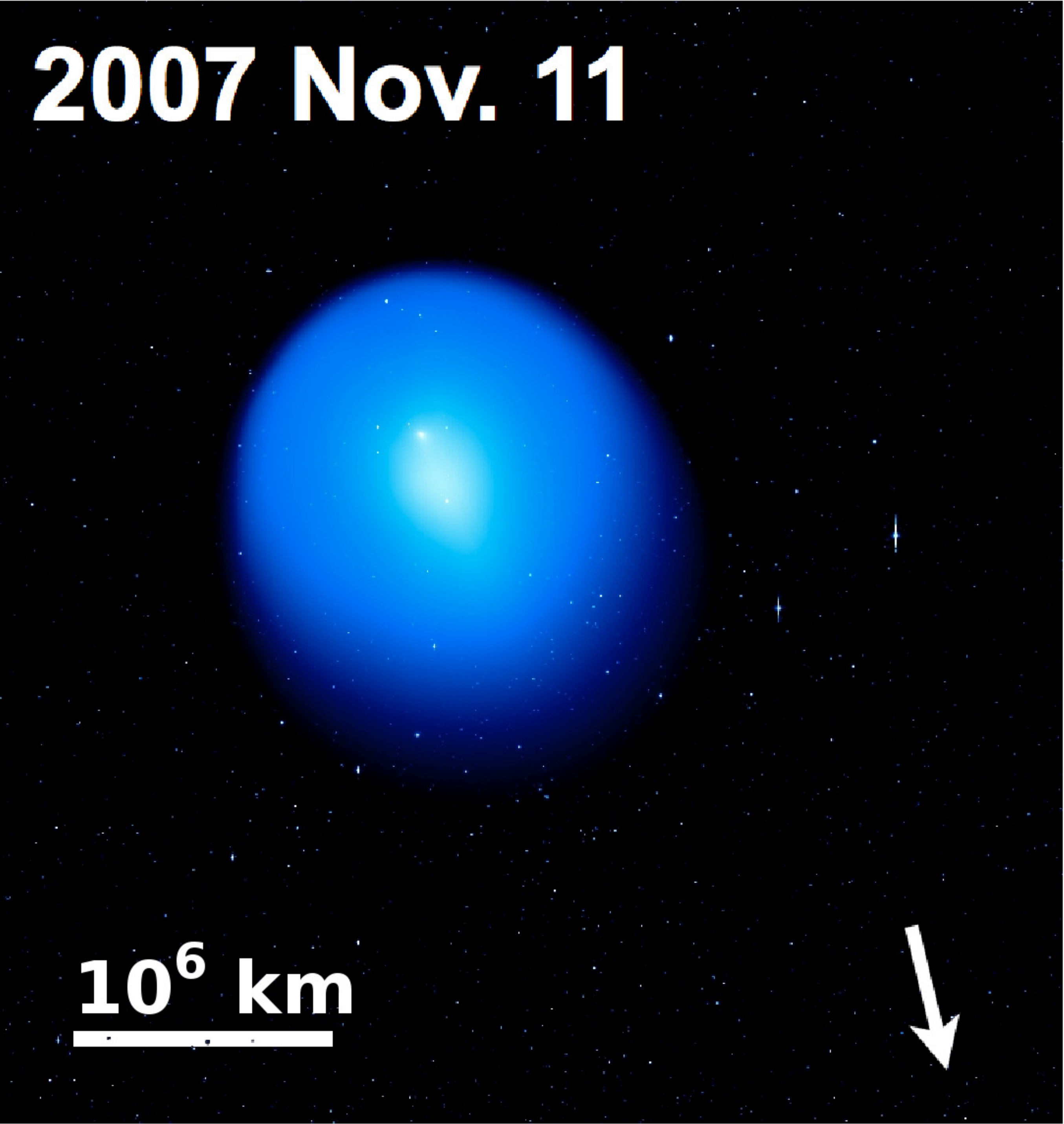}}
\hspace{0.5bp}
\subfloat{\includegraphics[width=0.2\textwidth]{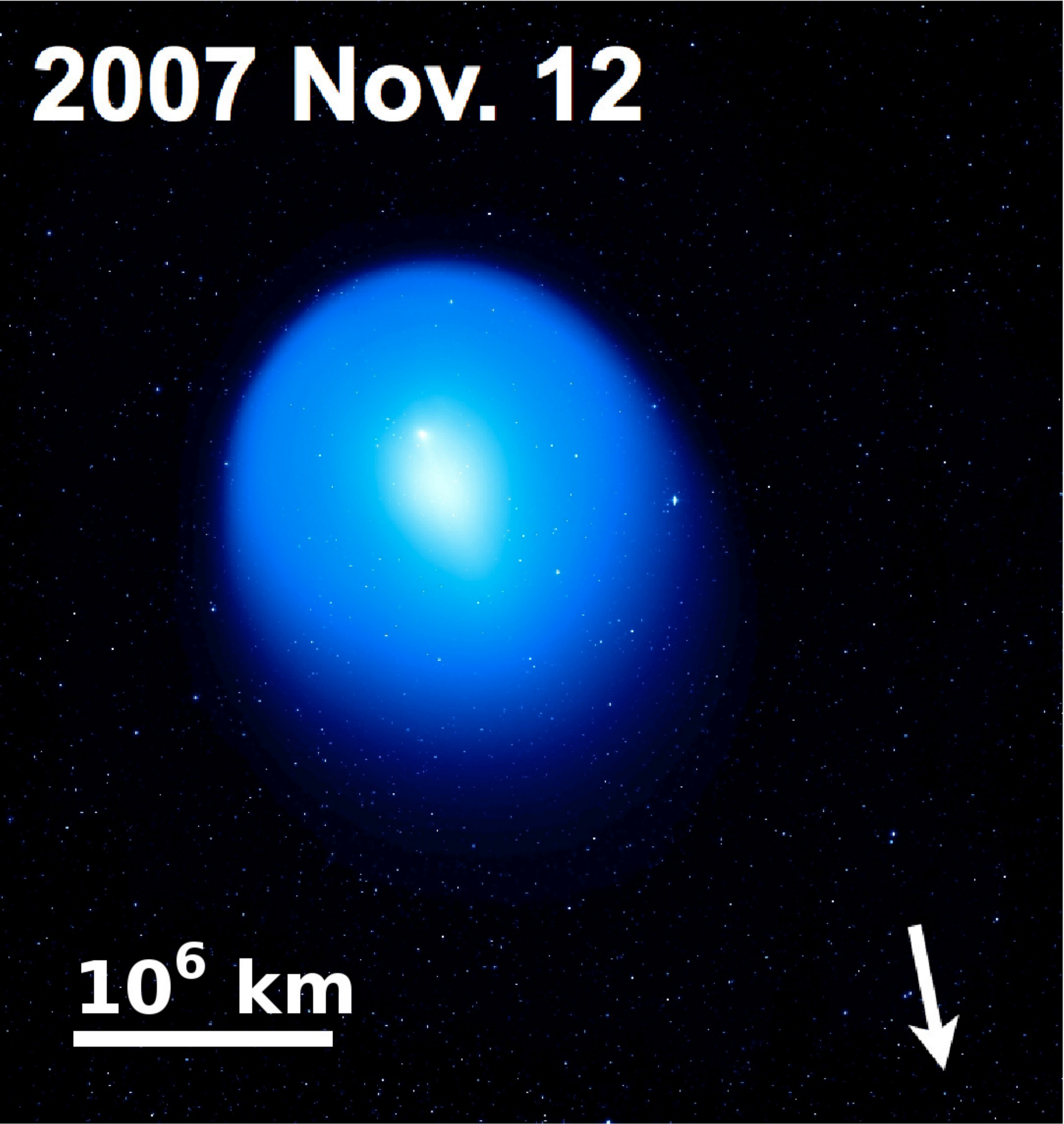}}
\hspace{0.5bp}
\subfloat{\includegraphics[width=0.2\textwidth]{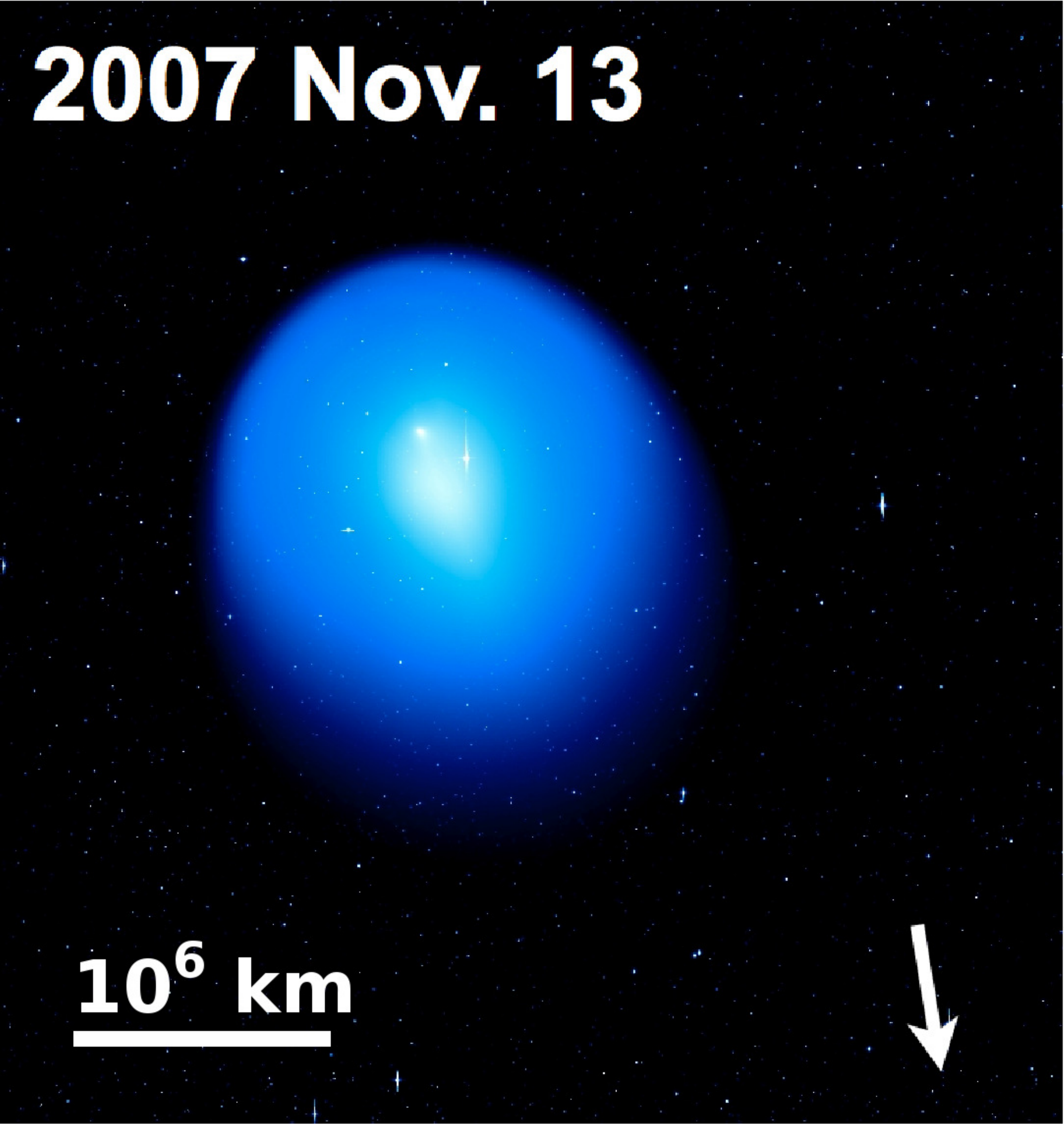}}
\hspace{0.5bp}
\subfloat{\includegraphics[width=0.2\textwidth]{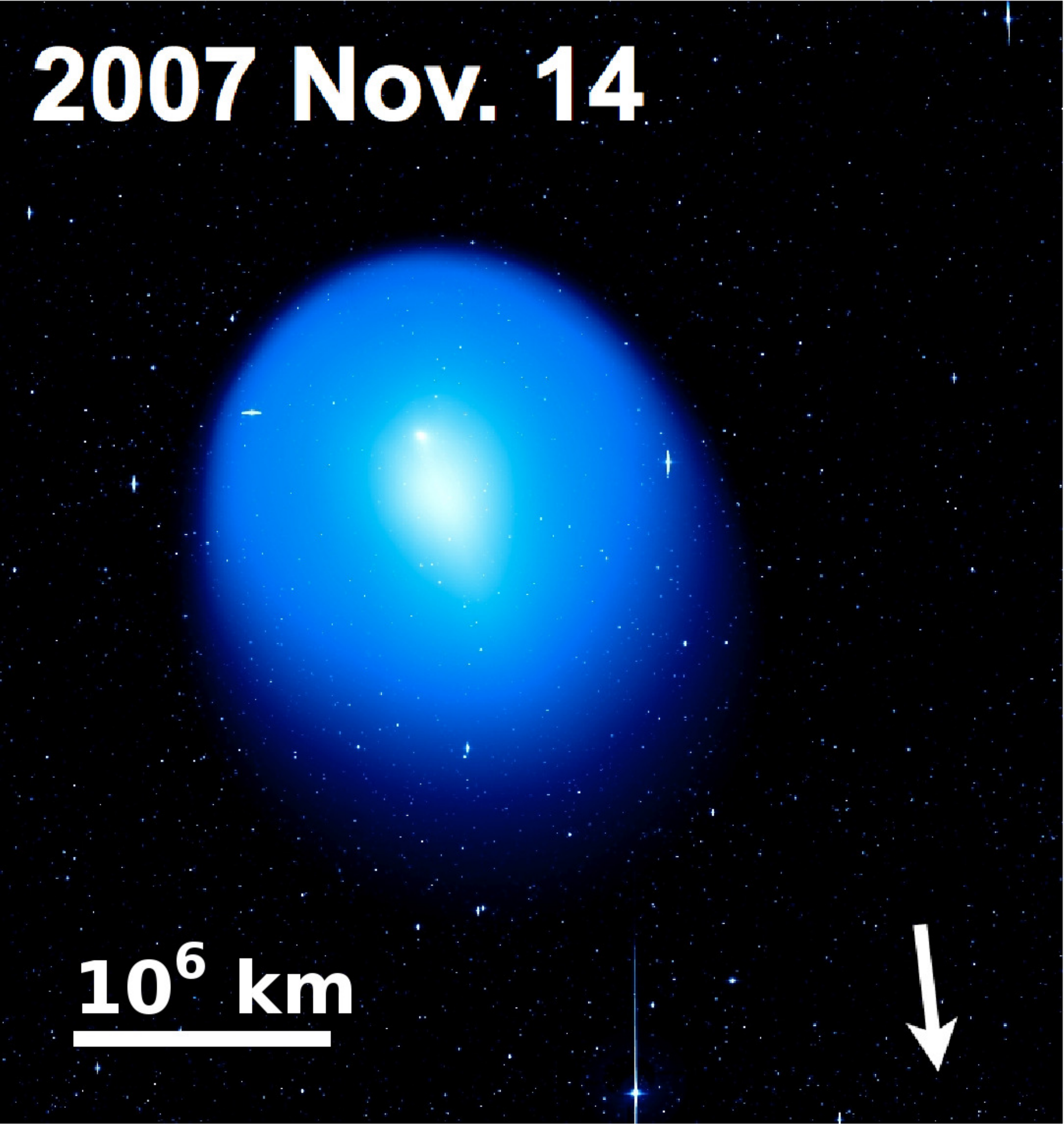}}
\vspace{2bp}
\subfloat{\includegraphics[width=0.2\textwidth]{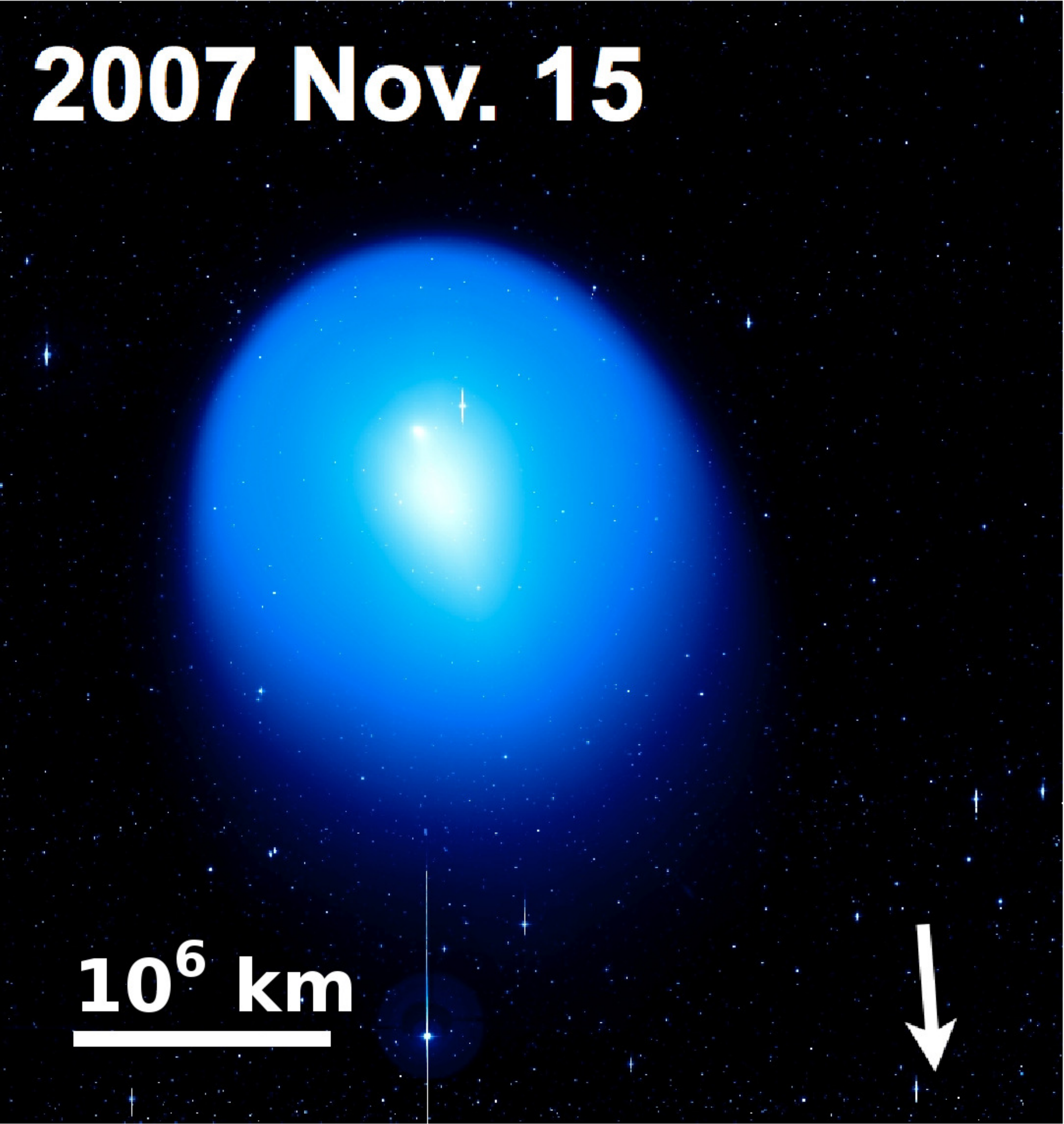}}
\hspace{0.5bp}
\subfloat{\includegraphics[width=0.2\textwidth]{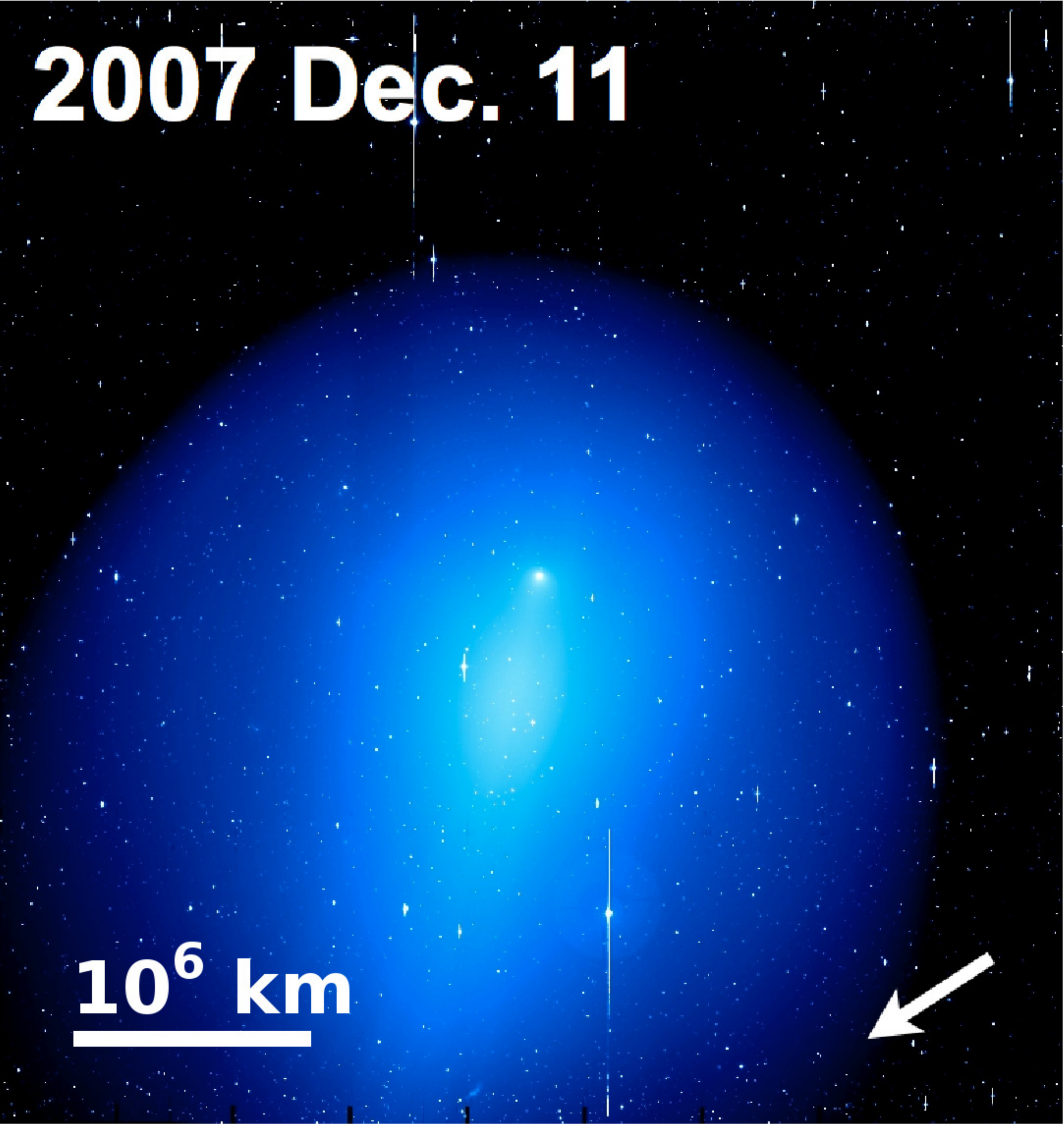}}
\hspace{0.5bp}
\subfloat{\includegraphics[width=0.2\textwidth]{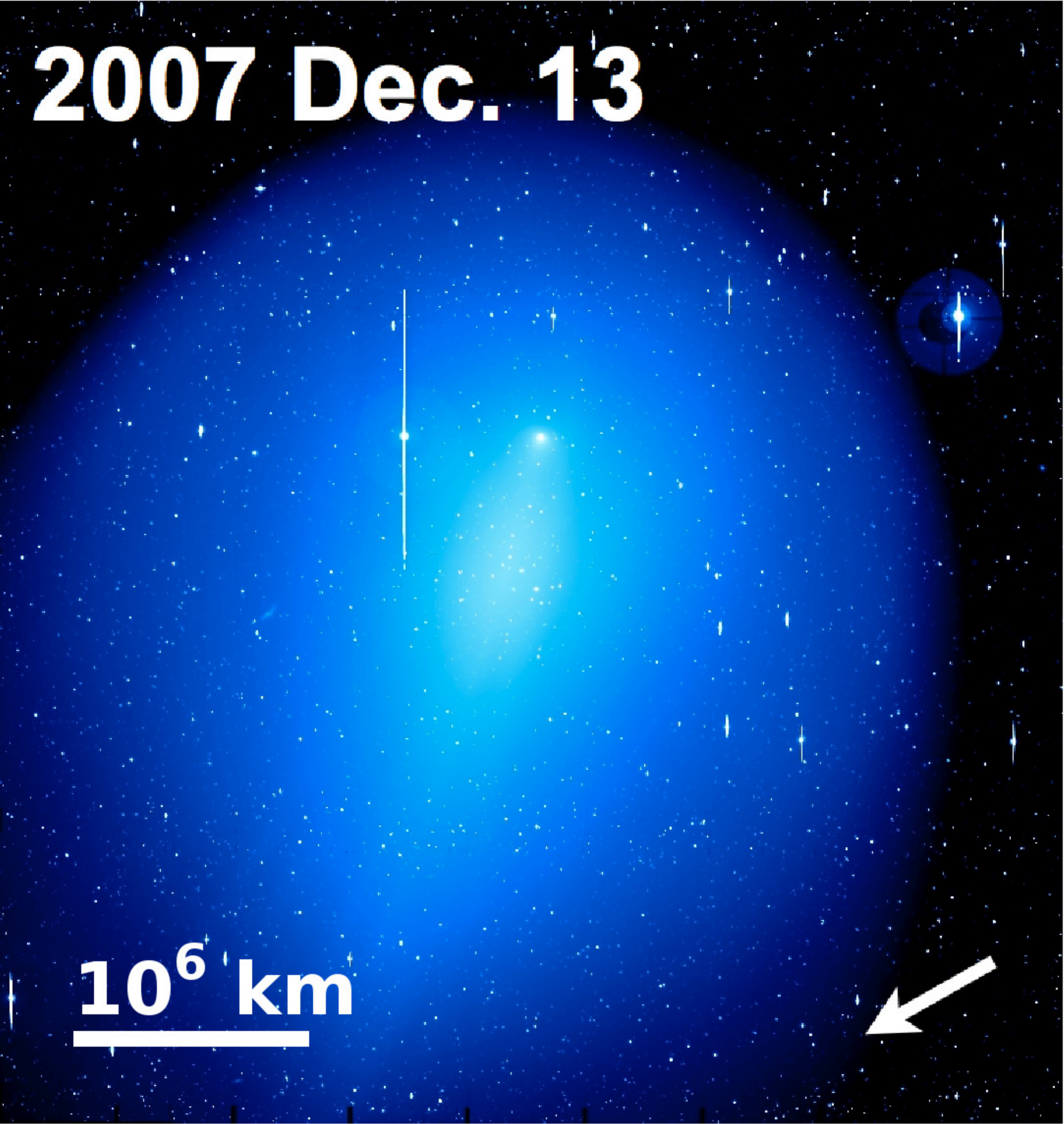}}
\hspace{0.5bp}
\subfloat{\includegraphics[width=0.2\textwidth]{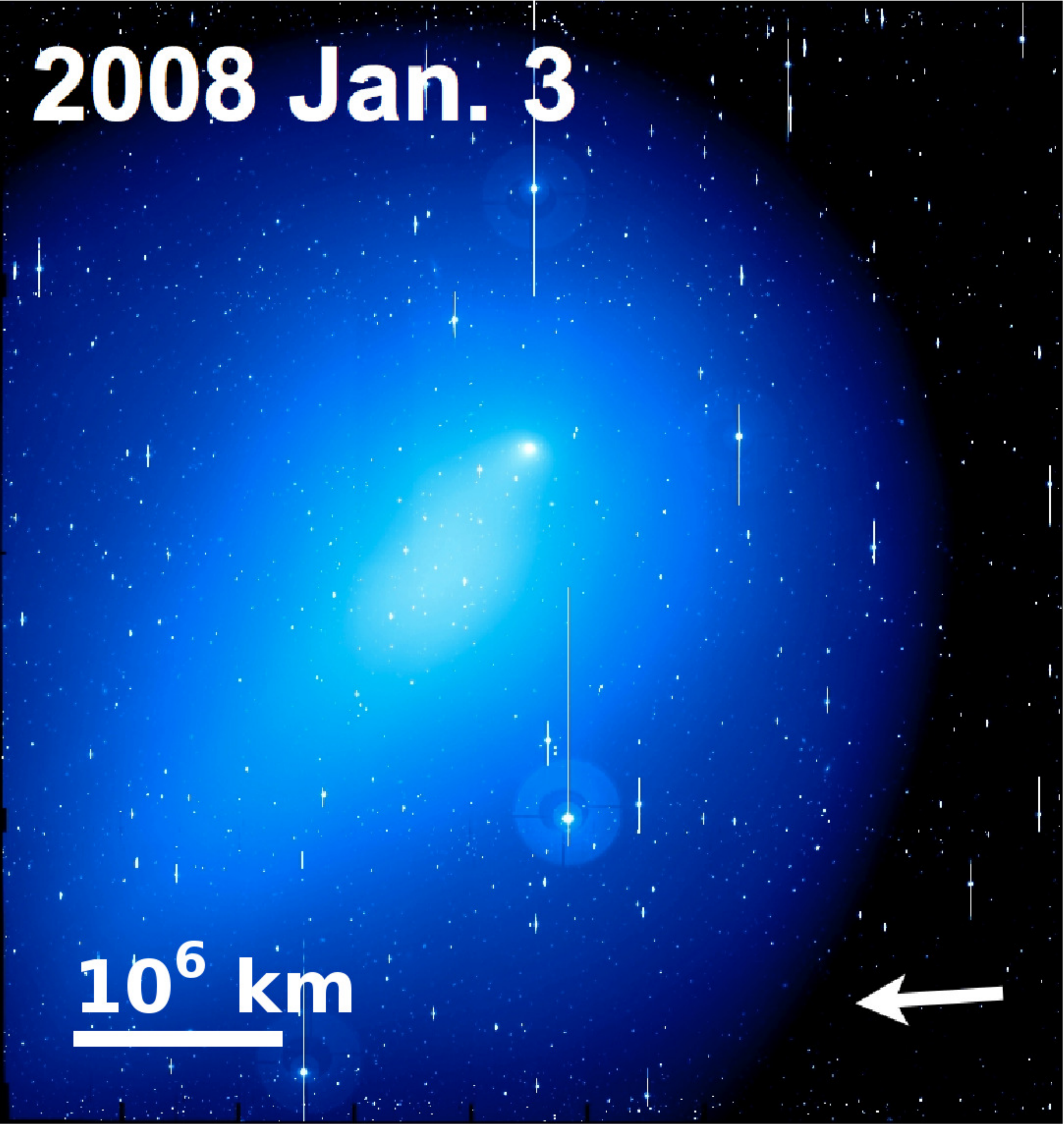}}
\vspace{2bp}
\subfloat{\includegraphics[width=0.2\textwidth]{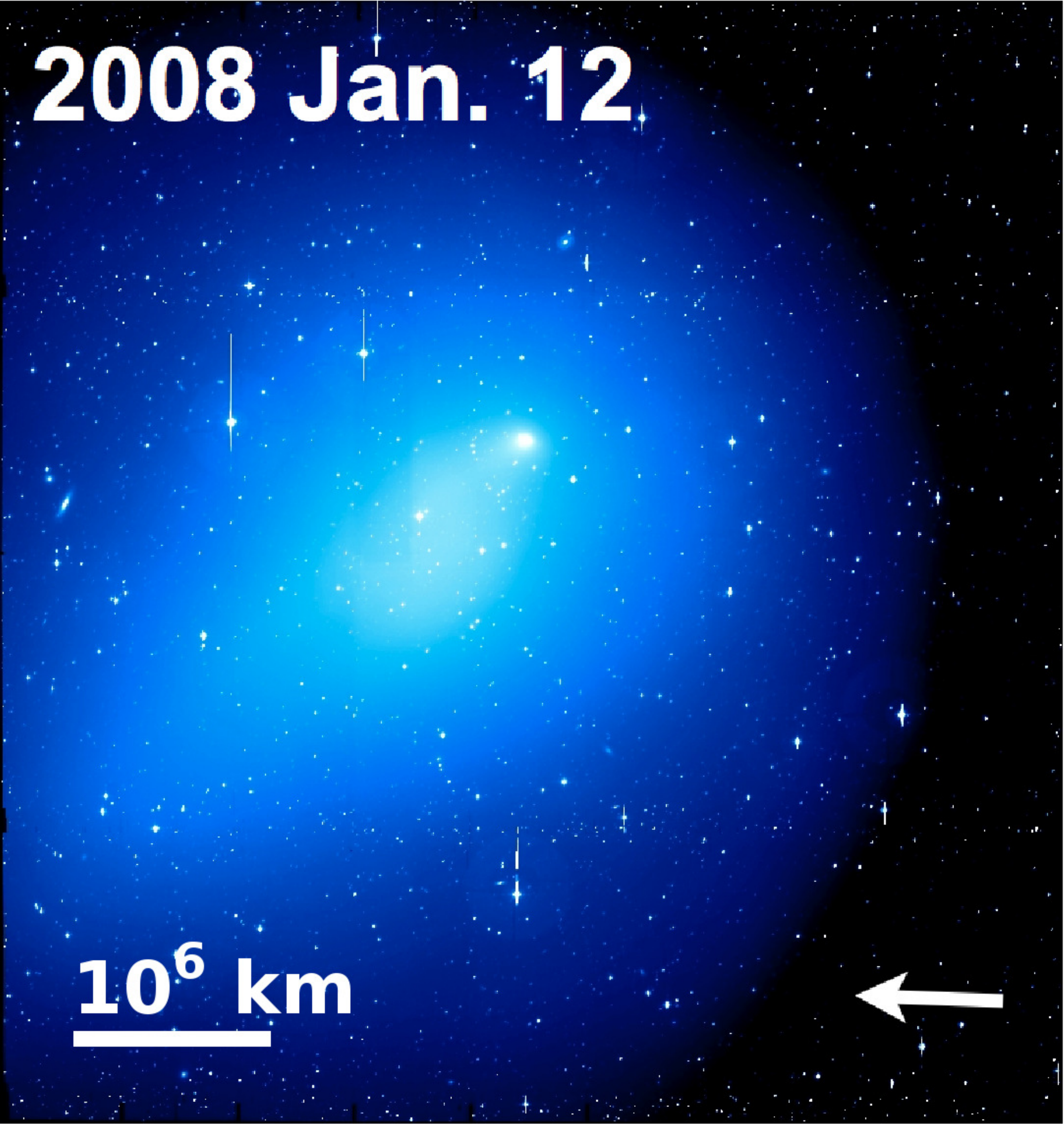}}
\hspace{0.5bp}
\subfloat{\includegraphics[width=0.2\textwidth]{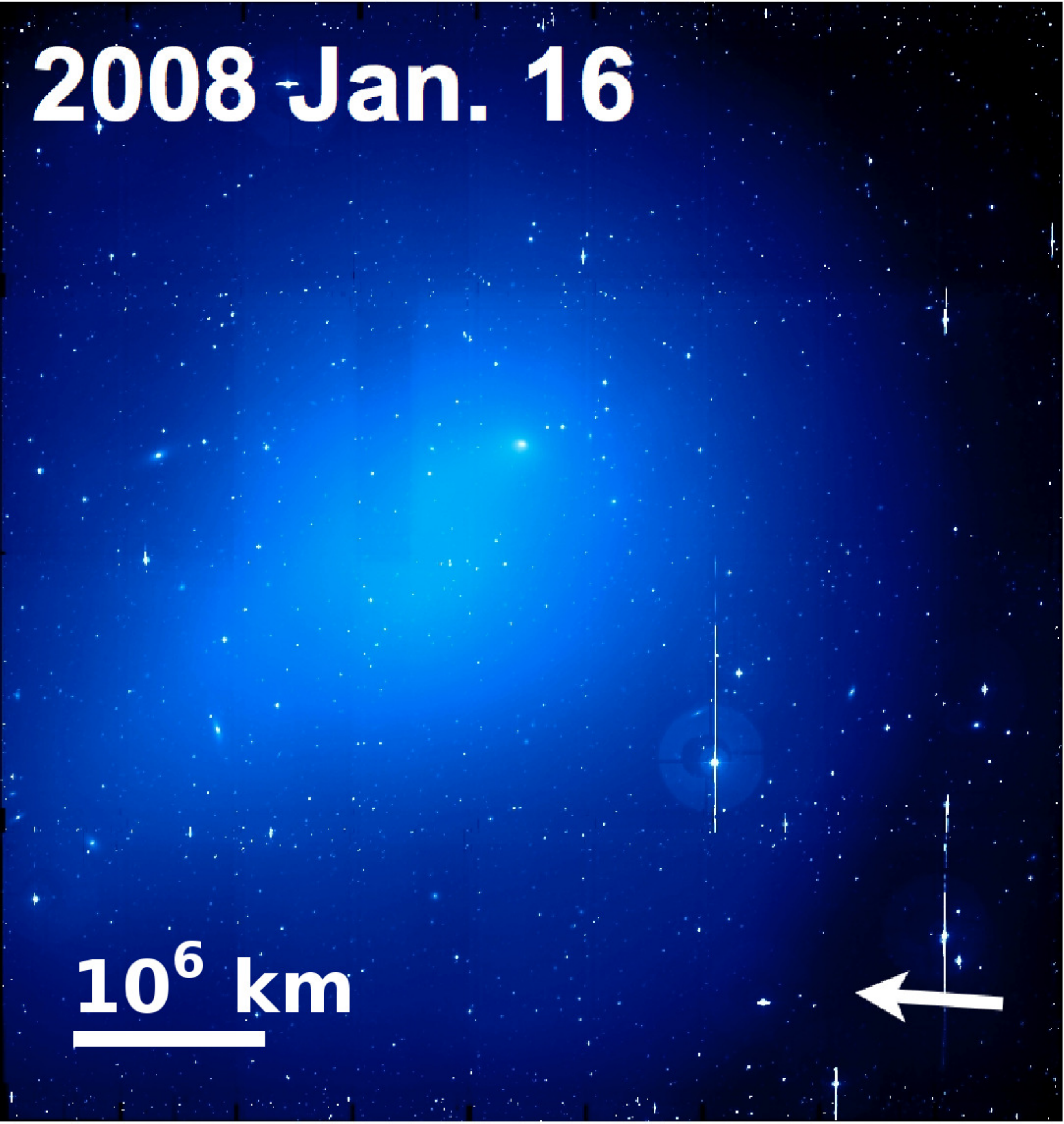}}
\hspace{0.5bp}
\subfloat{\includegraphics[width=0.2\textwidth]{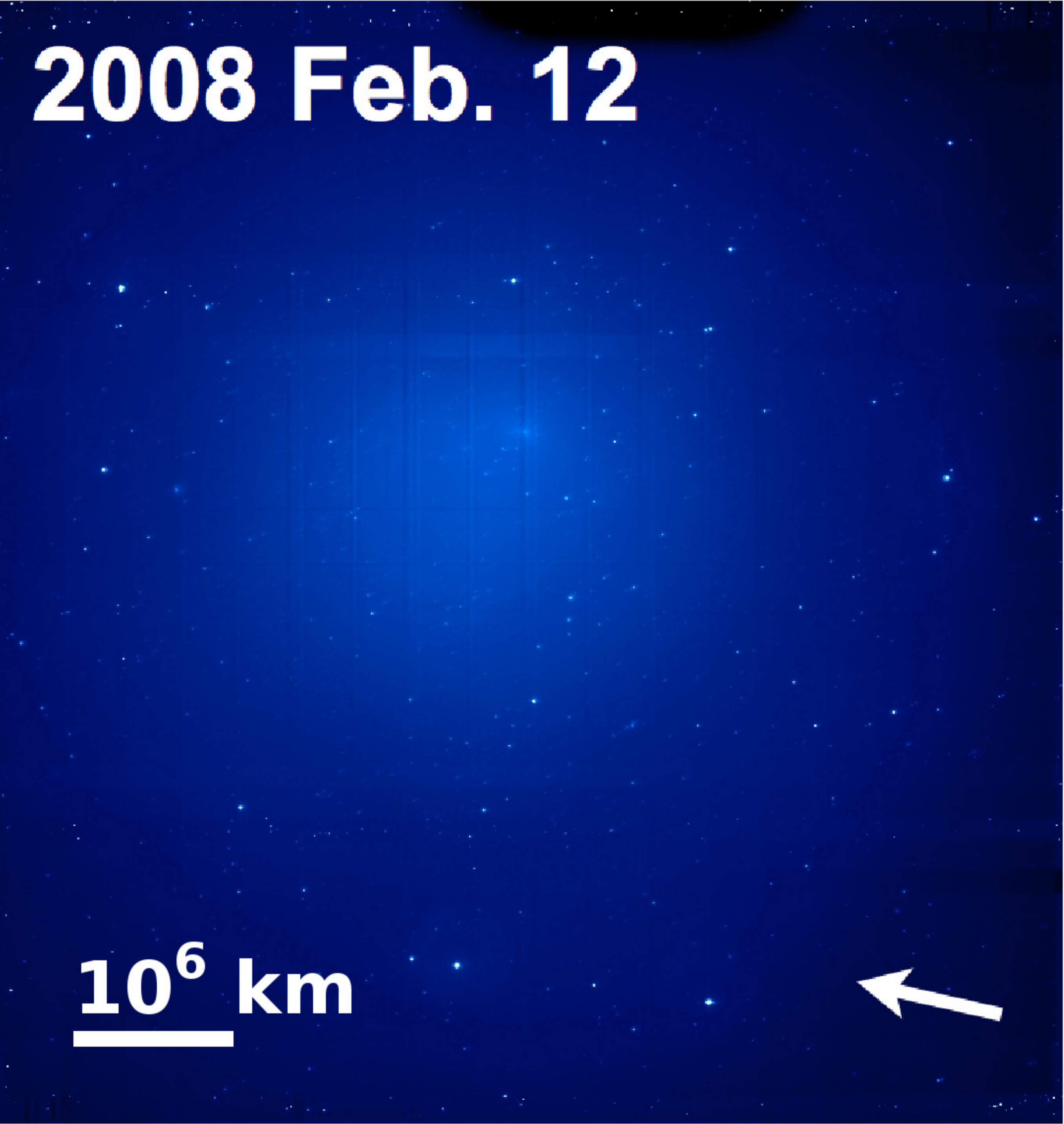}}
\caption{The full MegaCam field of view is shown for each observation of 17P/Holmes.  Each image is oriented with north up, east left.  The anti-solar vector, as projected on the plane of the sky, is denoted by a white arrow.  The final image is slightly larger due to the use of a larger dithering patten.  As the coma surrounding the nucleus expands the surface brightness drops and the coma becomes more diffuse.}
\label{fig:fullFoV}
\end{figure}

\begin{figure}
\centering
\subfloat{\includegraphics[width=0.2\textwidth]{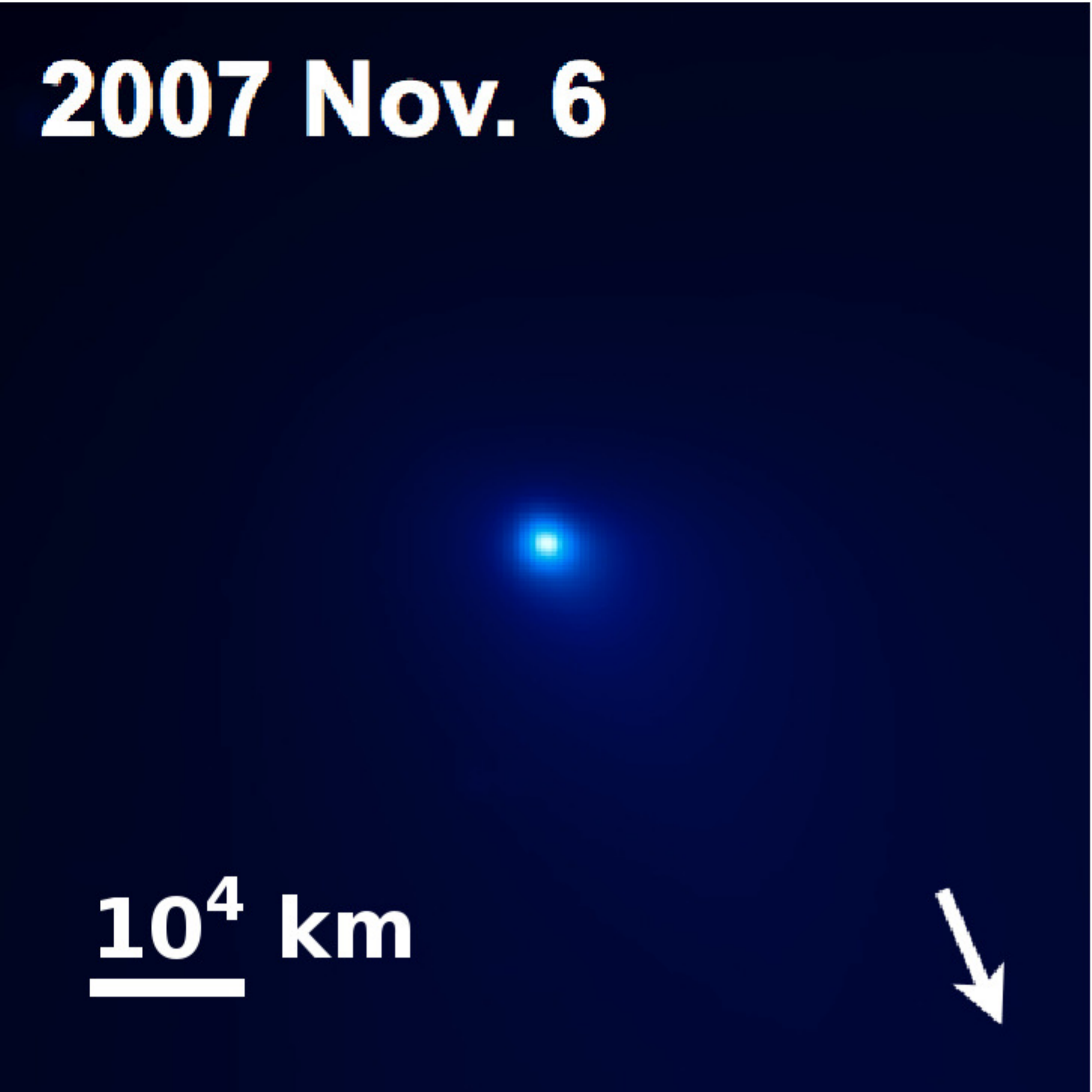}}
\hspace{2bp}
\subfloat{\includegraphics[width=0.2\textwidth]{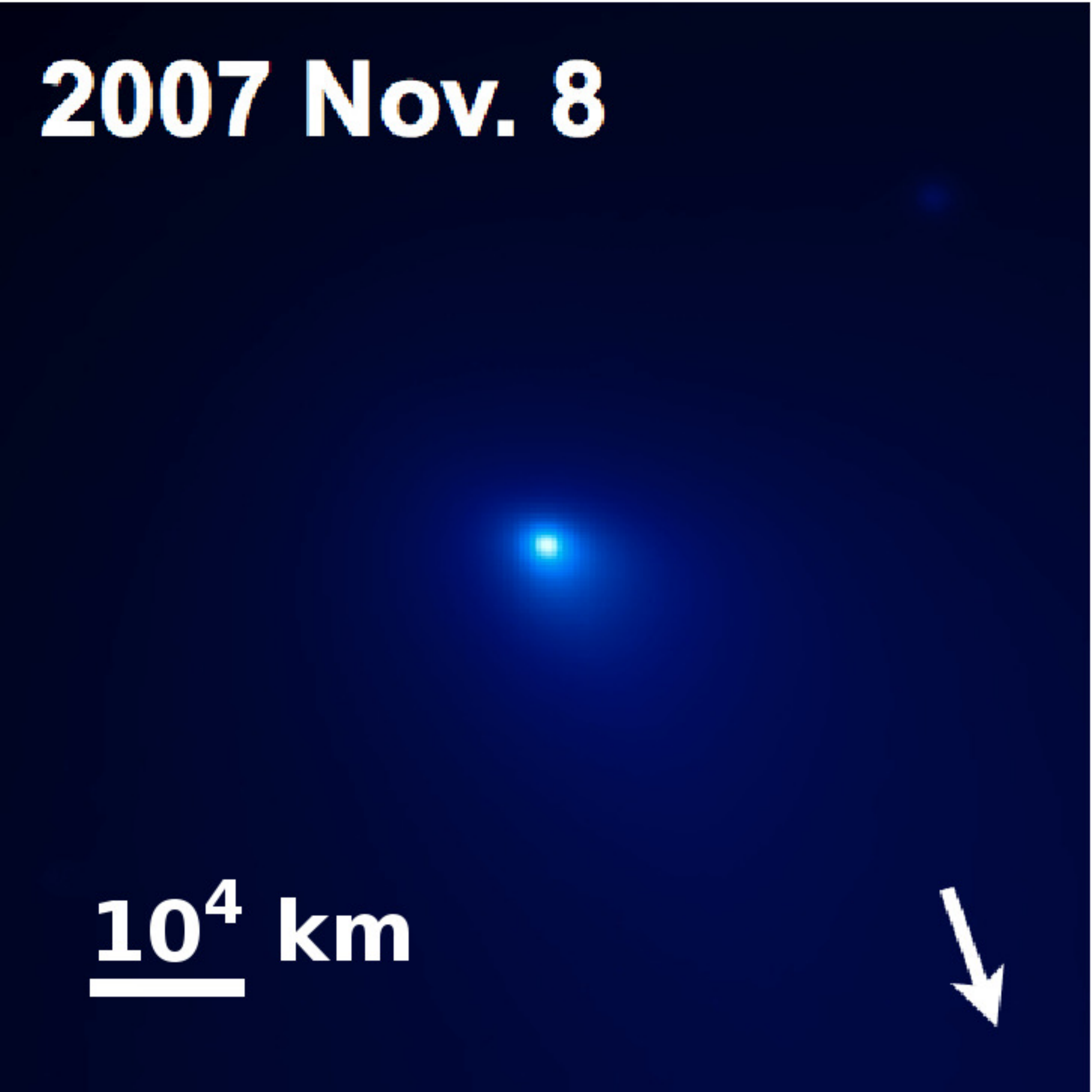}}
\hspace{2bp}
\subfloat{\includegraphics[width=0.2\textwidth]{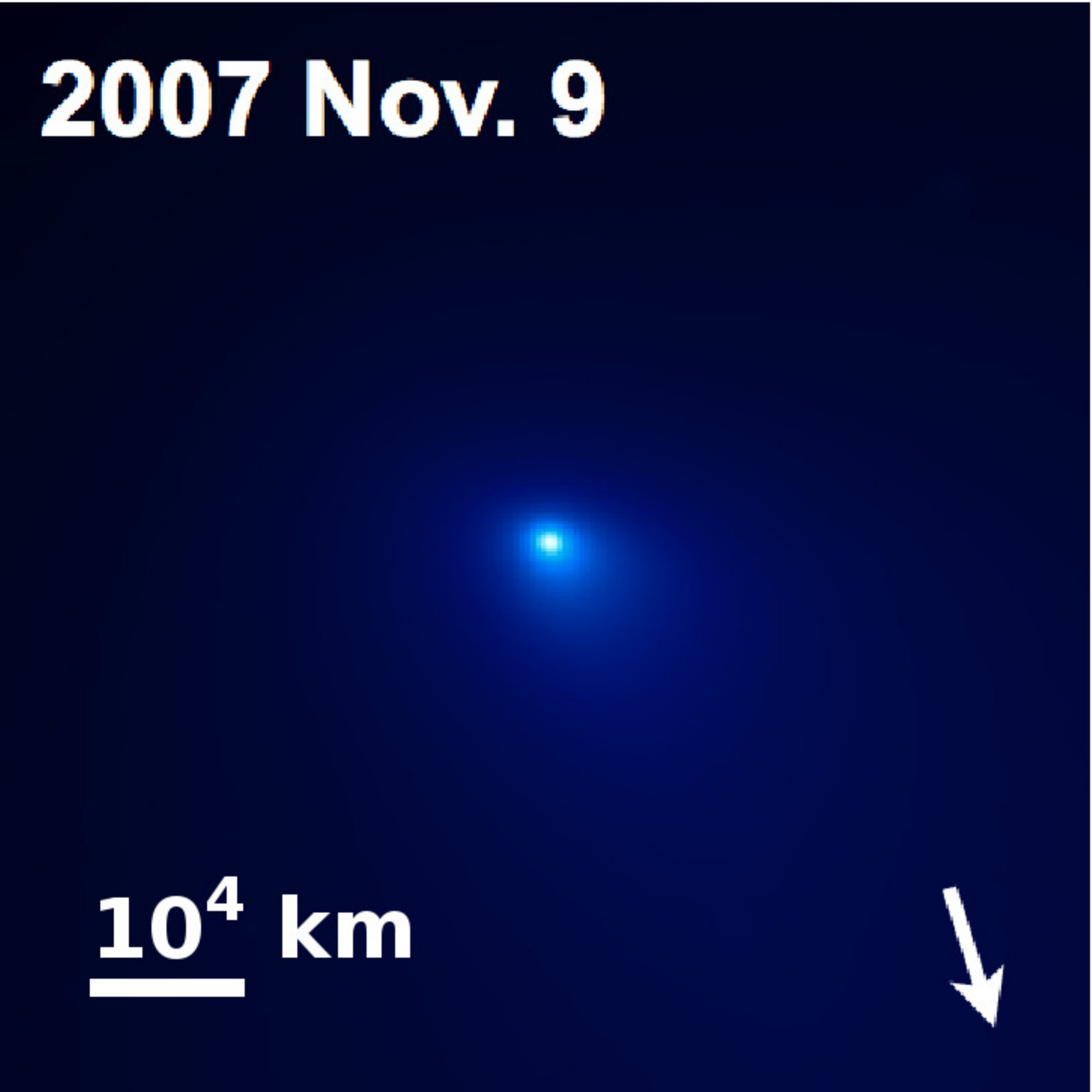}}
\hspace{2bp}\
\subfloat{\includegraphics[width=0.2\textwidth]{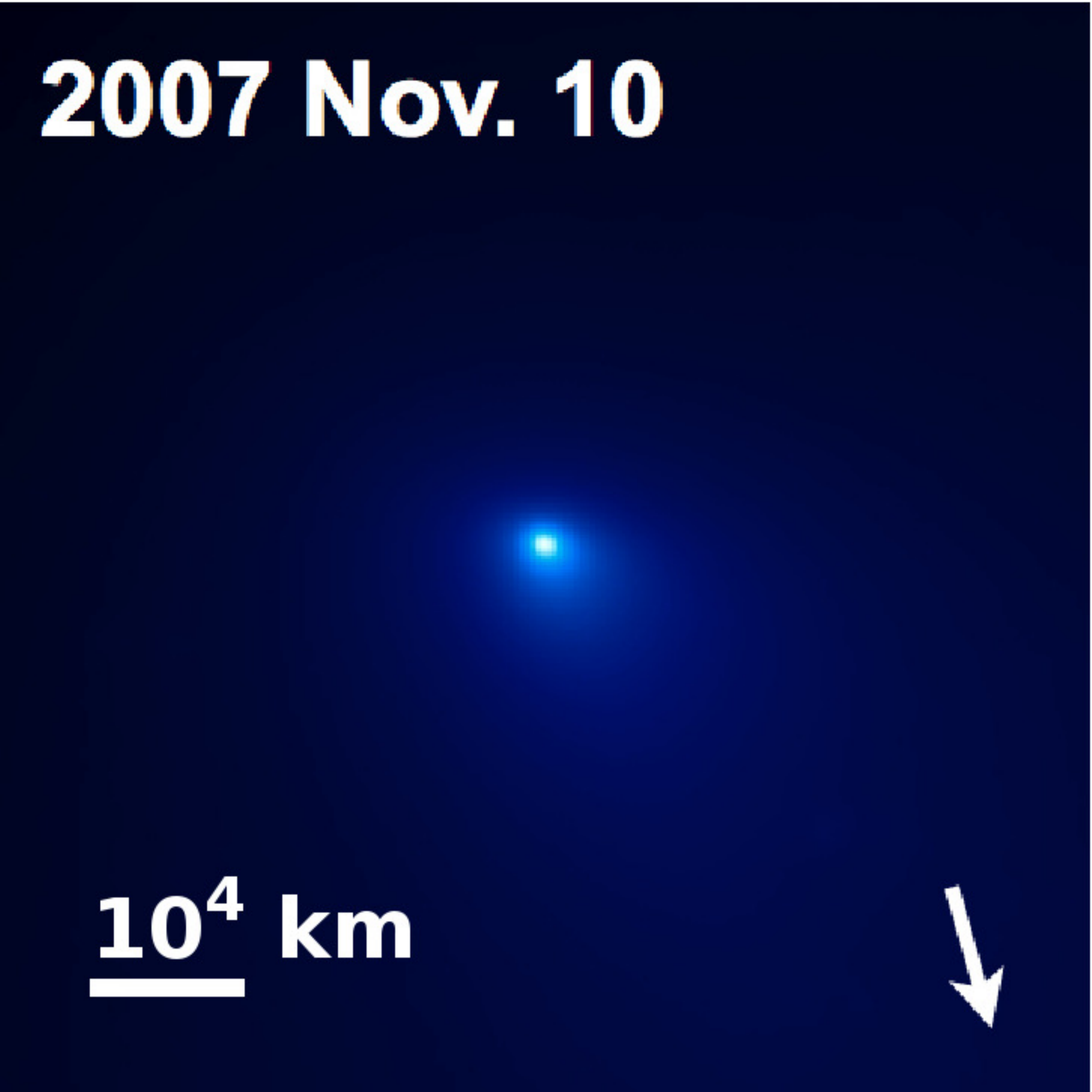}}
\vspace{2bp}
\subfloat{\includegraphics[width=0.2\textwidth]{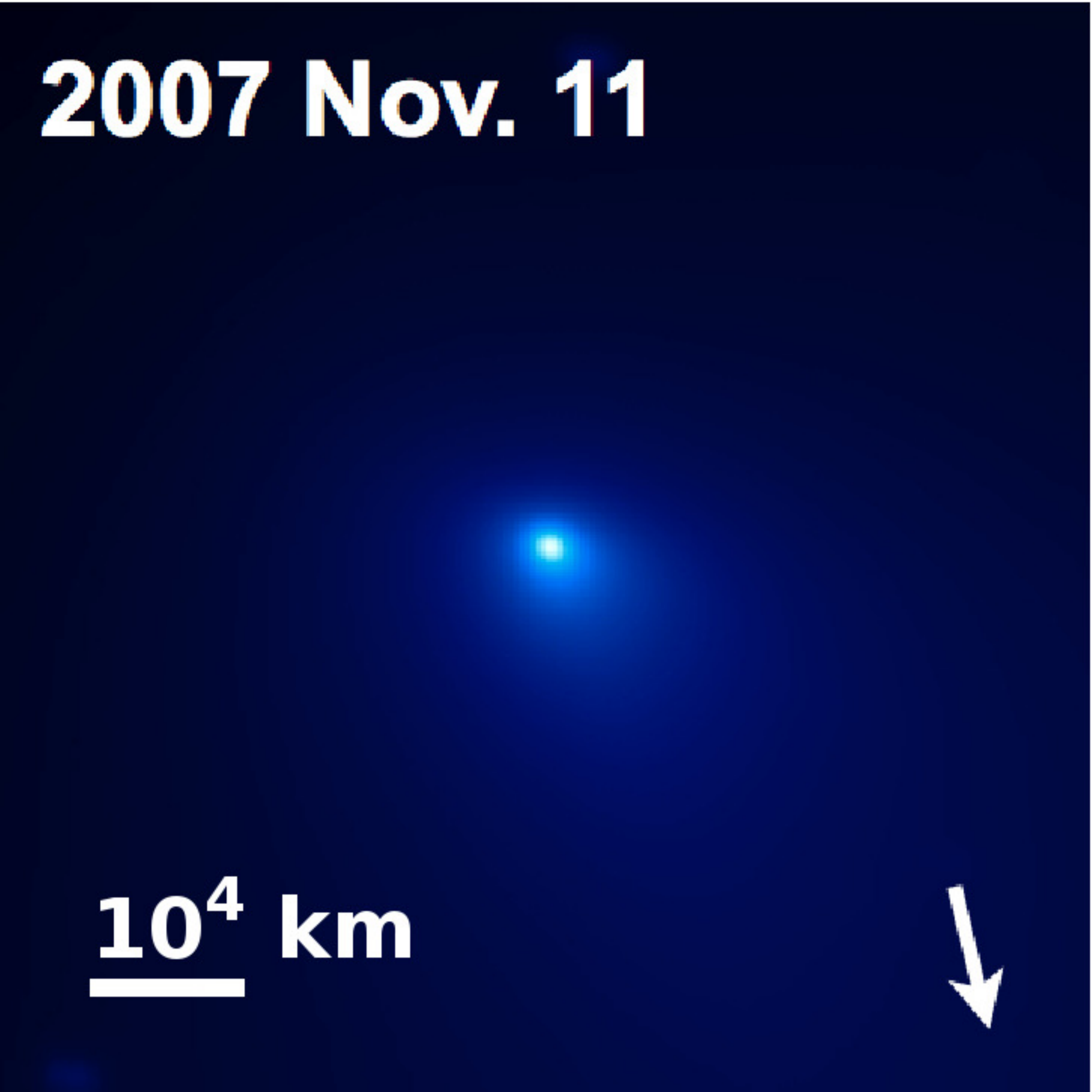}}
\hspace{2bp}
\subfloat{\includegraphics[width=0.2\textwidth]{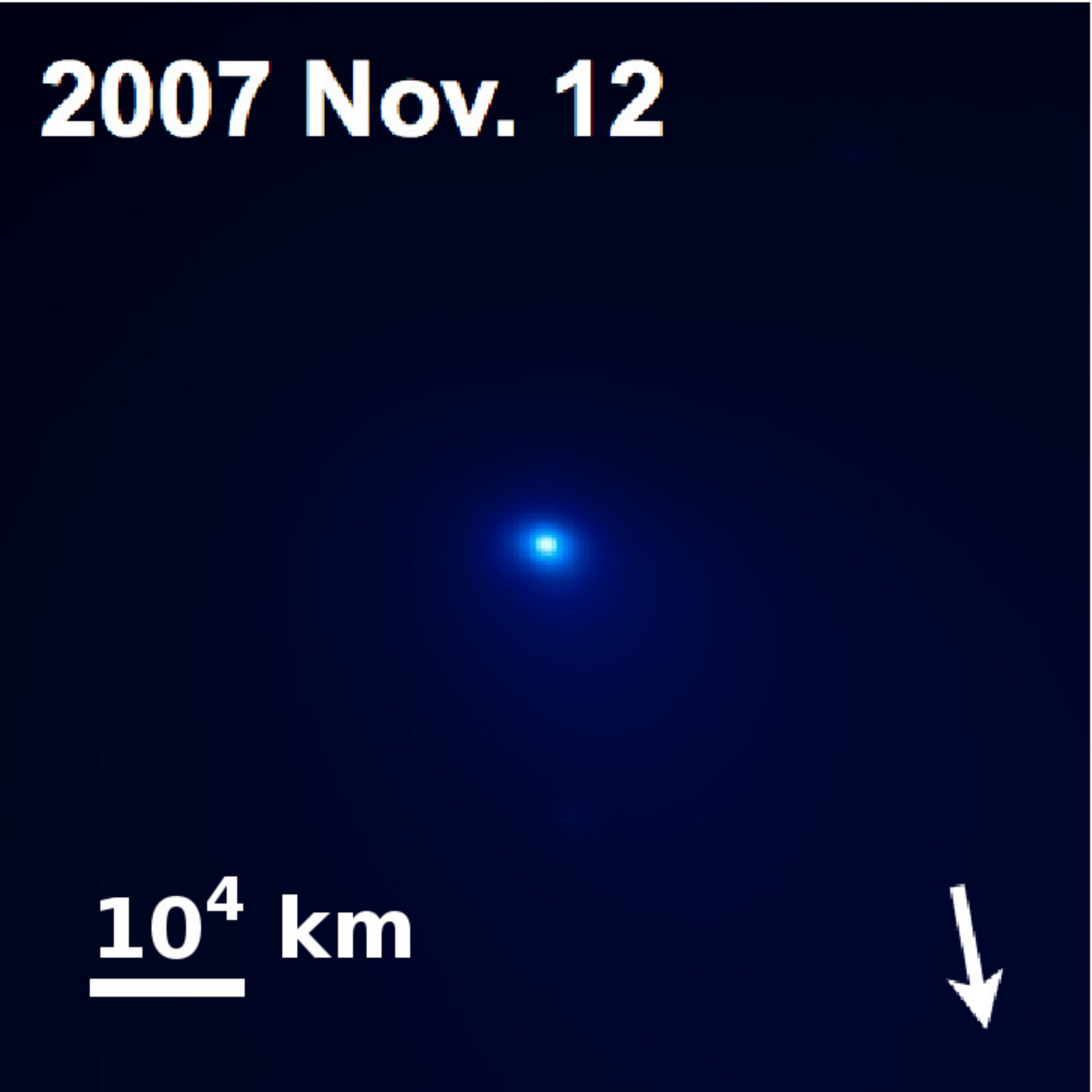}}
\hspace{2bp}
\subfloat{\includegraphics[width=0.2\textwidth]{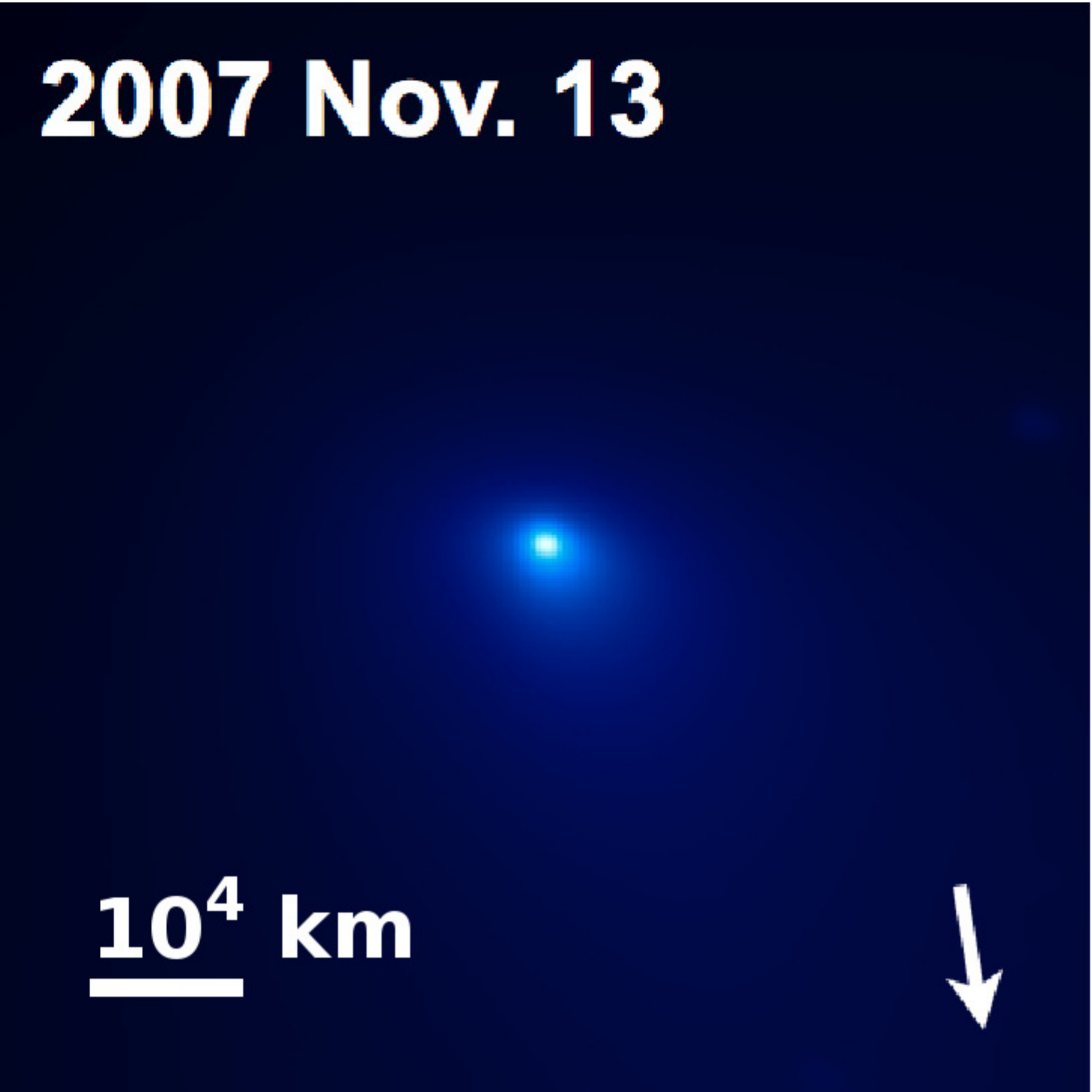}}
\hspace{2bp}
\subfloat{\includegraphics[width=0.2\textwidth]{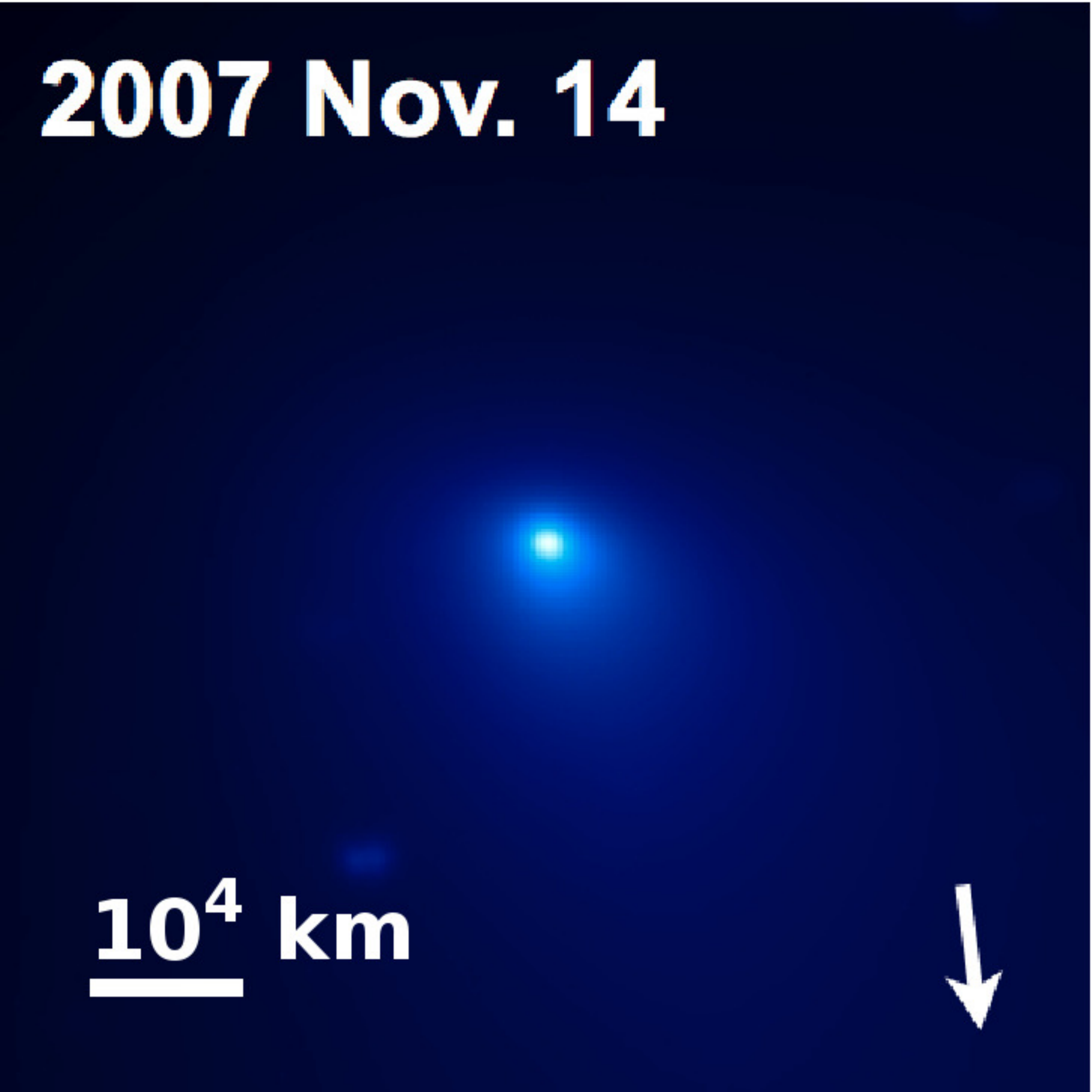}}
\vspace{2bp}
\subfloat{\includegraphics[width=0.2\textwidth]{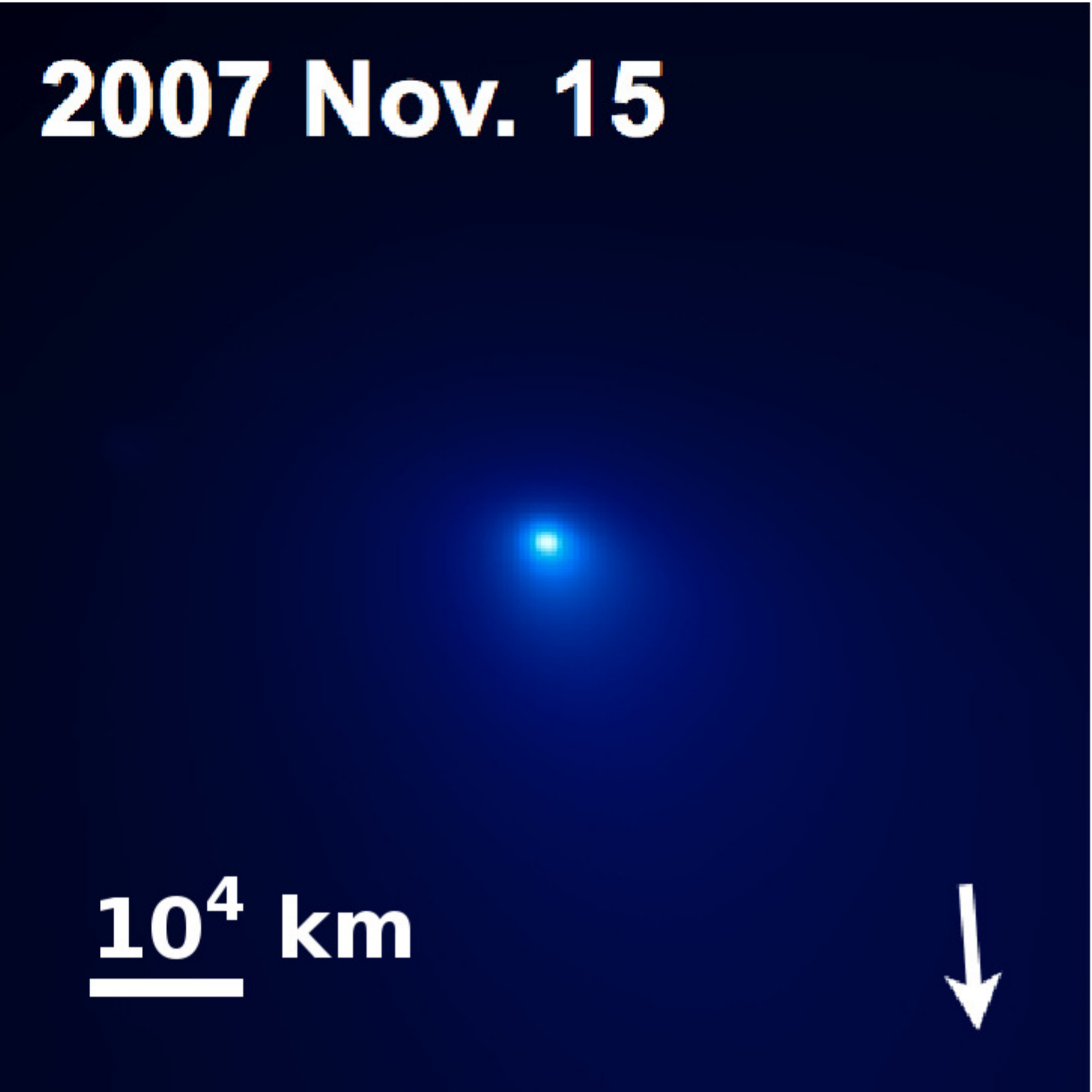}}
\hspace{2bp}
\subfloat{\includegraphics[width=0.2\textwidth]{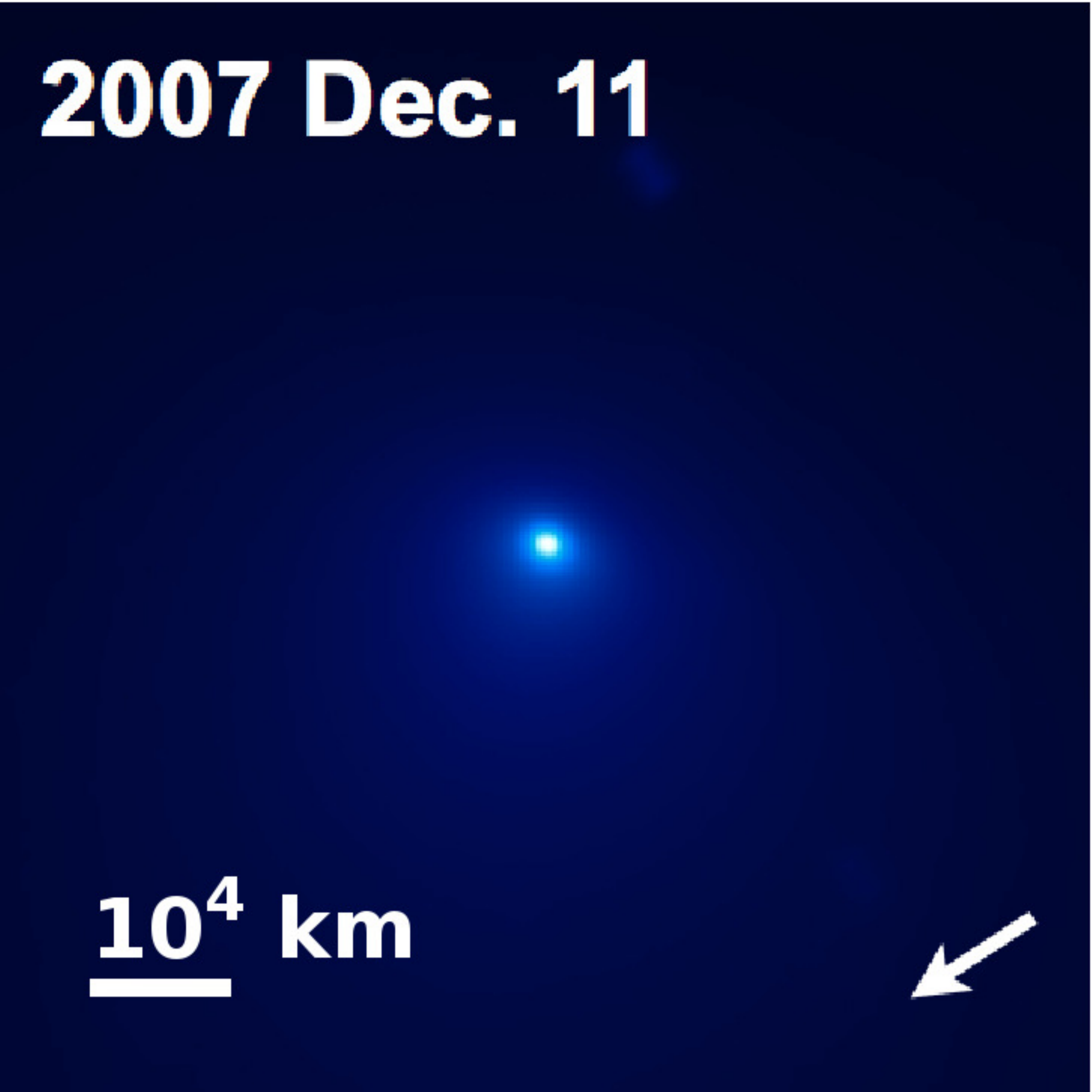}}
\hspace{2bp}
\subfloat{\includegraphics[width=0.2\textwidth]{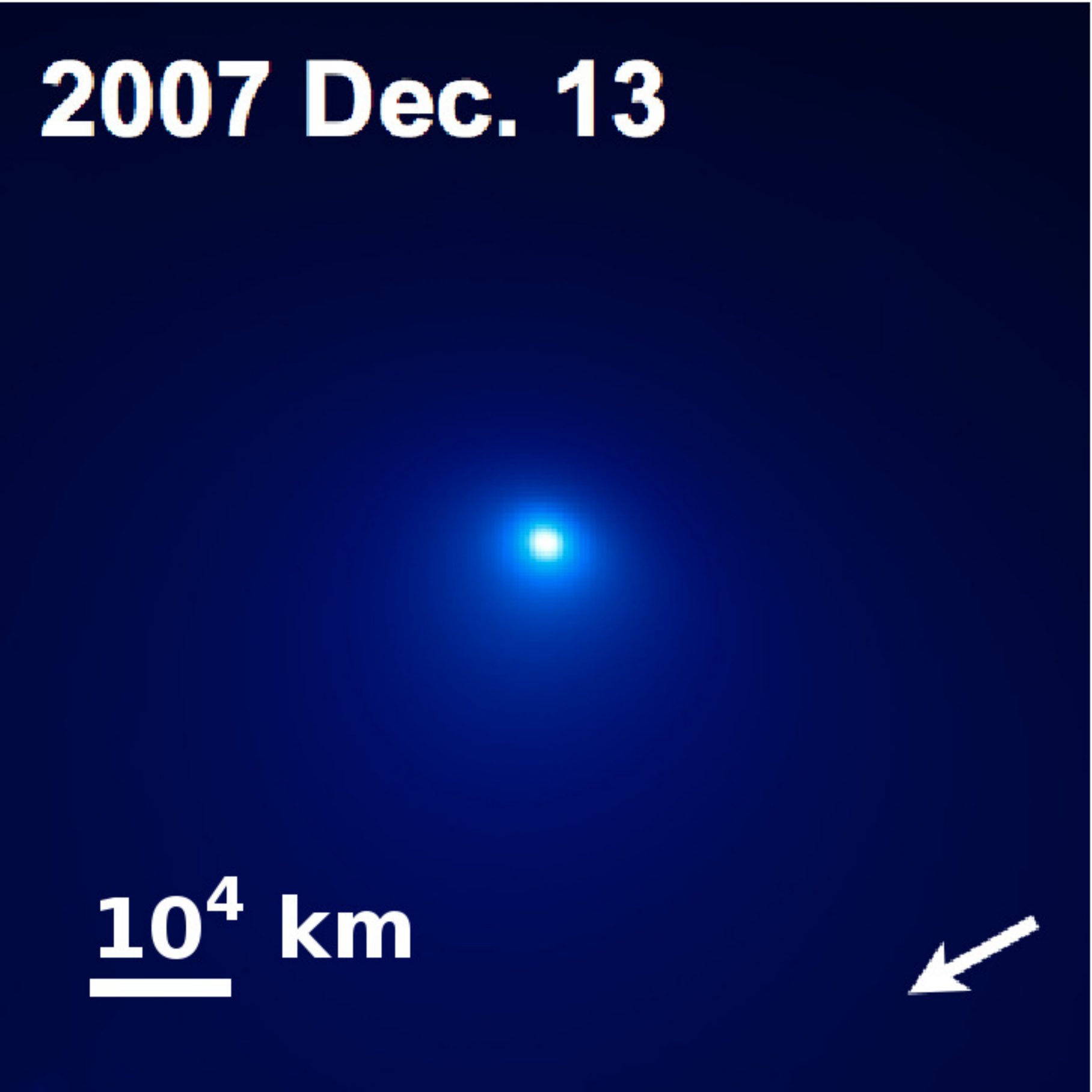}}
\hspace{2bp}
\subfloat{\includegraphics[width=0.2\textwidth]{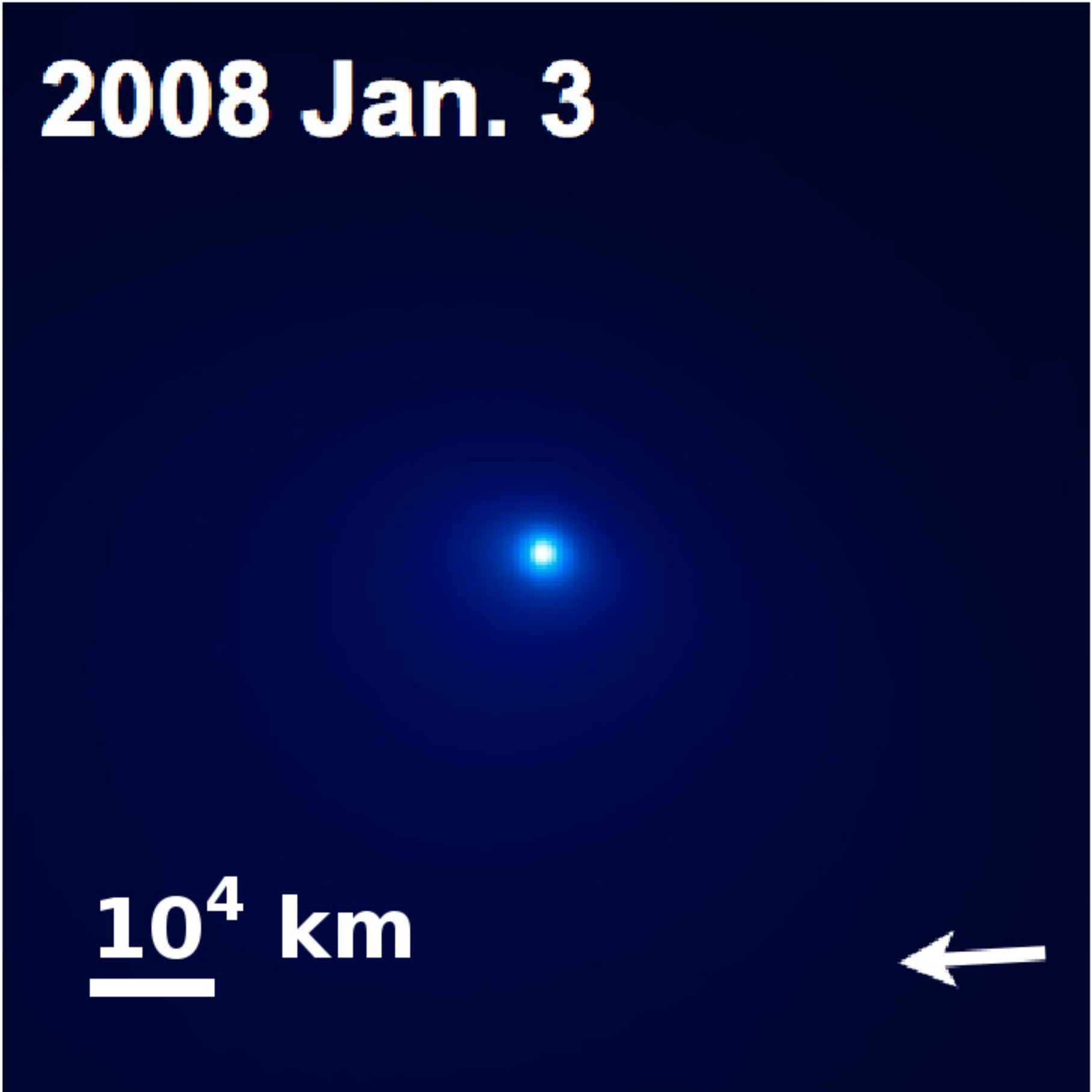}}
\vspace{2bp}
\subfloat{\includegraphics[width=0.2\textwidth]{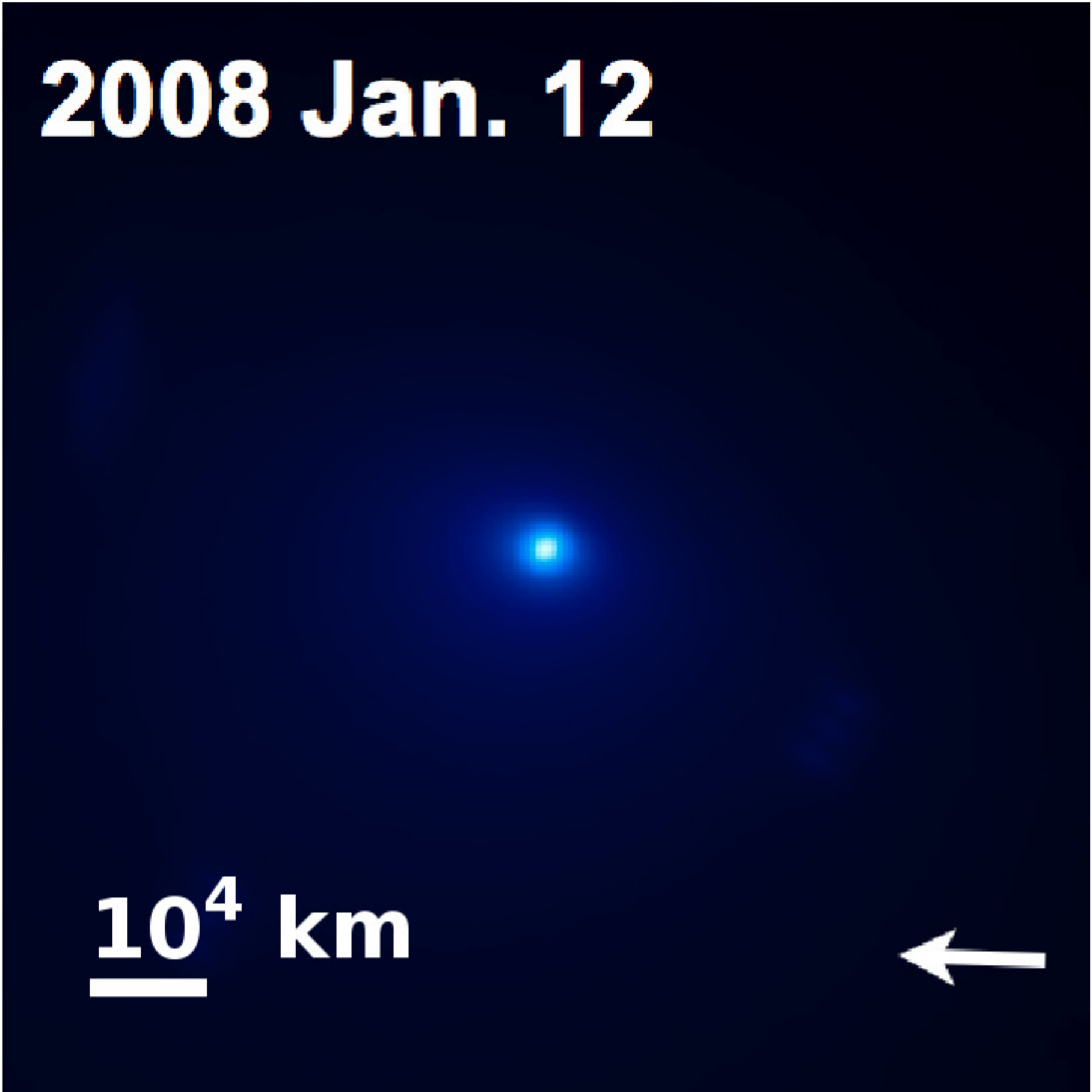}}
\hspace{2bp}
\subfloat{\includegraphics[width=0.2\textwidth]{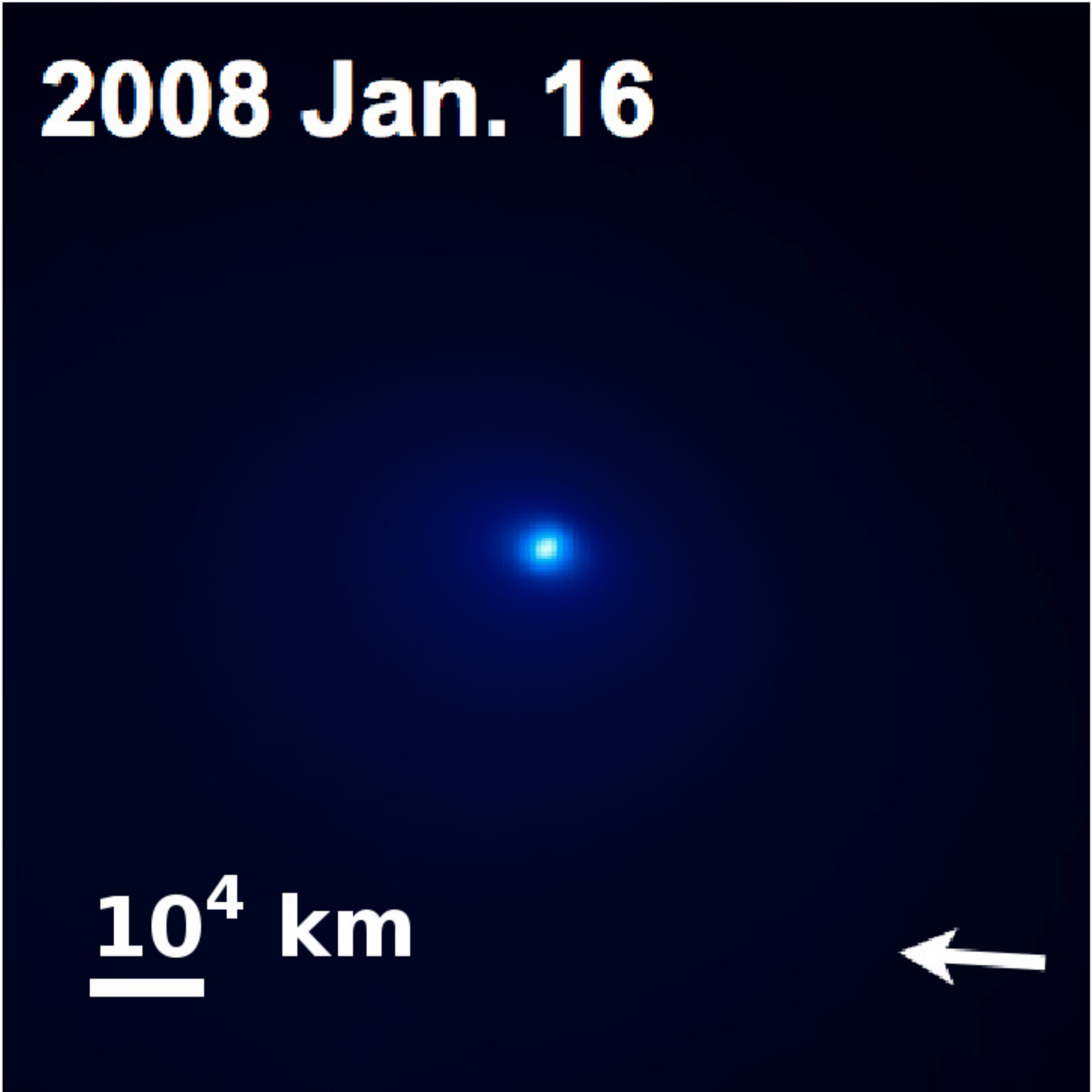}}
\hspace{2bp}
\subfloat{\includegraphics[width=0.2\textwidth]{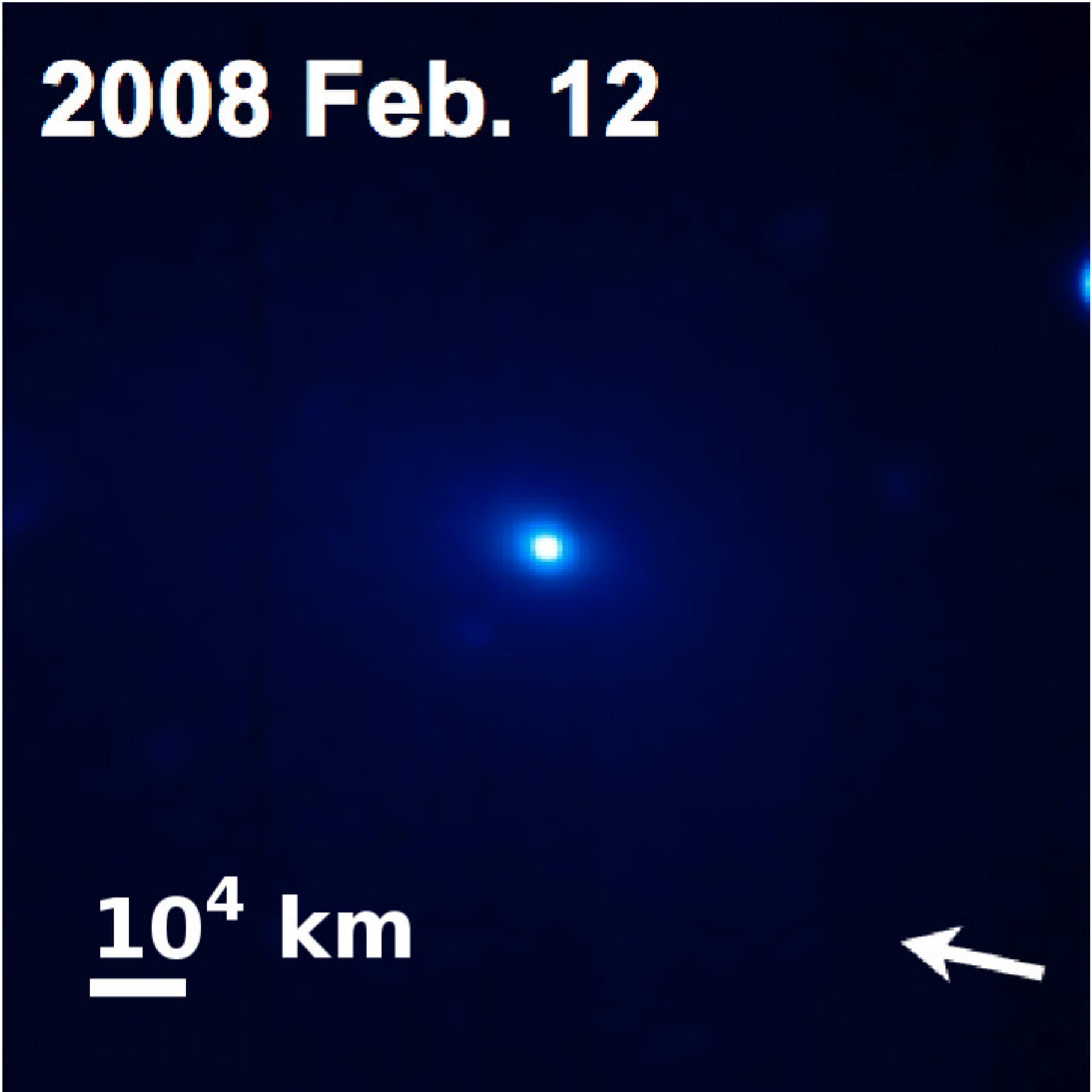}}
\caption{The inner coma of 17P/Holmes as seen throughout our observations.  Each image is 1$^{\prime}$ $\times$ 1$^{\prime}$ and is oriented with north up and east left.  The anti-solar vector, as projected on the plane of the sky, is denoted by a white arrow.}
\label{fig:nucpan}
\end{figure}

\begin{figure}
\centering
\subfloat{\includegraphics[width=0.2\textwidth]{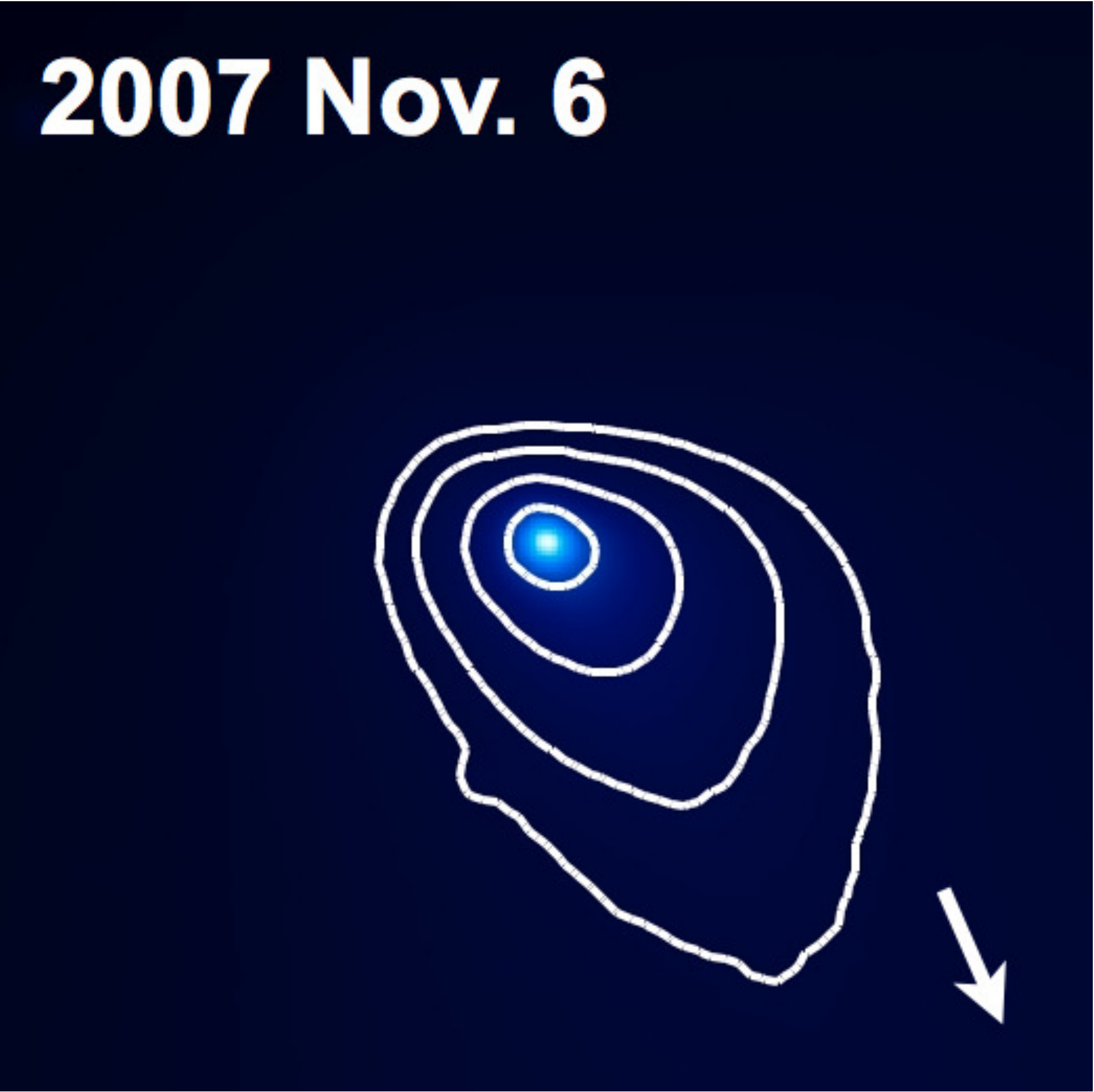}}
\hspace{2bp}
\subfloat{\includegraphics[width=0.2\textwidth]{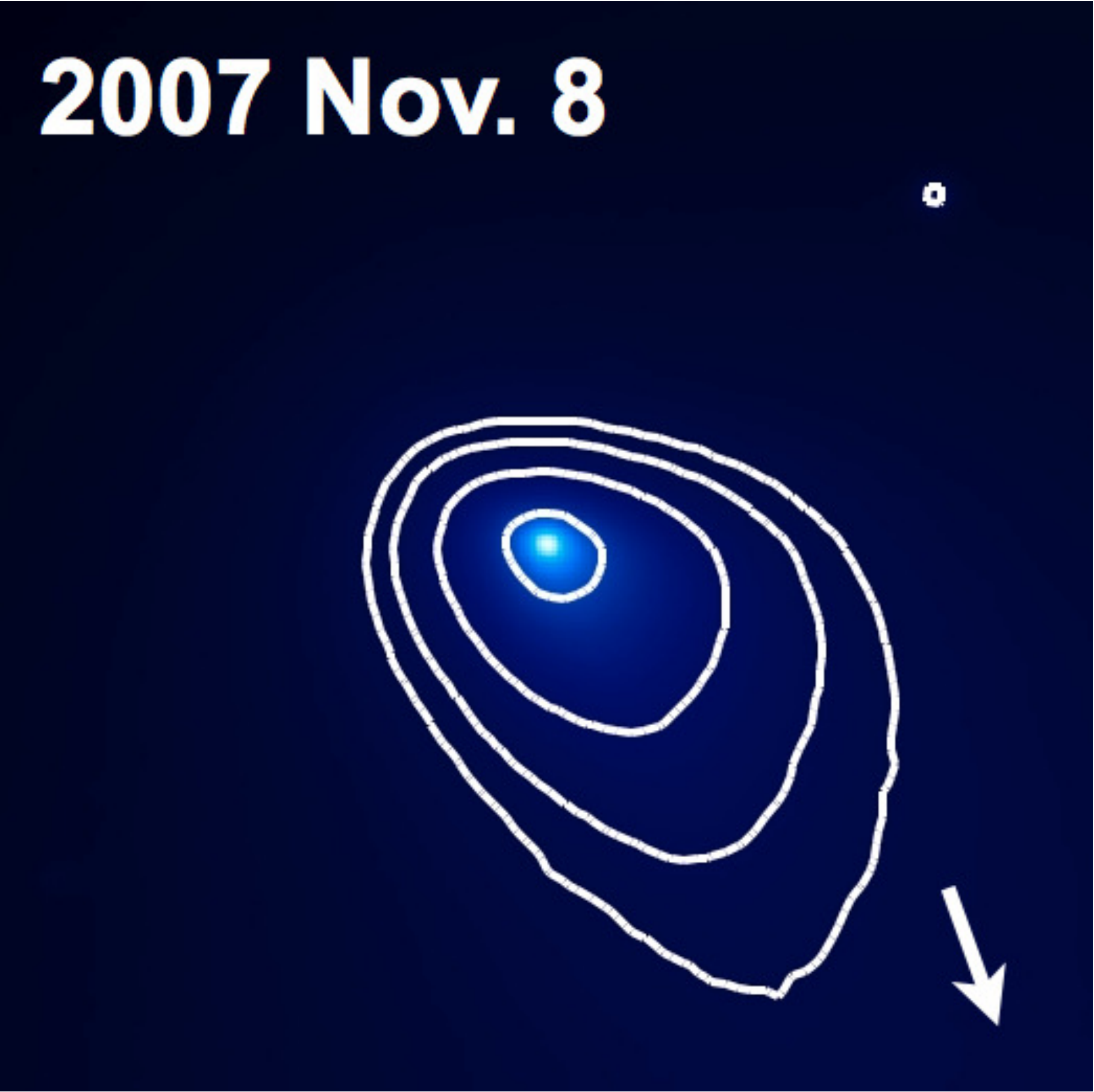}}
\hspace{2bp}
\subfloat{\includegraphics[width=0.2\textwidth]{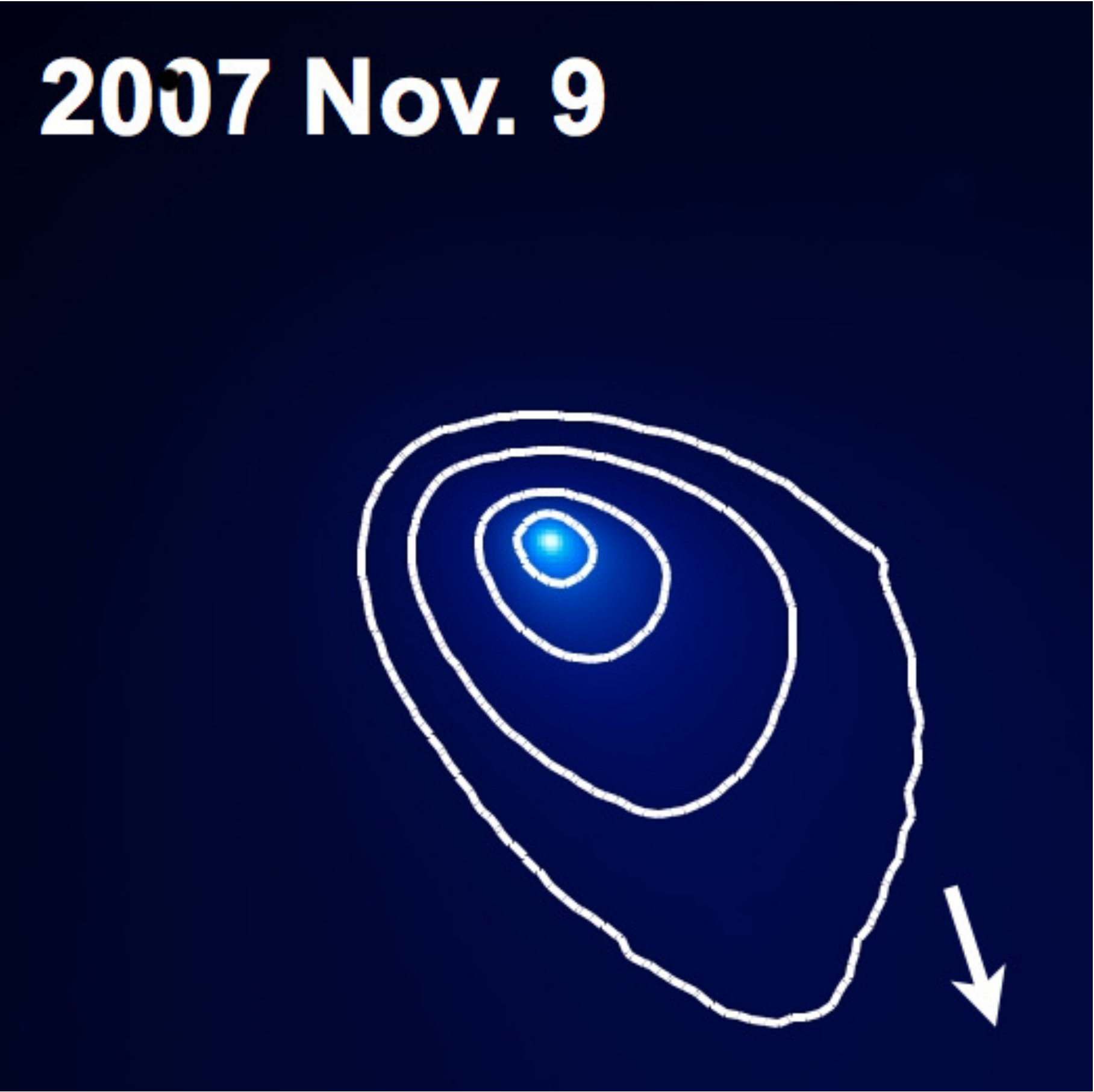}}
\hspace{2bp}
\subfloat{\includegraphics[width=0.2\textwidth]{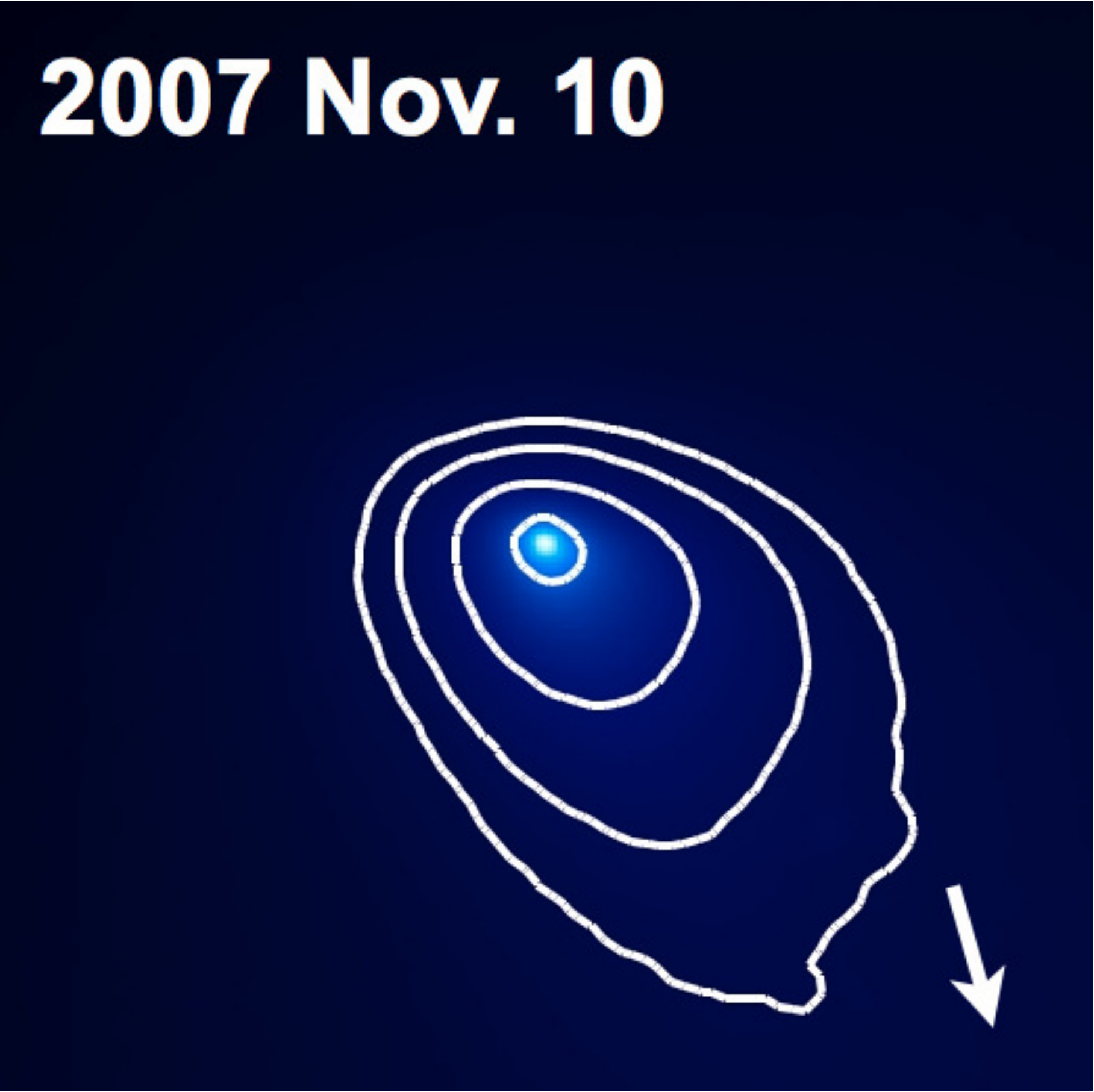}}
\vspace{2bp}
\subfloat{\includegraphics[width=0.2\textwidth]{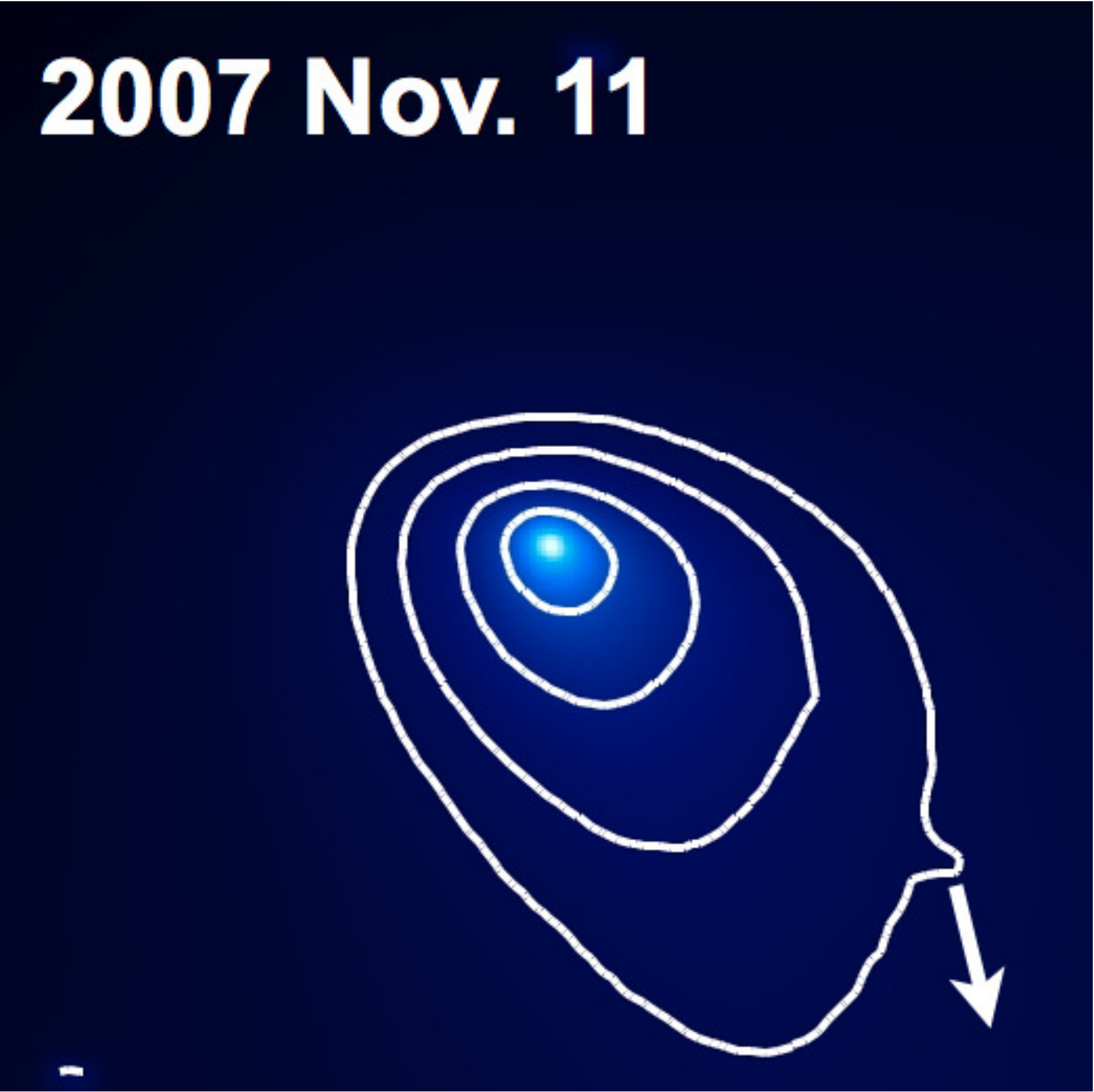}}
\hspace{2bp}
\subfloat{\includegraphics[width=0.2\textwidth]{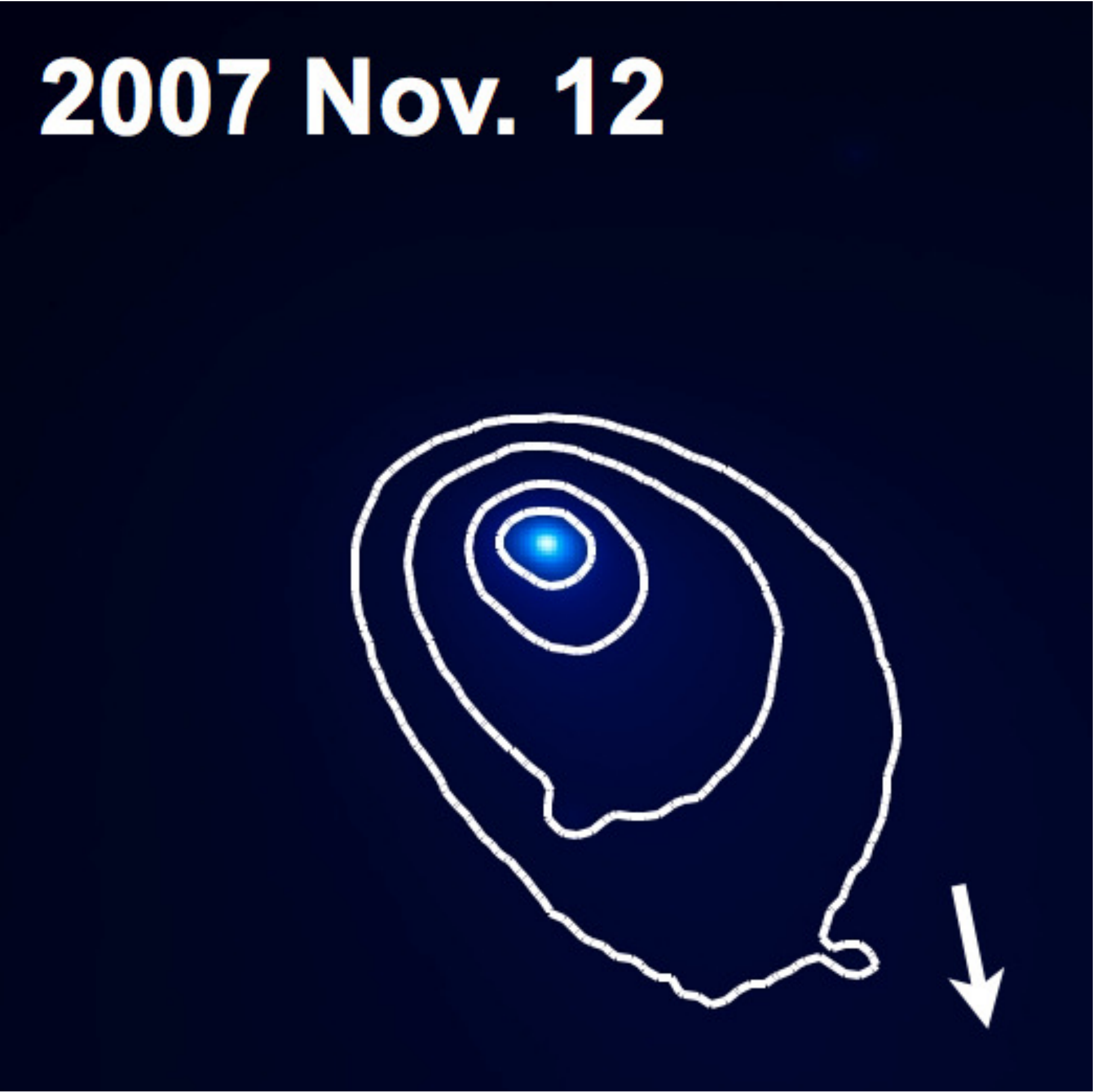}}
\hspace{2bp}
\subfloat{\includegraphics[width=0.2\textwidth]{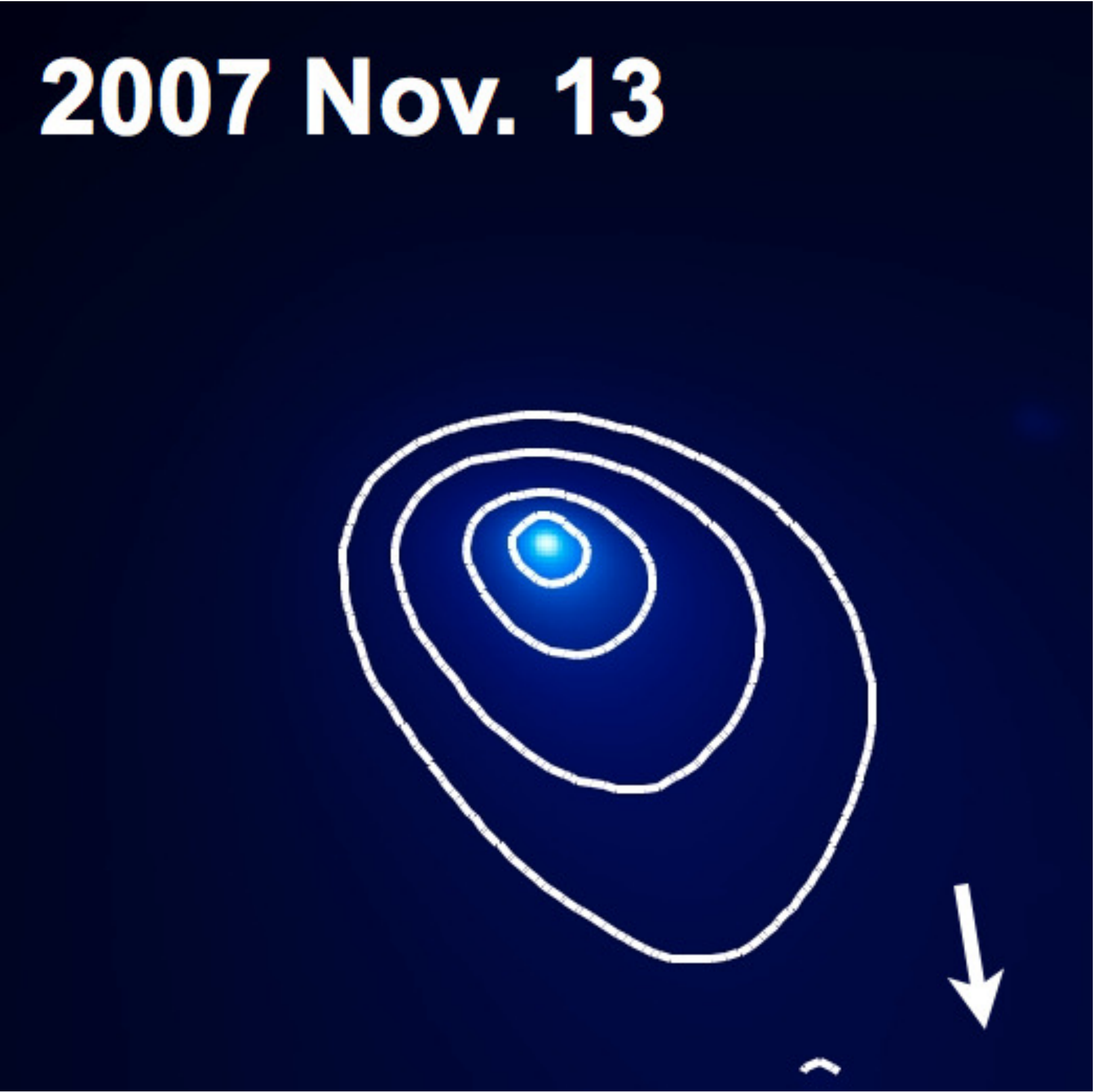}}
\hspace{2bp}
\subfloat{\includegraphics[width=0.2\textwidth]{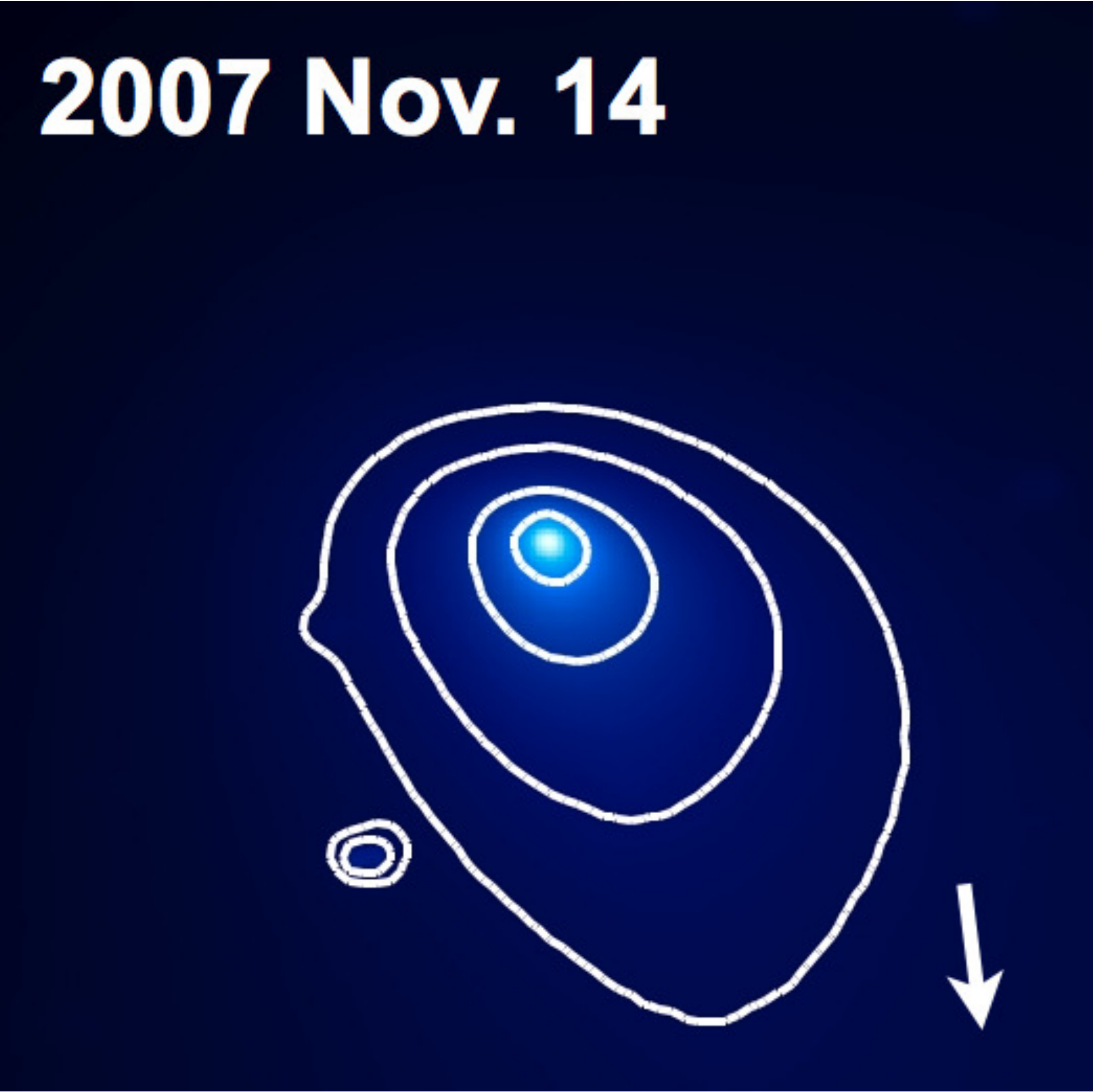}}
\vspace{2bp}
\subfloat{\includegraphics[width=0.2\textwidth]{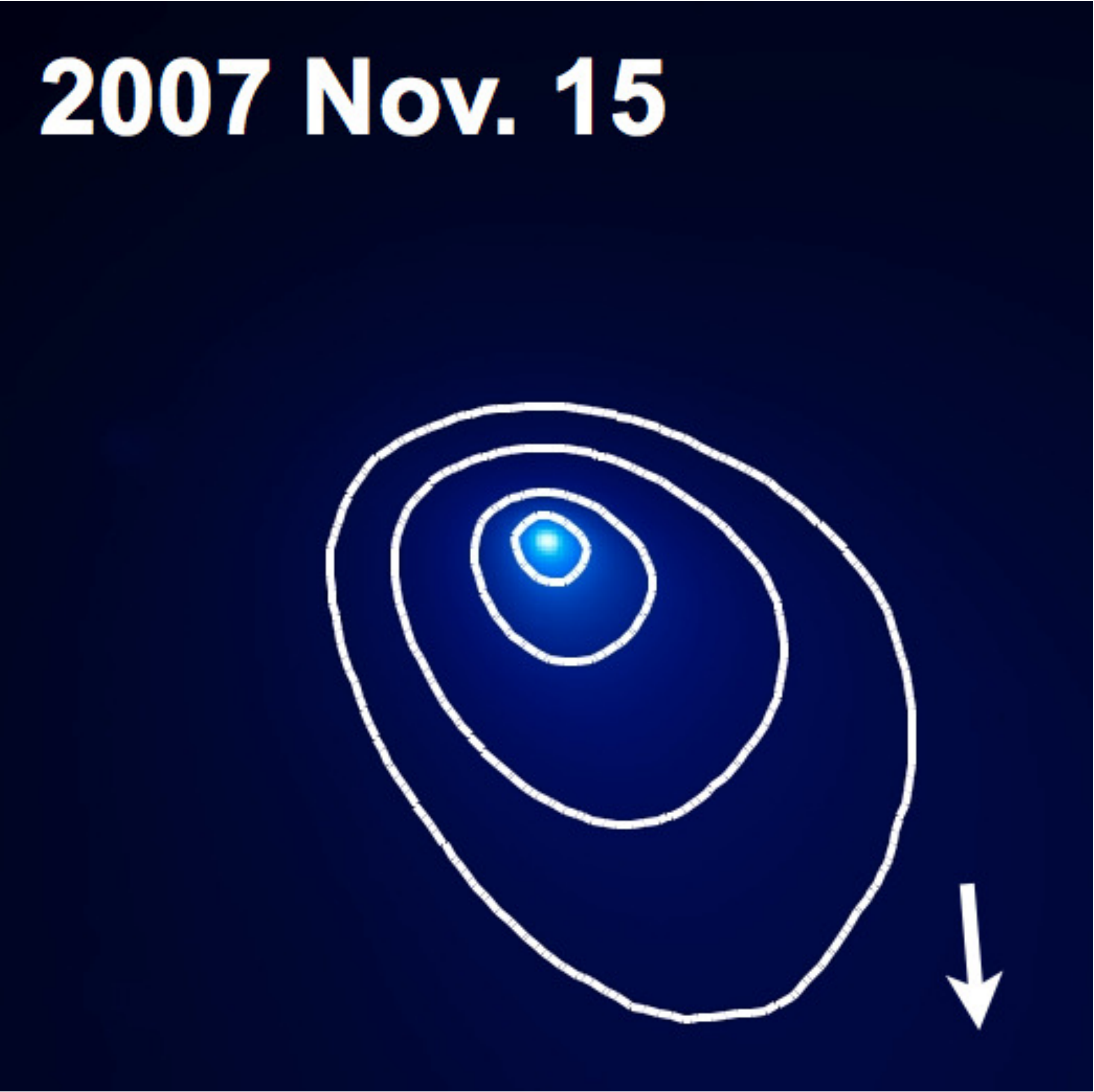}}
\hspace{2bp}
\subfloat{\includegraphics[width=0.2\textwidth]{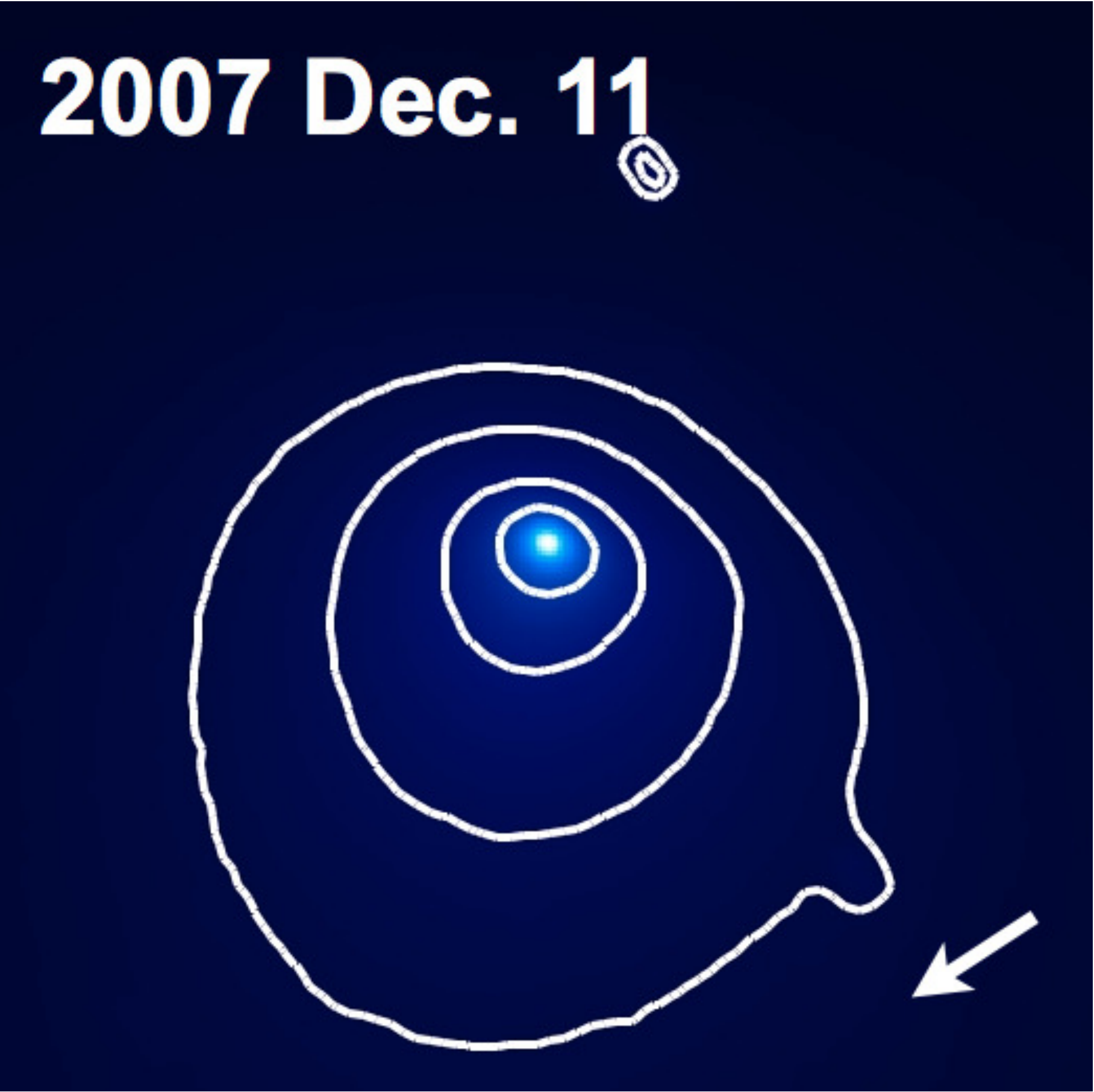}}
\hspace{2bp}
\subfloat{\includegraphics[width=0.2\textwidth]{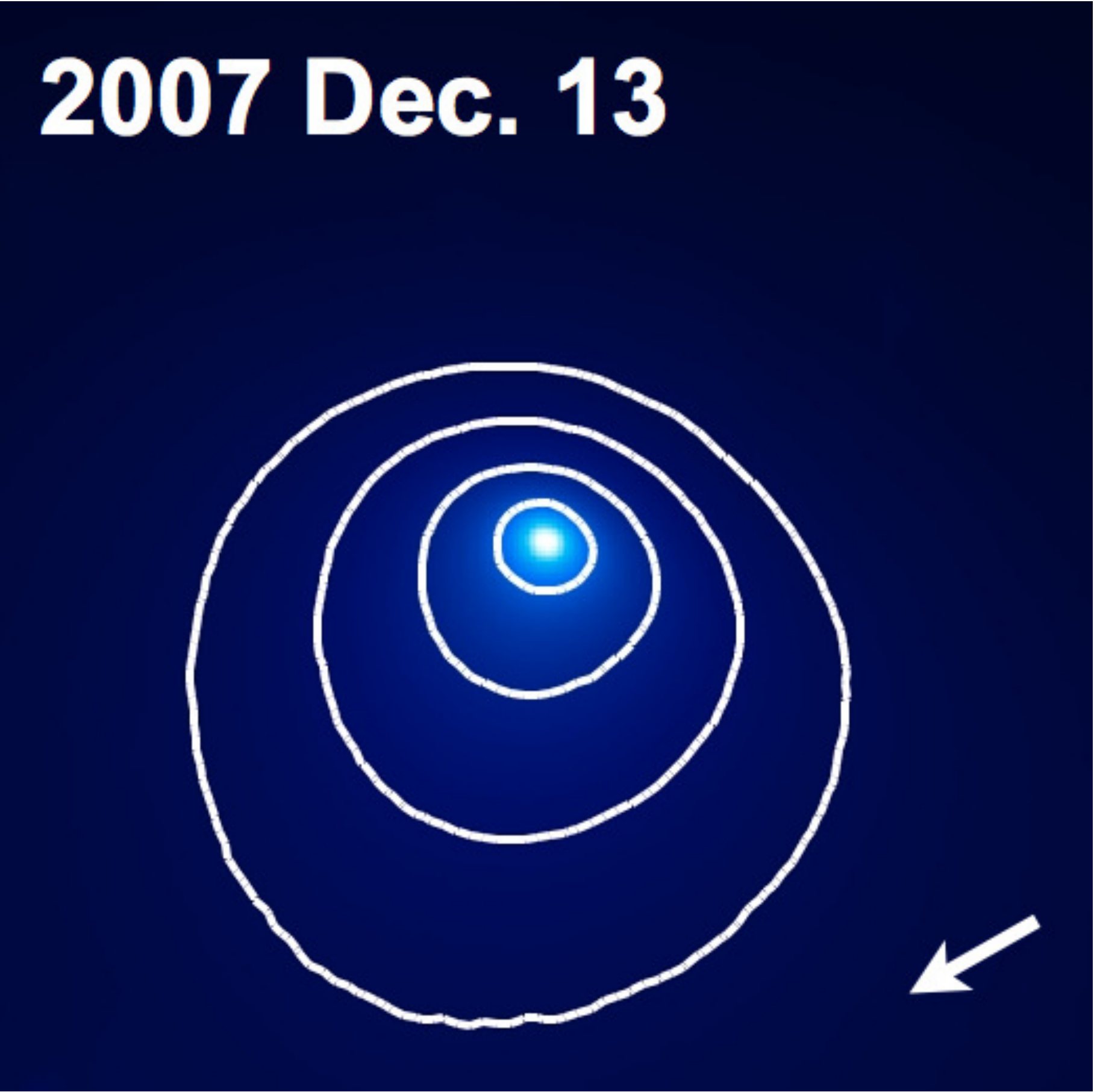}}
\hspace{2bp}
\subfloat{\includegraphics[width=0.2\textwidth]{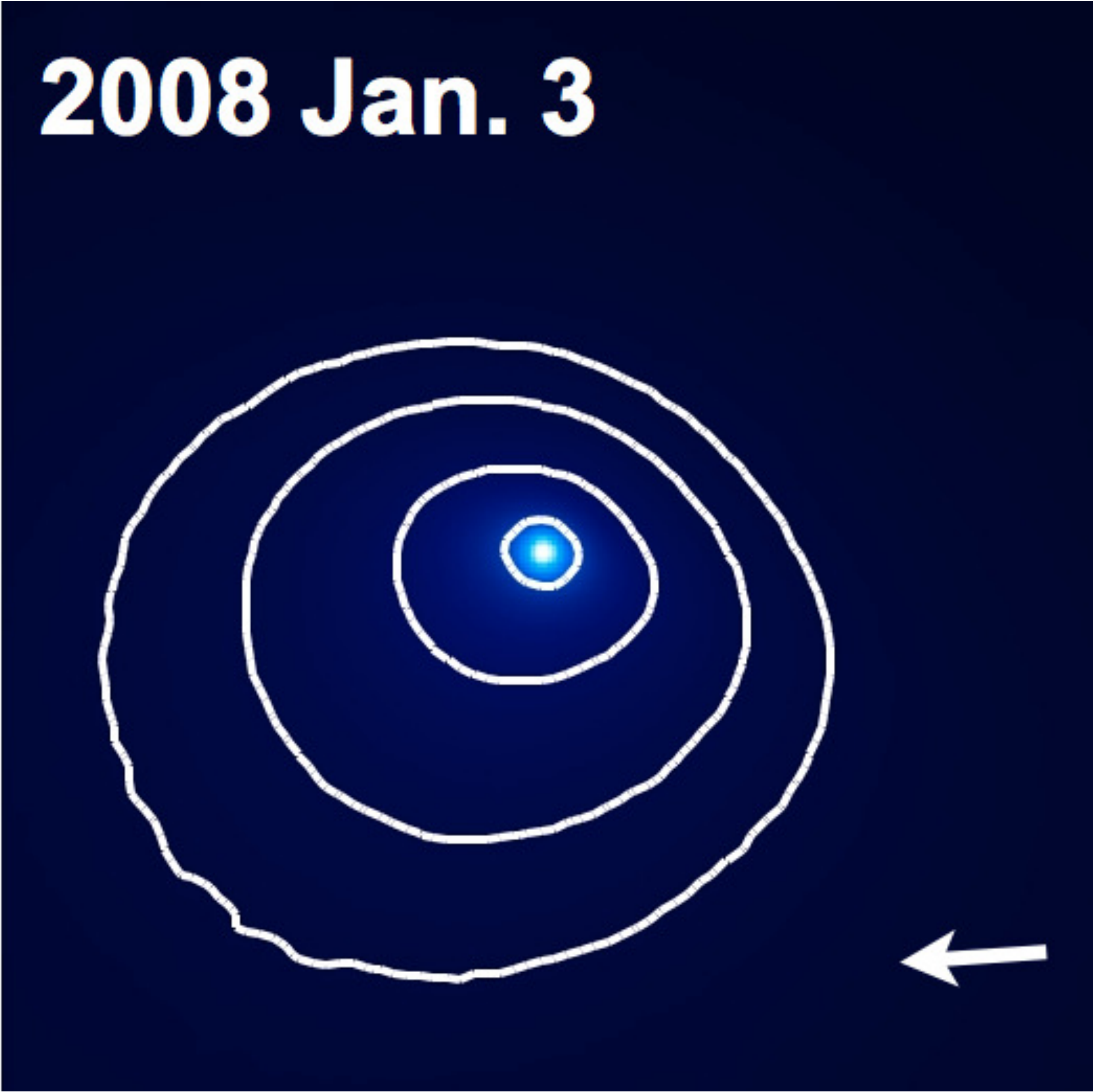}}
\vspace{2bp}
\subfloat{\includegraphics[width=0.2\textwidth]{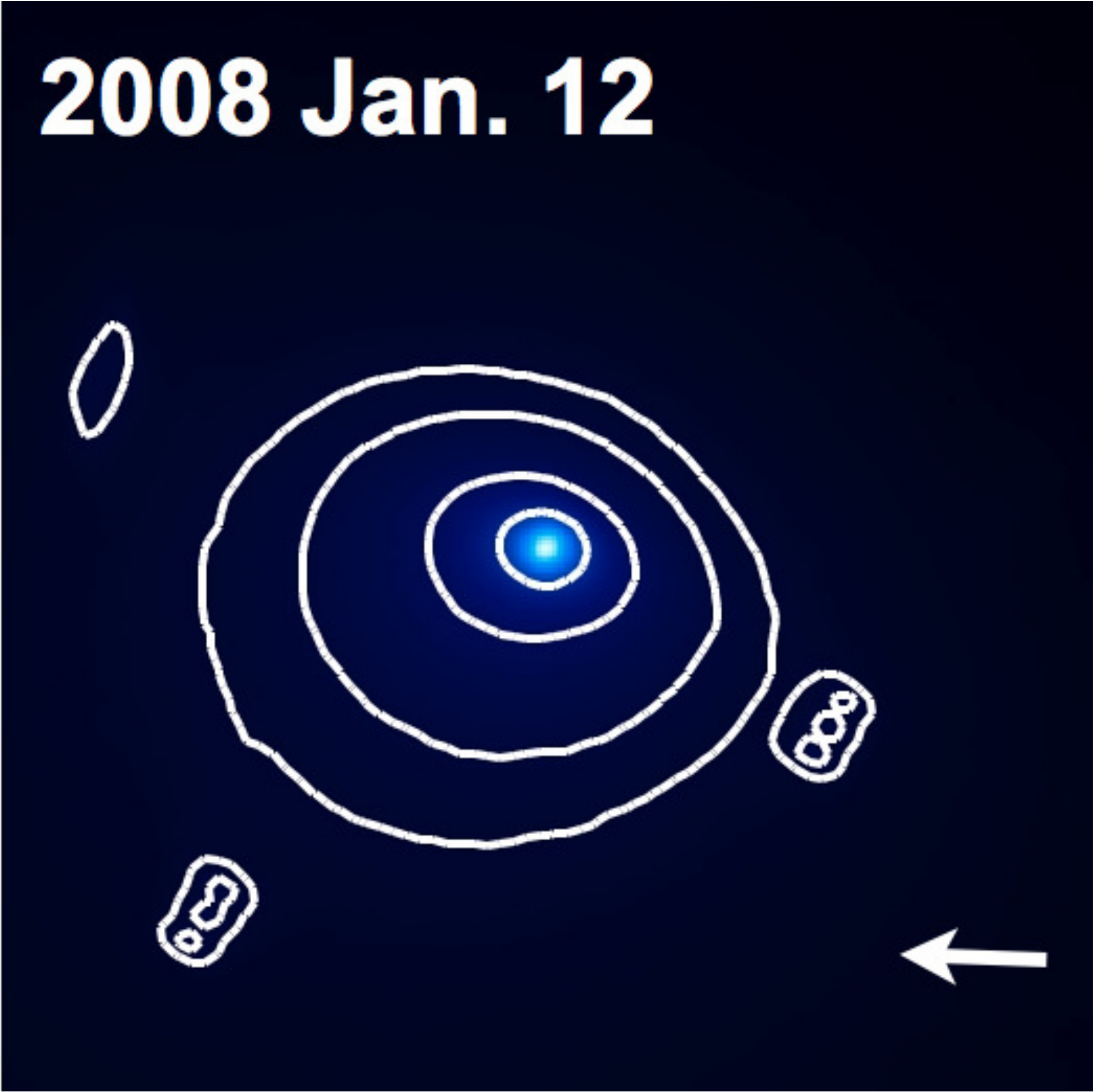}}
\hspace{2bp}
\subfloat{\includegraphics[width=0.2\textwidth]{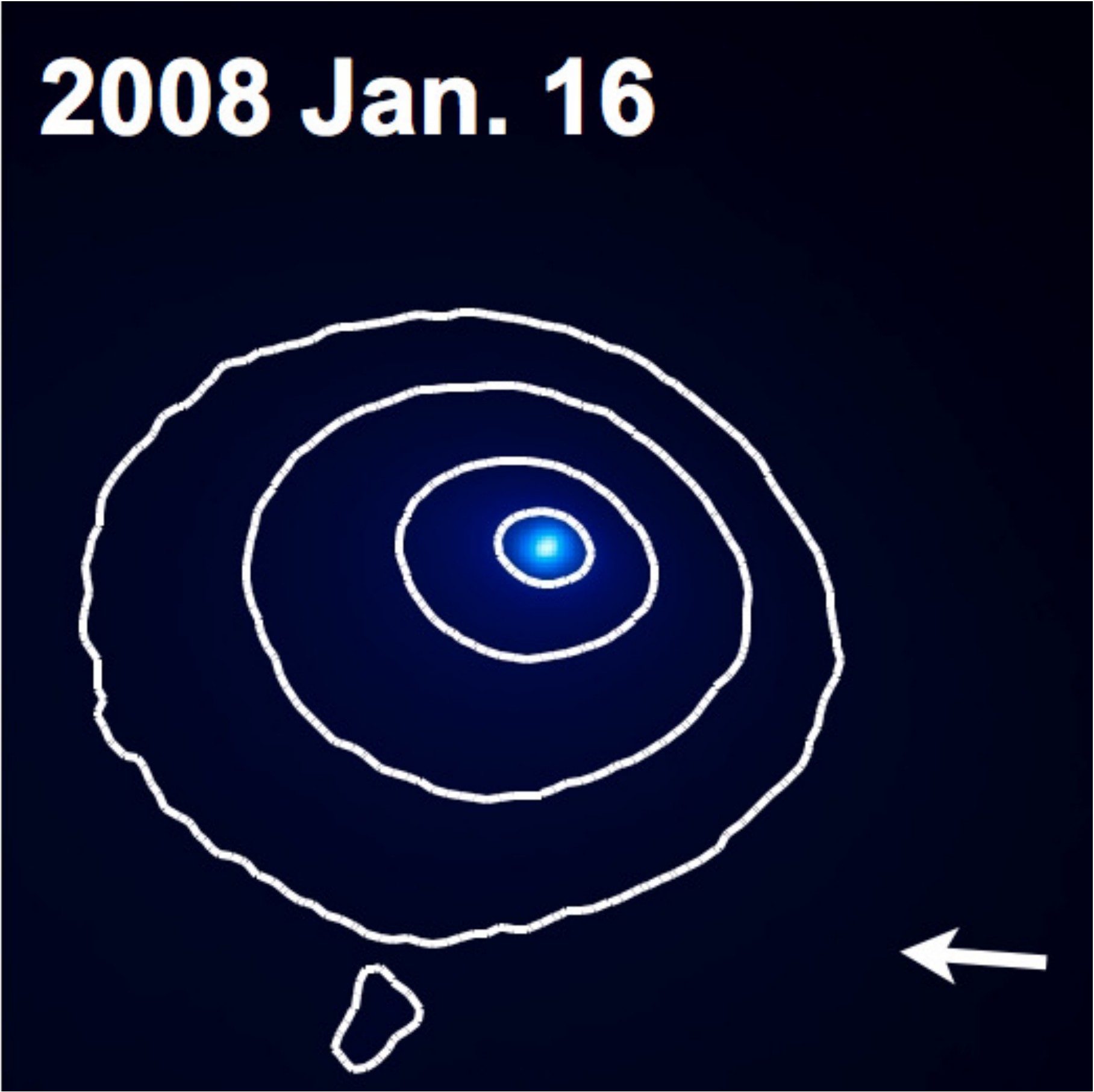}}
\hspace{2bp}
\subfloat{\includegraphics[width=0.2\textwidth]{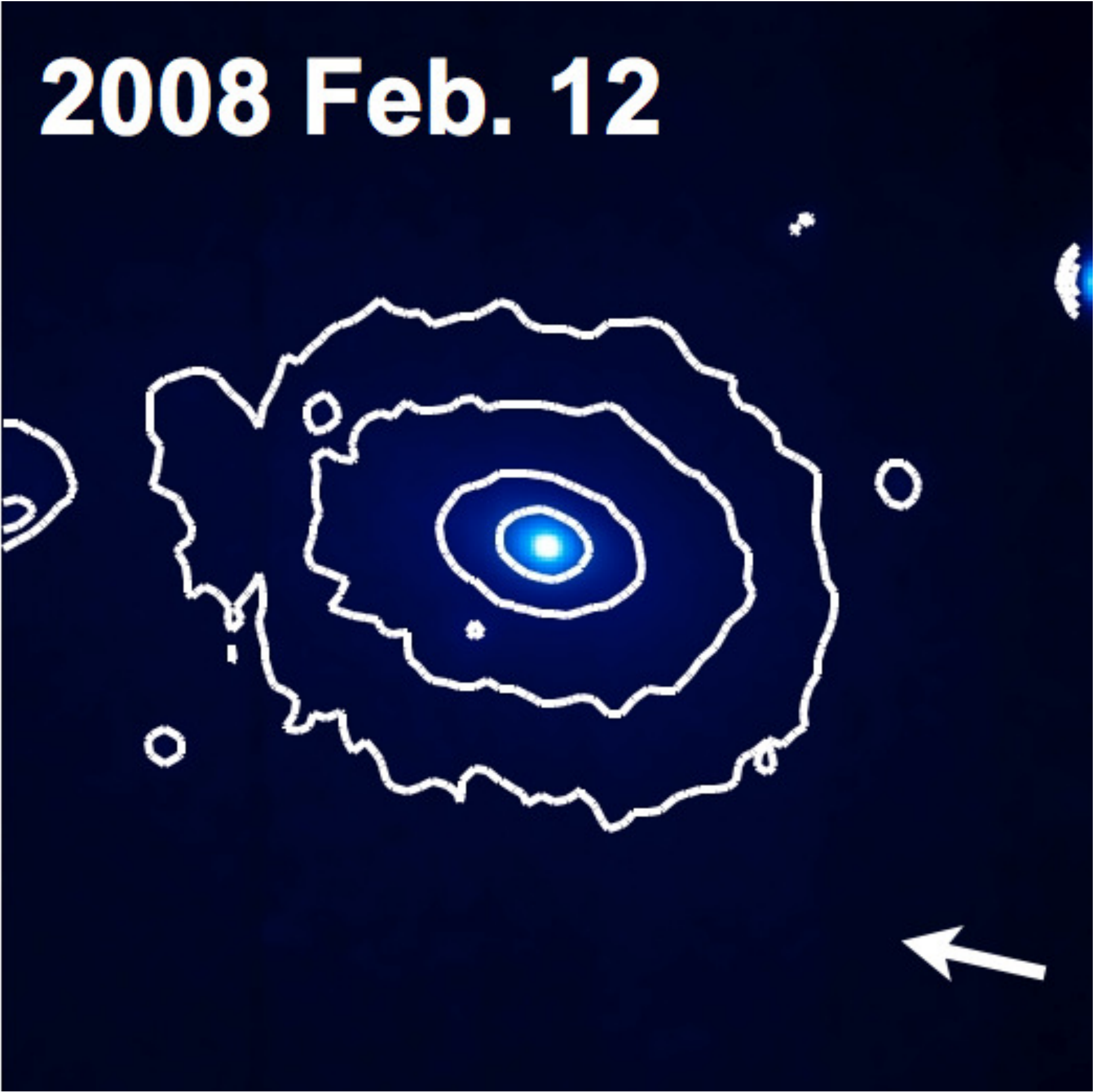}}
\caption{The central arcminute region of each image centered on the nucleus in north-up, east-left orientation with the projected anti-solar vector marked by a white arrow.  The isophotes are of arbitrary brightness and were chosen to highlight the morphological evolution of the coma.  The innermost coma was initially elongated but became more symmetrical towards the end of our observations.}
\label{fig:nuccon}
\end{figure}

All images were automatically pre-processed using the Elixir pipeline \citep{2004PASP..116..449M}, which uses bias frames and twilight flatfields to remove the instrumental signature, and calculates instrumental zero-points using Landolt standard star fields \citep{1992AJ....104..340L} observed on each night.  We used SWarp, software released by the Terapix data center at the Institut d'Astrophysique de Paris, to resample the images and attain $0^{\prime\prime}.2$ astrometric accuracy.  The dithered images from each night were median-combined to cover the 80$^{\prime\prime}$ gaps between CCDs and remove cosmic rays (Figure~\ref{fig:fullFoV}).  Each combined image was convolved with a two-dimensional Gaussian function with a full-width half-maximum of 1$^{\prime\prime}$ to minimize the effect of variations in seeing from night to night.  A panel showing the central square arcminute of each image is shown in Figure~(\ref{fig:nucpan}).  The same central regions are shown with isophotal contours in Figure~(\ref{fig:nuccon}) to better highlight the evolution of the coma morphology.  The inner coma initially appears elongated in the anti-solar direction but becomes more symmetrical at later dates.

The sky transparency was monitored using the CFHT Skyprobe and  we found that only observations made on UT 2008 Feb.\ 12 experienced significant variations in attenuation ($>$ 0.1 mag).  For all other nights the weather was photometric.  Systematic errors due to flatfielding or scattered light are well-characterized for MegaCam and do not exceed 5\% \citep{2004PASP..116..449M}.  The sky background was determined in a region as far from the observable coma as possible within each stacked image, using the median signal within a $\sim$ 4$^{\prime} \times$ 4$^{\prime}$ box.  Although we sought regions on the image devoid of any noticeable brightness gradient from which to estimate the background level, it is likely that some coma contamination persisted in our sky background estimations.  Specifically, the gas coma was observed to expand at a velocity of 1 km s$^{-1}$ \citep{2008LPICo1405.8340D}, meaning that the gas coma reached the projected edges of the MegaCam field of view by 2007 mid-November.  Fine dust grains dynamically well-coupled to the gas and ejected at high velocities could contaminate the outer regions within a month of the outburst, although our observations are less sensitive to the smallest grains since their scattering efficiency is small.  In Figures~(\ref{fig:fadingtot}) and~(\ref{fig:fadingnov}) we show the effects of a 10\% error on the sky background combined with the known sky transparency and systematic errors.  Plausible sky subtraction uncertainties are negligible in most of our data, given the high surface brightness of the coma.  Only in observations from January and February 2008 do sky subtraction errors rival systematic uncertainies due to atmosphere transparency and calibration uncertainties.




\section{Photometry of the Inner Coma}

The coma was initially observed as an approximately circular dust cloud with a bright condensation south-west of the nucleus.  As it expanded, the coma became more asymmetric with a parabolic nose building in the solar direction and a diffuse tail in the anti-solar direction.  The nucleus region was always identifiable in the inner coma and stars could be seen even through the innermost coma, implying a low optical depth.  We used aperture photometry to study the material closest to the nucleus with the aim of monitoring the comet's continued activity after its massive outburst.





The position of the nucleus was determined to sub-pixel accuracy using a two-dimensional derivative search and used as the center for all aperture photometry.  Since the geocentric distance increased from 1.62 AU on UT 2007 Nov.\ 6 to 2.62 AU on UT 2008 Feb.\ 12 (Table~\ref{table:obs}), an aperture of fixed angular radius would cover an increasing physical distance on the sky, and, correspondingly, more of the coma.  For this reason, we adjusted the angular radii of the photometry apertures used for each night to compensate for the increase in geocentric distance to 17P/Holmes, and instead used apertures of fixed physical radius at the comet.  Circular apertures of projected radii between 250 km and 25000 km in increments of 250 km were used.  

We chose not to subtract the flux of the nucleus from these measurements.  The expected flux contribution from the nucleus can be calculated using the observing geometry, a nucleus radius of 1.62 km, and a geometric albedo of 0.04 \citep{2006MNRAS.373.1590S,2004come.book..223L}.  The calculated nucleus contribution was typically less than 1\% of the flux measured within a 2500 km radius aperture.  In the worst case, the nucleus contribution amounted to 3\% of the measured flux but in all cases it was significantly smaller than other sources of error.


Trailing losses are negligible for observations between UT 2007 Nov.\ 6 and UT Dec.\ 11 since the motion of the comet during each exposure was less than the full-width-half-maximum of the image.  Later images were trailed by up to 7.7 pixels (1$^{\prime\prime}.4$).  To investigate the effect of trailing on our photometry we took an image with minimal trailing of 0.2 pixels (0$^{\prime\prime}$.4) and shifted and added duplicate images to mimic trailing.  We found that the loss of light from a 1.3$^{\prime\prime}$  radius aperture (2000 km for the image most badly affected by trailing) was on the order of 2\% and fell to less than 0.1\% for a 6$^{\prime\prime}$.3 (10$^{4}$ km) radius aperture.  We conclude that trailing has a negligible effect on the results from aperture photometry in our data.


We converted fluxes to apparent magnitudes, $m$, using the instrumental zero-points calculated by the Elixir pipeline, correcting for both airmass and the color of the filter.  To correct for changes in observing geometry we used Equation~(\ref{absmageq}) to determine absolute magnitudes, $H$ -
the magnitudes objects would have if observed at heliocentric and geocentric distances of 1 AU and a phase angle of 0$^{\circ}$.

\begin{equation}
H = m - 5\ log_{10} (r_{H} \Delta) - 2.5\ log_{10} \Phi_{\alpha}
\label{absmageq}
\end{equation}

\noindent where $r_{H}$ and $\Delta$ are the heliocentric and geocentric distances in AU respectively and $\Phi_{\alpha}$ is a phase function that corrects for the dependency of the brightness on the phase angle ($\alpha$).  The phase function of the observed dust coma is difficult to measure given the need for accurate photometry of the dust over a wide range of phase angles when the coma is bright enough to detect.  We use an approximation of the phase function for 1P/Halley based on a fit to data presented in \cite{1998Icar..132..397S} \citep{2011ApJ...728...31L}:

\begin{equation}
2.5\ log_{10} \Phi_{\alpha} = 0.045 \alpha - 0.0004 \alpha^{2}
\label{phieq}
\end{equation}

The phase corrections that result from Equation~(\ref{phieq}) match well with other commonly-used phase functions since the change in phase angle was relatively small ($<$ 8$^{\circ}$).  We acknowledge that this is a slight misuse of the concept of absolute magnitude given that Equation~(\ref{absmageq}) refers to unresolved point sources rather than distributed sources such as cometary comae.  An angular aperture will subtend a different distance at 1 AU than at other heliocentric distancs, resulting in an altered amount of material within the aperture.  However, since we used apertures of fixed physical distances from the nucleus we expect this effect to have little impact on our results.  Table~(\ref{table:char}) shows the results from aperture photometry for each night during our observing campaign.

\begin{figure}
\centering
\includegraphics[totalheight=15cm]{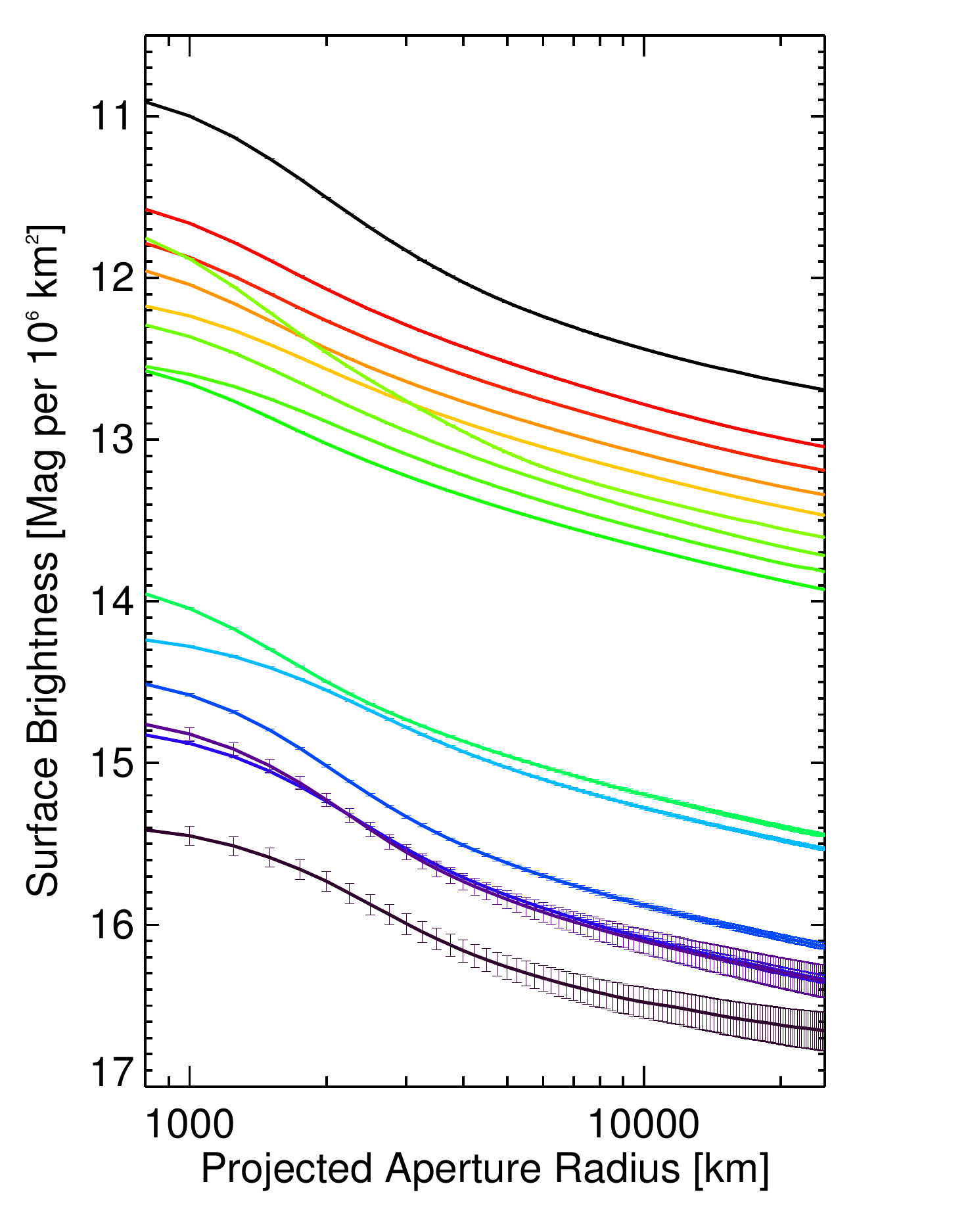}
\caption{The surface brightness profiles of the inner coma, centered on the nucleus, from 2007 Nov. 6 UT (top) to 2008 Feb. 12 UT (bottom).  Error bars are plotted and represent a 10\% uncertainty in the sky background estimate.  Each profile has a visible bump near the center and has a slope of $\sim$ -0.2 to -0.3 between the projected distances of 10000 km and 25000 km. The surface brightness of the inner coma, as observed at 25000 km, decreases monotonically with time.  The surface brightness profile within $\sim$ 5000 km on 2007 Nov.\ 12 is brighter than expected, which is consistent with a second outburst occurring at the nucleus shortly before our observation.}
\label{fig:SBall}
\end{figure}

The surface brightness was calculated from the total flux within concentric annuli centered on the nucleus and divided by the area measured in square kilometers.  The flux per unit area was converted to absolute magnitudes per million square kilometers at the distance of the comet, which accounted for changes in the observing geometry during the course of our observing campaign.  Figure~(\ref{fig:SBall}) shows the surface brightness profiles after the images were convolved with a 2-dimensional $1^{\prime\prime}$ Gaussian function.


\section{Results and Discussion}

\subsection{The Fading of Comet 17P/Holmes}
\label{sec:fading}
\begin{figure}
\centering
\includegraphics[totalheight=13cm]{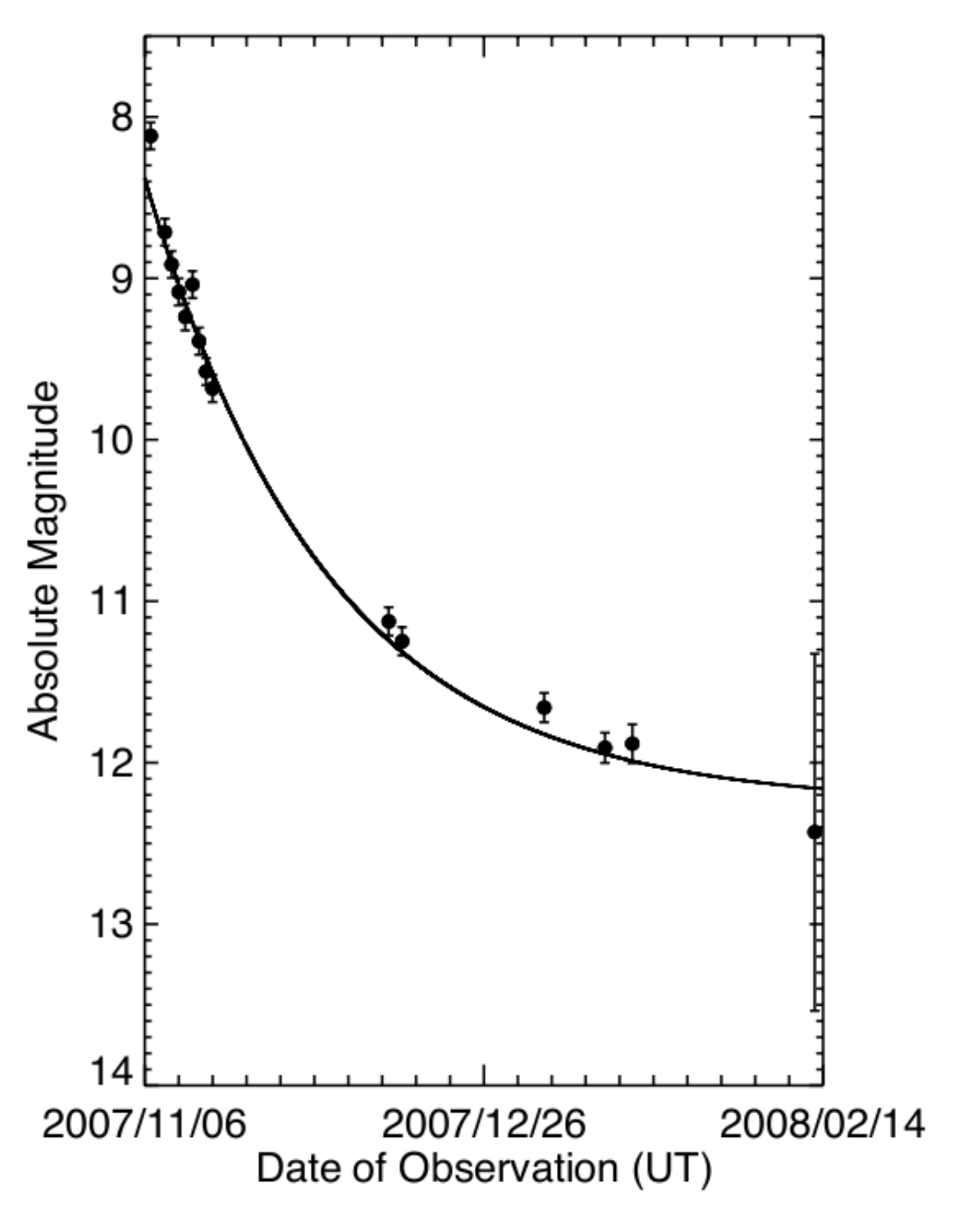}
\caption{The fading of 17P/Holmes over four months as observed using an aperture with a projected radius of 2500 km.  The absolute magnitudes were calculated using Equation~(\ref{absmageq}) and have been fitted with an exponential function.  The brightness of the inner coma dropped by a factor of $\sim$ 50 during our observing campaign.}
\label{fig:fadingtot}
\end{figure}

The absolute magnitude within a 2500 km radius aperture increases from 8.1 mag on UT 2007 Nov.\ 6 to 12.4 mag on UT 2008 Feb.\ 12, indicating a $\sim$ 50-fold decrease in brightness (Figure~\ref{fig:fadingtot}).  Both the initial fading observed throughout 2007 November and the fading during the entire observing campaign can be reasonably well fitted with an exponential function of the form 
$H = a + b e^{c t}$
where $H$ is absolute magnitude, $t$ is time since the first observation in days, and $a$, $b$, and $c$ are constants.  The best fits, as determined via a non-linear least-squares fit, are shown in Figures (\ref{fig:fadingtot}) and~(\ref{fig:fadingnov}).  We note that there is no clear physical reason for the decrease in brightness to be fitted with an exponential function in log-linear space.  However, \cite{2009AJ....138.1062S} found that similar functions matched the changing production rates for OH, CN, NH, C$_{2}$, C$_{3}$, and the continuum over a period of several months.


The expansion of the dust coma was studied by numerous observers and the expansion velocity was found to be $\sim$ 550 m s$^{-1}$ \citep{2007CBET.1123....1G,2008ICQ....30....3S,2008LPICo1405.8278S,2010MNRAS.407.1784H}, although \cite{2008A&A...479L..45M} reported a lower value of 200 m s$^{-1}$.  For this paper we use the relationship between grain velocity, $v$ (m s$^{-1}$), and radius, $a_{\mu}$ ($\mu$m), derived by \cite{1951ApJ...113..464W} and assume that a 1 $\mu$m radius dust grain has a velocity of 550 m s$^{-1}$ so that:

\begin{equation}
v = \frac{550}{\sqrt{a_{\mu}}}.
\label{veleq}
\end{equation}

The high measured velocities imply that the majority of dust particles ejected would exit even a 25000 km radius aperture in $\sim$ 1 day.  Even 1 cm sized particles ($a_{\mu}$ = 10$^4$) would leave a 25000 km radius aperture in about 50 days by Equation~(\ref{veleq}).  Given the brightness observed for several months following the outburst, the nucleus must have continued to supply the inner coma with fresh material after the 2007 Oct.\ outburst.  This result initially seems to be in conflict with the production curve calculated by \cite{2011ApJ...728...31L}, who found that significant activity persisted for only a few days, after which relatively little new material was injected into the coma.  However, their analysis refers to the entire coma of 17P/Holmes, which spanned several hundred thousand km, whereas our work focuses on a very small fraction of the coma close to the nucleus, and is consequently much more sensitive to small, localized changes.


The apparent magnitude, $m$, can be used to estimate the total cross-section of the dust within an aperture using:

\begin{equation}
\sigma = \frac{2.25\times10^{22}}{g_{\lambda}  \Phi_{\alpha}} ~ \pi~ r_{H}^{2} ~\Delta^{2} ~ 10^{-0.4 (m - m_{\Sun})}
\label{crosssectioneq}
\end{equation}

where $\sigma$ is the total geometric cross-section of the material (m$^{2}$), $g_{\lambda}$ is the geometric albedo, and $m_{\sun}$ is the apparent magnitude of the Sun.  All other variables have the same definitions as in previous equations.  The variables $r_{H}$ and $\Delta$ are given in AU.  The albedo of the ejected material was measured to be between 0.03 $\pm$ 0.01 and 0.12 $\pm$ 0.04  at a phase angle of 16$^{\circ}$ \citep{2010ApJ...714.1324I}.  The corresponding geometric albedos - as would be observed at a phase angle of 0$^{\circ}$ - are estimated to be 0.05 and 0.20 after using Equation~(\ref{phieq}) to calculate a phase correction.

\begin{figure}
\centering
\includegraphics[totalheight=15cm]{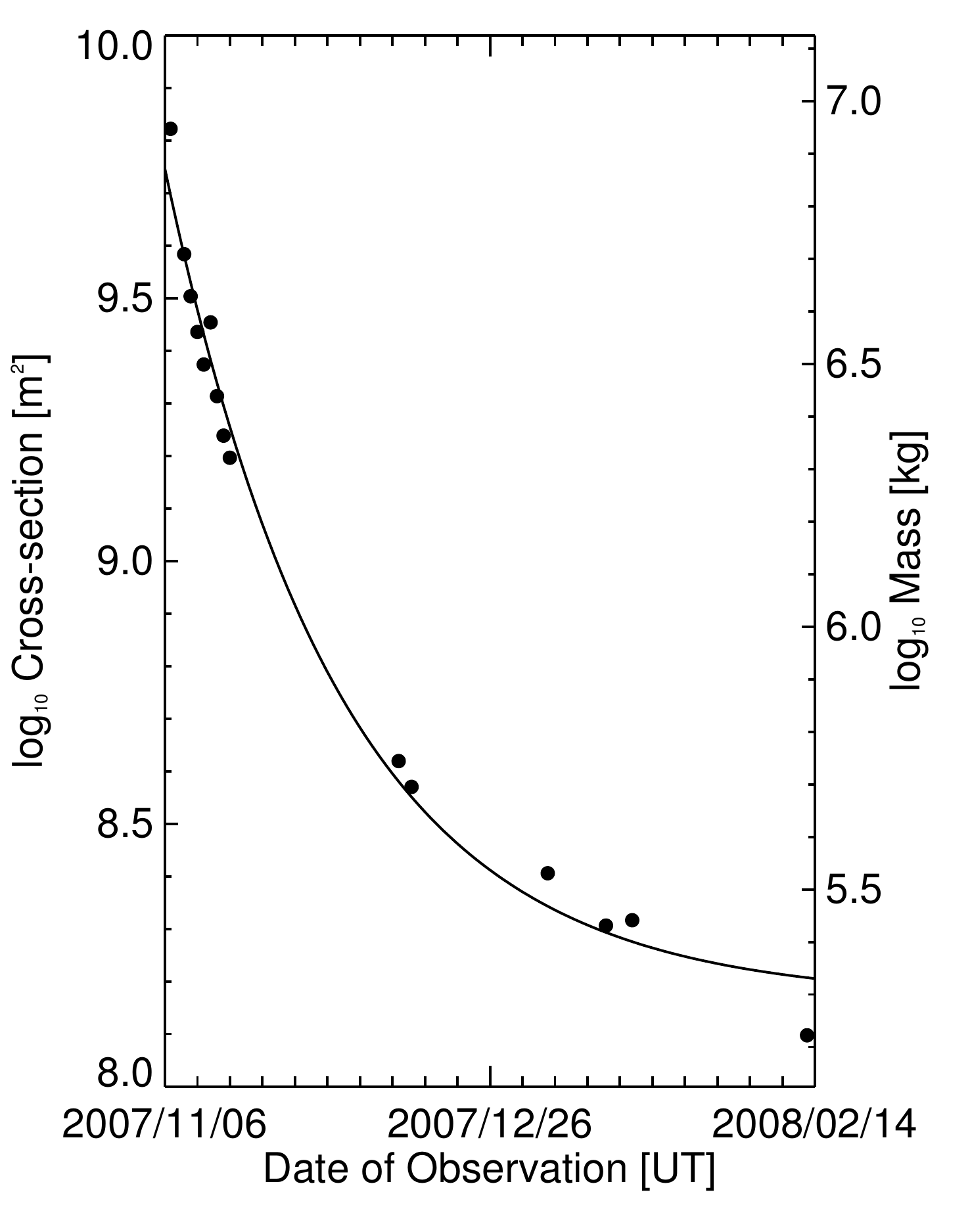}
\caption{The estimated cross-section and mass of material within 2500 km of the nucleus, calculated using Equation~(\ref{crosssectioneq}) and the assumption that the material consists of dust grains with radii of 1 $\mu$m with an albedo of 0.1 and a density of 1000 kg m$^{-3}$.}
\label{fig:sigmass}
\end{figure}

Given the uncertainties, we adopt a nominal value of 0.1 for the geometric albedo of dust grains and accept that the actual value may differ from this by a factor of two.  The apparent magnitude of the Sun in the SDSS r$^{\prime}$ filter is -26.95 mag \citep{2001AJ....122.2749I}.  We then convert the cross-section into a crude mass, assuming that spherical dust grains with radii of 1 $\mu$m and a bulk density of 1000 kg m$^{3}$ are scattering the light.  Figure~(\ref{fig:sigmass}) shows the change in cross-section and mass of the material within a 2500 km radius aperture between UT 2007 Nov.\ 6 and UT 2008 Feb.\ 12.  The cross-section of material contained within the aperture decreases from 6.6 $\times$ 10$^{9}$ m$^{2}$ to 1.3 $\times$ 10$^{8}$ m$^{2}$.  These cross-sections correspond to effective circular radii of 46 km and 6 km and masses of 8.9 $\times$ 10$^{6}$ kg and 1.7 $\times$ 10$^{5}$ kg, respectively, which was only a tiny fraction of the total mass ejected (2 - 90 $\times$ 10$^{10}$ kg; \citealt{2011ApJ...728...31L}).  The mass estimates are strongly dependent on the assumed size distribution of the particles, and are used here for order of magnitude estimates only.



For small apertures we can derive mass loss rates using the aperture crossing times of the dust grains.  Large apertures will contain a mix of old, slow-moving large grains and fresh, high-velocity small grains, while the reflected light contained in small apertures is dominated by new material that may be only hours old.  
We continue to use our earlier assumptions that micron-sized particles traveling at 550 m s$^{-1}$ dominate the cross-section of observed material.  We infer a mass loss rate of 4900 kg s$^{-1}$ at the beginning of our observations, which had decreased to 90 kg s$^{-1}$ by UT 2008 Feb.\ 12.  Peak mass loss rates of 3 $\times$ 10$^{5}$ kg s$^{-1}$ and 6-7 $\times$ 10$^{5}$ kg s$^{-1}$ were estimated by \cite{2011ApJ...728...31L} and \cite{2008ICQ....30....3S}, respectively.  Our results are lower by several orders of magnitude, indicating a rapid decrease following the initial outburst.

\subsection{A Second Outburst}

\begin{figure}
\centering
\includegraphics[totalheight=13cm]{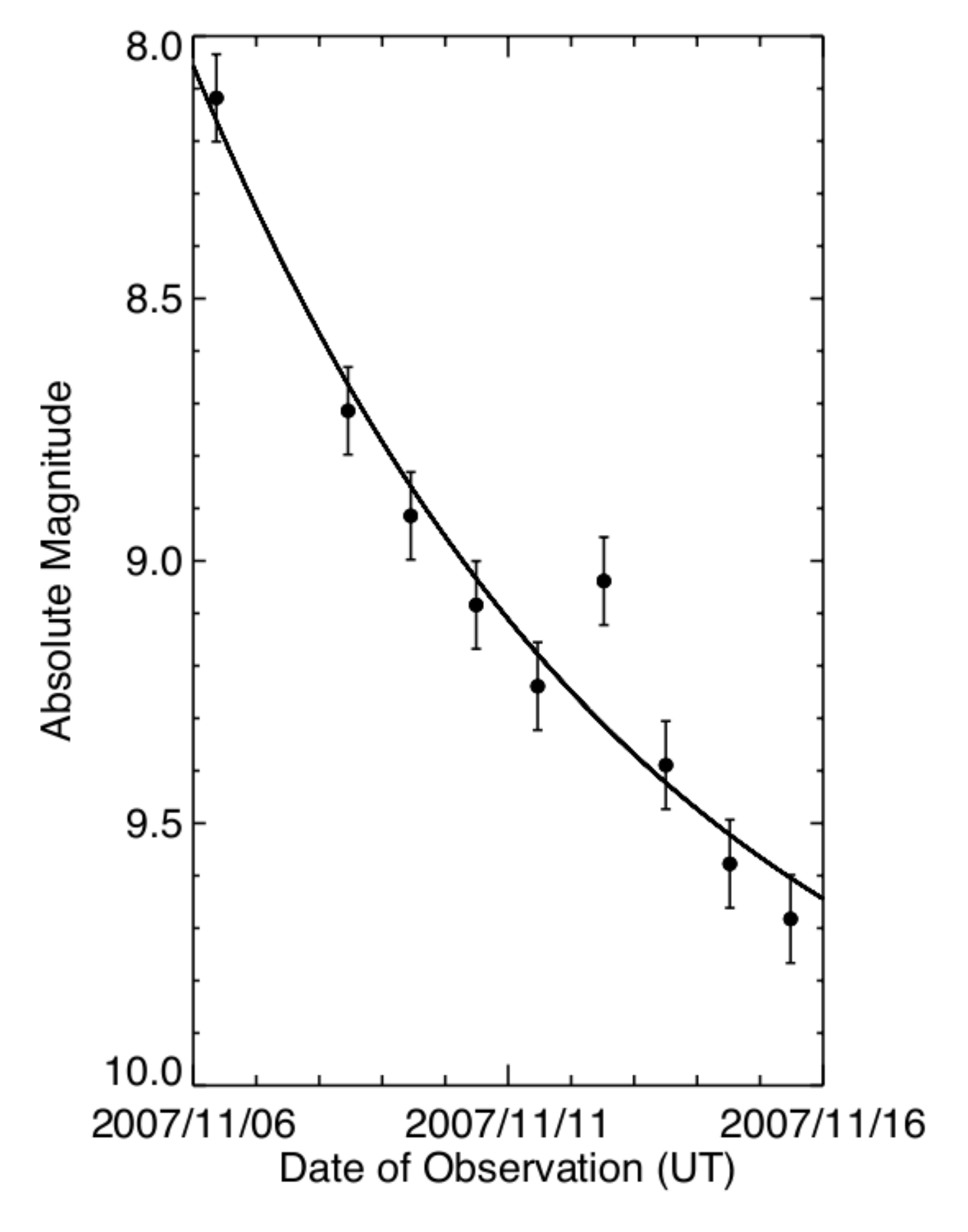}
\caption{The photometric evolution of the inner coma of 17P/Holmes during 2007 Nov.\ as measured with a 2500 km radius aperture.  The best fit to the data was found by fitting an exponential function and is shown by the solid line.  While there was an overall decrease in the absolute magnitude of 17P/Holmes during this time, an apparent brightening can be seen to occur on 2007 Nov.\ 12.  The deviation of the observed magnitude from the value expected according to the best fit is 0.3 mag, corresponding to a 3-$\sigma$ deviation.}
\label{fig:fadingnov}
\end{figure}

The fading inner coma of 17P/Holmes brightened unexpectedly on UT 2007 Nov.\ 12 by $\sim$ 25\% (Figure~\ref{fig:fadingnov}).  The expected absolute magnitude, as determined from a fading function fit to the November data of the exponential form discussed previously, is 9.3 mag, whereas the absolute magnitude calculated from the observed brightness was 9.0 mag.  Given the associated photometric uncertainties, this is a 3.0 $\sigma$ (99.8\% probability) result.  We note that this significance increases to 3.7 $\sigma$ (99.97\% probability) if the UT 2007 Nov.\ 12 data point is excluded from the fit.  
A background object bright enough to cause the increase would need to have an apparent magnitude of 14.6 mag; an inspection of the STScI Digitized Sky Survey showed that there was no such object near 17P/Holmes.  We attribute the brightening to a second, albeit much smaller, outburst on the surface of the nucleus.  The subsequent observation of 17P/Holmes on 2007 Nov.\ 13 also lies above the fit shown in Figure~(\ref{fig:fadingnov}), suggesting that some of the material released during the second outburst remained in the aperture at this time.  

The surface brightness profiles of UT 2007 Nov.\ 11, 12 and 13 provide independent evidence for a second outburst (Figure~\ref{fig:SBall}).  The coma within $\sim$ 5000 km of the nucleus is brighter than expected on UT 2007 Nov.\ 12 when compared to the surface brightness profiles of 17P/Holmes on the previous and following nights.  We can estimate the onset time of the second outburst if we assume that material ejected was dominated by dust grains traveling at $\sim$ 550 m s$^{-1}$.  We find that this material could have traveled 5000 km in 2.5 hours, leading us to conclude that the outburst started at approximately 2007 Nov.\ 12.4.  It seems likely that the outburst shut down within hours, given that subsequent observations revealed no unusual activity.


The difference between the expected magnitude and the observed magnitude, as shown in Figure~(\ref{fig:fadingnov}) using a 2500 km radius aperture, amounts to 0.3 mag.  Using Equation~(\ref{crosssectioneq}) and assuming that the material consists of dust grains with radii of 1 $\mu$m, albedos of 0.1, and densities of 1000 kg m$^{-3}$, the extra material has a scattering cross-section of 6 $\times$ 10$^{8}$ m$^{2}$, and a mass of 8 $\times$ 10$^{5}$ kg, corresponding to a sphere of material with a radius of 6 m.  

The short duration of the Nov.\ 12 event suggests that the loss of material did not cause significant exposure of volatile-rich patches on the surface of the nucleus.  Ejection of a surface layer by gas pressure from within the nucleus would likely have left an icy crater exposed that would have remained active for days or weeks as volatile deposits diminished.
The event could have involved the dislodging of a fragment that was only weakly bound to the surface.  The fragment would then have broken apart into a collection of particles with a much greater total cross-section.  Rotational spin-up upon ejection, or the sublimation of volatiles that were holding the fragment together could have caused the break-up.
The nucleus may have remained unstable for a significantly longer period of time given that \cite{2009SASS...28...51M} detected a 0.85 mag brightening on UT 2009 Jan.\ 4.  It is likely that 17P/Holmes will continue to undergo minor outbursts for the next few orbital periods as unstable regolith settles and freshly-exposed volatiles sublimate.





\subsection{Surface Brightness Profile}

The distribution of material in the inner coma can be studied using the surface brightness profile.  We fitted the gradient of each surface brightness profile between 10000 km and 25000 km using a power law.  The gradient, $m_{p}$, is given by

\begin{equation}
m_{p} = \frac{d\ ln\ B(R)}{d\ ln\ R}
\label{meq}
\end{equation}

where $B(R)$ is the flux per 10$^{6}$ km$^{2}$, and $R$ (km) is the distance from the nucleus.  A dust coma generated by an isotropic, steady source would have a slope $m_{p}$ = -1 \citep{1991ASSL..167...19J}.  Instead, the measured slope varied between -0.2 and -0.3 with no significant temporal trend (Table~\ref{table:mtable}).  
We can be confident that the shallow slopes are not a result of inaccurate sky subtraction caused by coma-contamination.  If the estimated sky value actually included a contribution from the dust coma, the images would be over-subtracted, resulting in an artificially steep slope.  The shallow slope suggests that either the dust production rate varied in such a way as to temporarily produce a smaller scattering cross-section close to the nucleus, or that the cross-section increased with increasing nucleo-centric distances, possibly through a process such as dust fragmentation.  The likelihood of the activity on the nucleus varying with time in such a way as to consistently produce a shallow surface brightness profile for several months is extremely low. Another process is needed to account for the shallow surface brightness gradients.

Cometary ejecta is likely to include agglomerate grains, which are clusters of smaller particles that are loosely held together by electrostatic forces or volatile material \citep{2004JGRE..10912S03C}.  Heating of these tenuous aggregates by incident solar radiation could cause them to break apart, greatly increasing their scattering cross-section.  For example, the brightness of a 1 cm radius cluster of micron-sized dust grains could increase by a factor of $\sim$ 10$^{4}$ if it broke apart into its constituent particles.   Thus, fragmenting or disaggregating particles could cause a shallow surface brightness profile where the scattering cross-section of material increases with distance from the nucleus.  \cite{2009AJ....137.4538Y} identified populations of both hot and cold grains in the inner coma of 17P/Holmes using infrared spectra.  They postulated that the cold ice grains and hot dust grains could have originally been contained within a porous aggregate that broke apart when exposed to sunlight.  We conjecture that fragmenting aggregates were responsible for the shallow profile at small nucleo-centric distances ($<$ 25000 km), \cite{2010MNRAS.407.1784H} suggested that this process continued out to distances of greater than 10$^{5}$ km.

\subsection{Ice Grain Halo}











\begin{figure}
\centering
\includegraphics[totalheight=12cm]{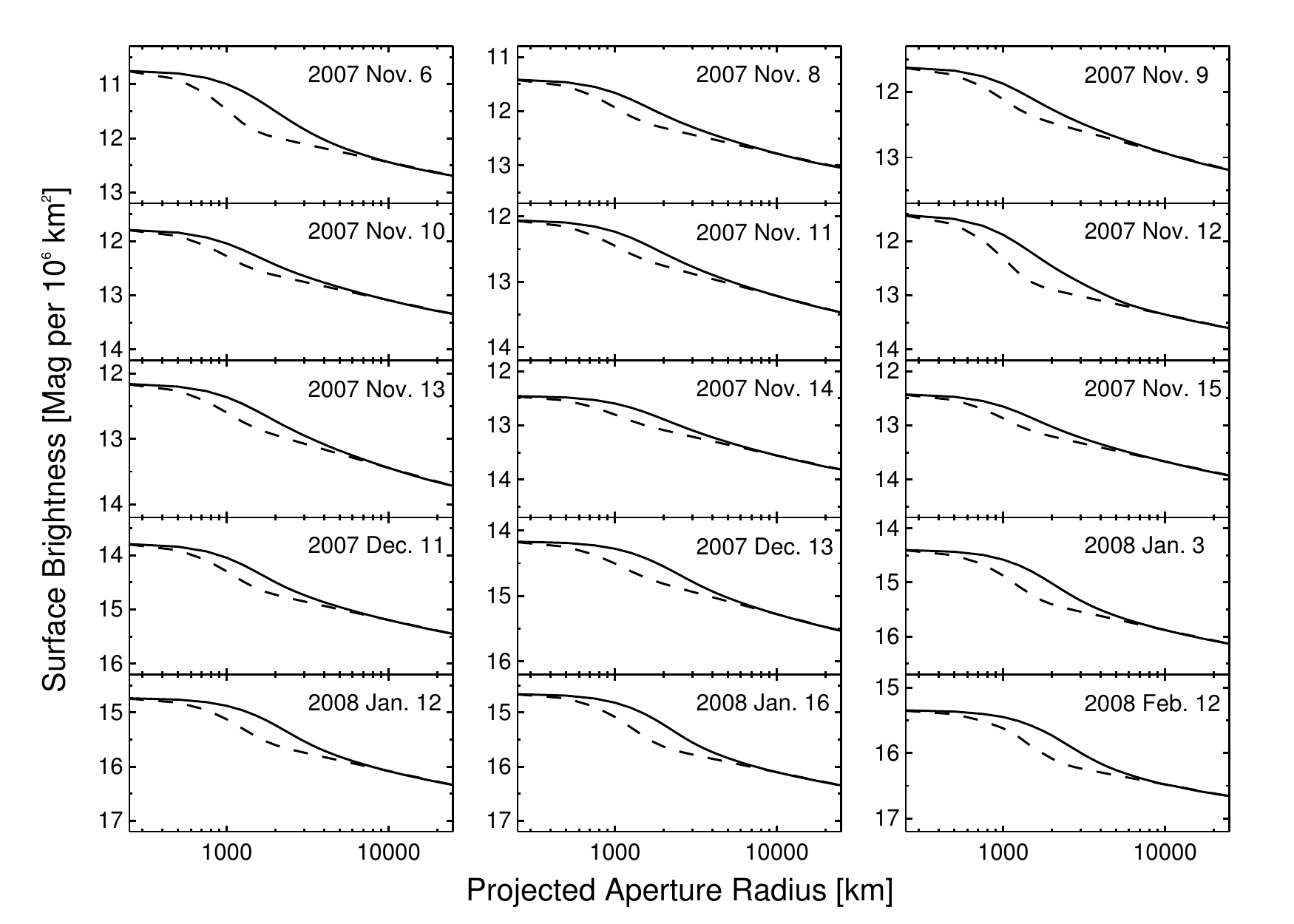}
\caption{The surface brightness profile of 17P/Holmes' inner coma (solid line) compared to a generated profile (dashed line).  The generated profile consists of an unresolved central source with coma that has the same slope as the 17P/Holmes coma between 10000 km and 25000 km.  Both profiles have been convolved with a 1$^{\prime\prime}$ Gaussian function to eliminate the effects of variable seeing from night to night.  The generated profile fails to simulate enough flux in the innermost coma of 17P/Holmes, leading to the possibility that an ice grain halo persisted at small nucleo-centric distances.}
\label{fig:SBcomp}
\end{figure}

The surface brightness profiles consistently show a bump in the innermost region before the profiles straighten out to a decreasing slope (Figure~\ref{fig:SBall}).  To investigate the nature of this bump we created a model surface brightness profile for each night with a slope matching that measured at large angular radii in the CFHT data, plus an unresolved central source.  The model was forced to match those brightness levels observed at 250 km and 10000 km, and then convolved with a 1$^{\prime\prime}$ Gaussian function to simulate the effects of seeing.  Figure~(\ref{fig:SBcomp}) compares the results of our model (dashed line) to the data (solid line).  The comparison shows that the brightness of the central region cannot simply be attributed to a smooth coma with a contribution from the nucleus.  One possible interpretation is that a halo of icy sublimating grains persisted around the nucleus.  Ice grain halos in comets were suggested decades ago \citep{1970P&SS...18..717D} and evidence for halos has been reported in several comets since then, including  C/1996 B2 Hyakutake \citep{1997Sci...277..676H} and 103P/Hartley 2 \citep{2011ApJ...734L...1M}.  \cite{2009AJ....138.1062S} found that derived water production rates for 17P/Holmes during the month following the outburst were too large to be explained by sublimation of water ice from the surface of the nucleus.  He concluded that a population of icy grains must have been sublimating near the nucleus, which fits well with the ice grain halo scenario.

The extent of an ice grain halo is dependent on the sublimation rate of the ice grains, which in turn depends on several factors including the size, albedo, temperature, and density of the grains.  To explore these factors, we calculate the sublimation rate assuming thermal equilibrium.  The sublimation rate is expressed in meters per second, representing the rate at which an icy surface would retreat under these conditions.  We assume two cases: isothermal grains where the heat absorbed is radiated isotropically back into space (``cold grains''), and grains at the temperature of the sub-solar point (``hot grains'').  The sublimation rate is strongly dependent on the Bond albedo, which is the fraction of total power incident on a body that is reflected from the surface.  High-albedo pure ice grains have low sublimation rates since they absorb little heat and remain cold.  However, adding even a small amount of refractory material to the ice lowers the albedo significantly, resulting in greater absorption of heat and an increase in the temperature of the grain \citep{1982Icar...49..244C}.  

The radius of the ice grain halo, $L$, is related to the grain velocity, $v(a)$ (m s$^{-1}$), radius, $a$ (m), and sublimation rate, $\dot{a}$ (m s$^{-1}$), by:

\begin{equation}
L = v(a) \frac{a}{\dot{a}}
\label{extenteq}
\end{equation}

\begin{figure}
\centering
\includegraphics[totalheight=16cm,angle=90]{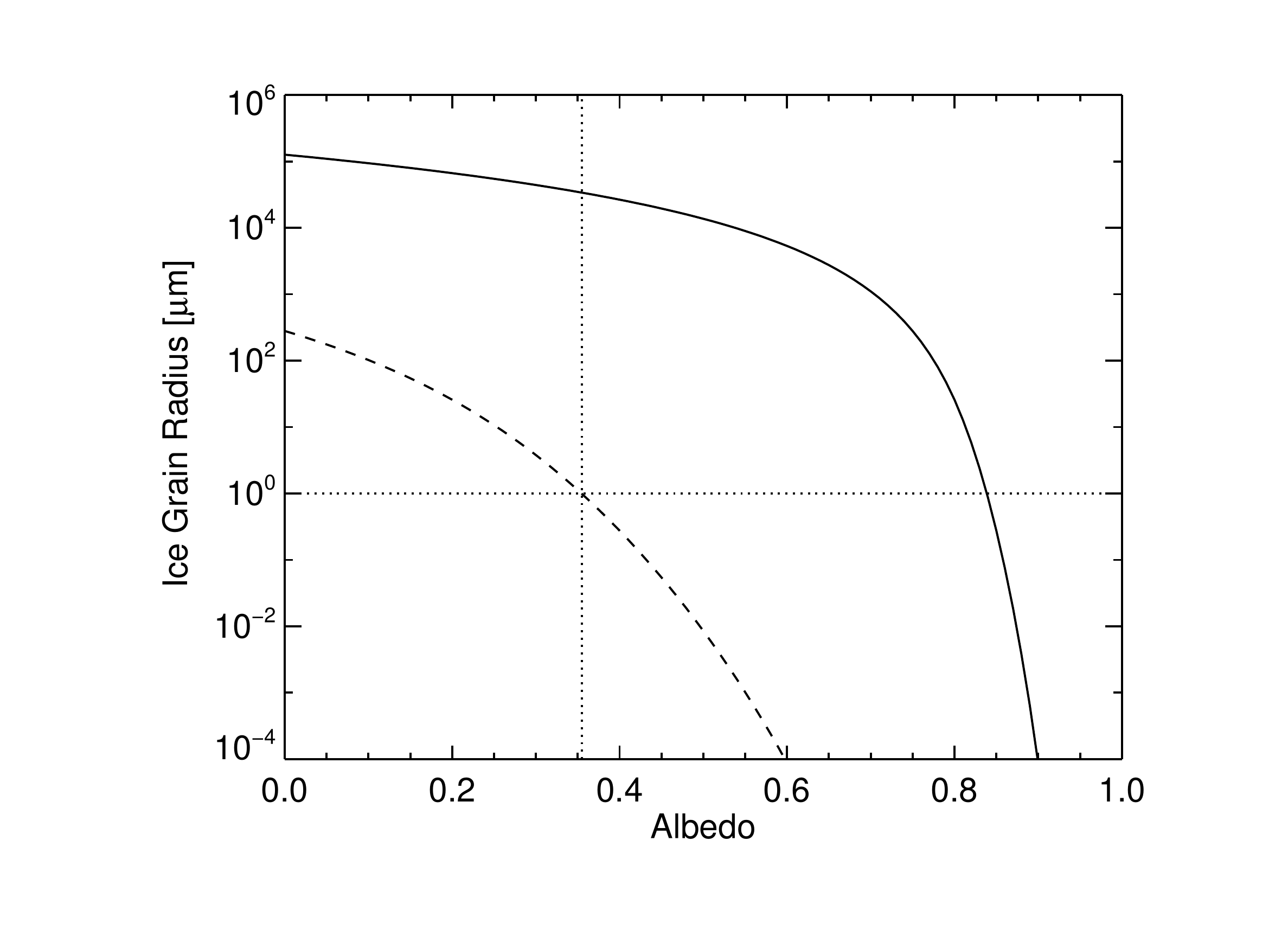}
\vspace{-1.2cm}
\caption{Grain radius versus albedo relation for ice grains having a sublimation lifetime sufficient to travel 3450 km - the median of our estimated ice grain halo radii.  The dashed and solid lines depict results for low and high grain temperature models, as described in the text.  Cold micron-sized ice grains must have had an albedo of $\sim$ 0.35 (dotted lines) in order to move 3450 km from the nucleus at 550 m s$^{-1}$, which suggests that the ice was contaminated with albedo-lowering regolith and was not pure.}
\label{fig:avspv}
\end{figure}

We utilized the velocity-radius relationship given in Equation~(\ref{veleq}) and measured $L$ by comparing the surface brightness profile on each night to the model profile.  We defined the edge of the halo as where the brightness fell to within 10\% of the model.  We acknowledge that this approach is model-dependent but use it to explore the concept of an ice grain halo in this context.  The radius determined varied between $\sim$ 2900 km and 4900 km (Table~\ref{table:mtable}).  The median radius was 3450 km, consistent with results from \cite{2009AJ....137.4538Y} who found that ice grains persisted up to 3500 km from the nucleus in 2007 Oct.\  We plot the radius-albedo relations for hot and cold ice grains that can survive across a 3450 km radius aperture in Figure~(\ref{fig:avspv}).  Cold ice grains with a radius of 1 $\mu$m (c.f. \citealt{2009AJ....137.4538Y}) must have a Bond albedo of $\sim$ 0.35 in order to survive, while micron-sized grains in the high temperature limit must have a higher Bond albedo, of $\sim$ 0.85.  Of these two approximations, the cold grain case is probably more relevant to ice grains in 17P/Holmes, given their small size and the likelihood that they were rotating rapidly.  

The results of \cite{2010ApJ...714.1324I} suggest that the geometric albedo of the coma was between $\sim$ 0.05 and 0.20.  We note that the results from \cite{2010ApJ...714.1324I} are based on measurement of the flux reflected from all material in a 5300 km radius aperture, which would have included a significant contribution from lower-albedo dust grains as well as high-albedo ice grains.  In this sense, our result that the Bond albedo of ice grains in the halo must have been $\sim$ 0.35 appears to be consistent with their findings.

Comparatively little contaminating material is required to significantly darken icy material.  \cite{1982Icar...49..244C} found that contaminating 1 mm snow grains with 0.1 $\mu$m soot particles that amounted to only 10$^{-5}$ of the sample weight lowered the albedo from 0.94 to 0.40.  Results from \cite{2009AJ....137.4538Y}  suggested that only ice grains with $\lesssim$ 10\% impurities could survive long enough to travel 3500 km from the nucleus before sublimating.  \cite{2006Icar..180..473B} found that micron-sized pure ice grains beyond $\sim$ 2 AU should survive for weeks or even months, while dirty ice grains (10\% dirt by mass) should survive for only hours, which agrees well with our results.

\begin{figure}
\centerline{\hbox{
\includegraphics[totalheight=17cm,angle=90]{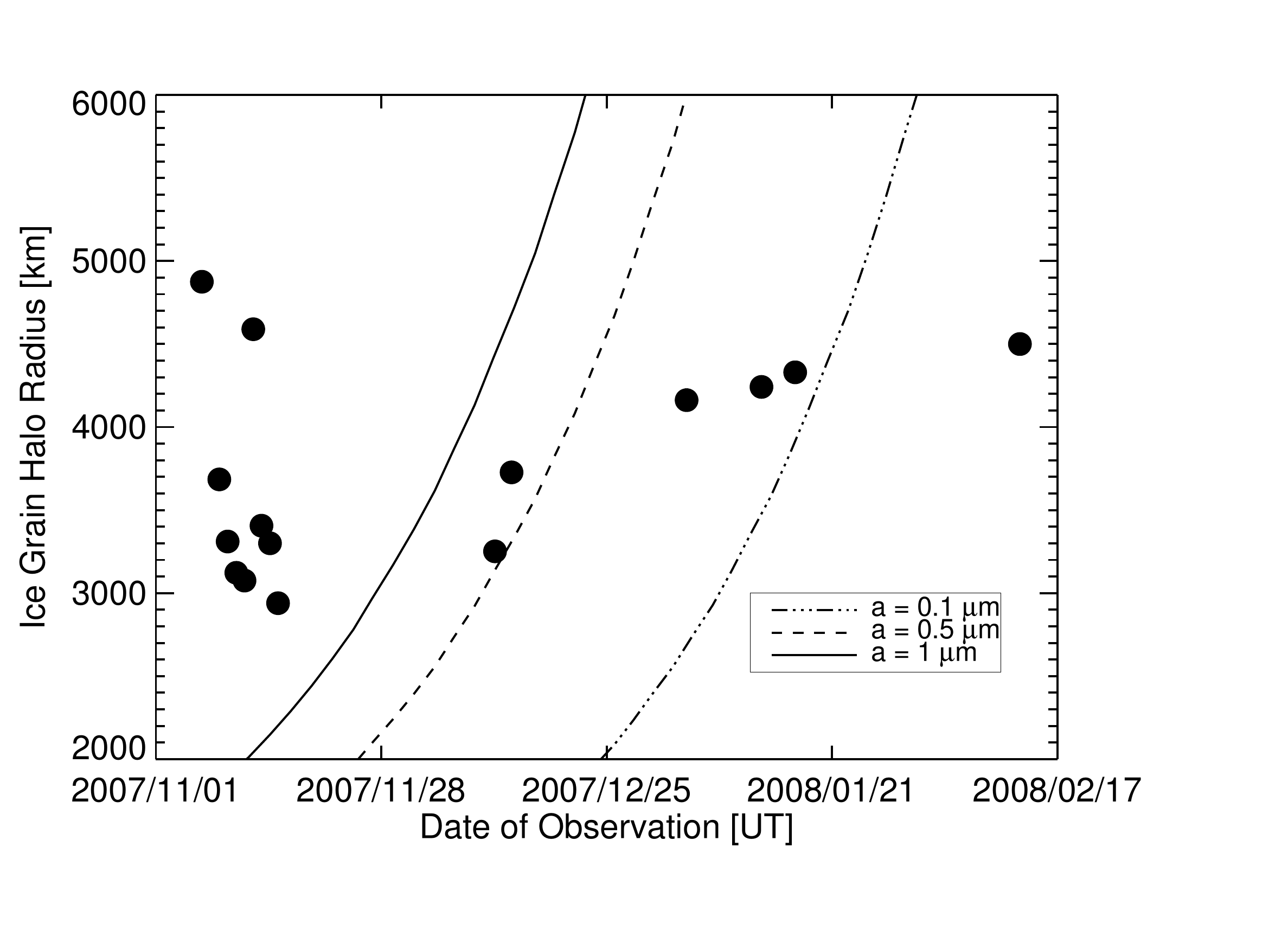}}}
\vspace{-1.5cm}
\caption{The radius of the observed ice grain halo (filled circles, from Figure 8) compared to the expected radius of an ice grain halo given the varying heliocentric distance of 17P/Holmes.  At larger heliocentric distances, and correspondingly lower temperatures, ice grains can survive for longer periods of time so we expect the radius of the halo to increase as 17P/Holmes moves away from the Sun.  We show the radius expected for halos with ice grains of varying sizes and constant albedos of 0.3.  None of our scenarios provide a good fit to the data.  We suggest that changes in other parameters, such as the ejection velocity of the ice grains, or the albedo, may play an important part in the maximum extent of the ice grain halo.}
\label{fig:extentfit}
\end{figure}

The ice grain halo radius varied during our observing campaign, although it is difficult to discern any obvious trends (Figure~\ref{fig:extentfit}).  We note that the halo was larger after periods of known activity, such as on 2007 Nov.\ 6 - our observation closest to the original outburst - and on 2007 Nov.\ 12 after the much smaller second outburst.  The radius appeared to decrease after these periods of activity until the heliocentric distance of 17P/Holmes began to change appreciably.  As the heliocentric distance increased from 2.63 AU on 2007 Dec.\ 11 to 2.90 AU on 2008 Feb.\ 12, the temperature of the ice grains must have dropped, causing a decrease in sublimation rates and thus, an increase in the survival time of the grains.  We attempted to fit the increase of the ice grain halo radius using the thermal equilibrium model described previously.  Figure~(\ref{fig:extentfit}) shows the data with results from models of the halo radii expected during our observations, assuming the halo consists of isothermal grains.  The increase in heliocentric distance by itself cannot account for the halo size variations.  We suggest that changes in the level of activity, size and albedo of the ice grains ejected, or the grain velocities may also play important roles in the extent of the ice grain halo.

\begin{figure}
\centering
\includegraphics[totalheight=12cm]{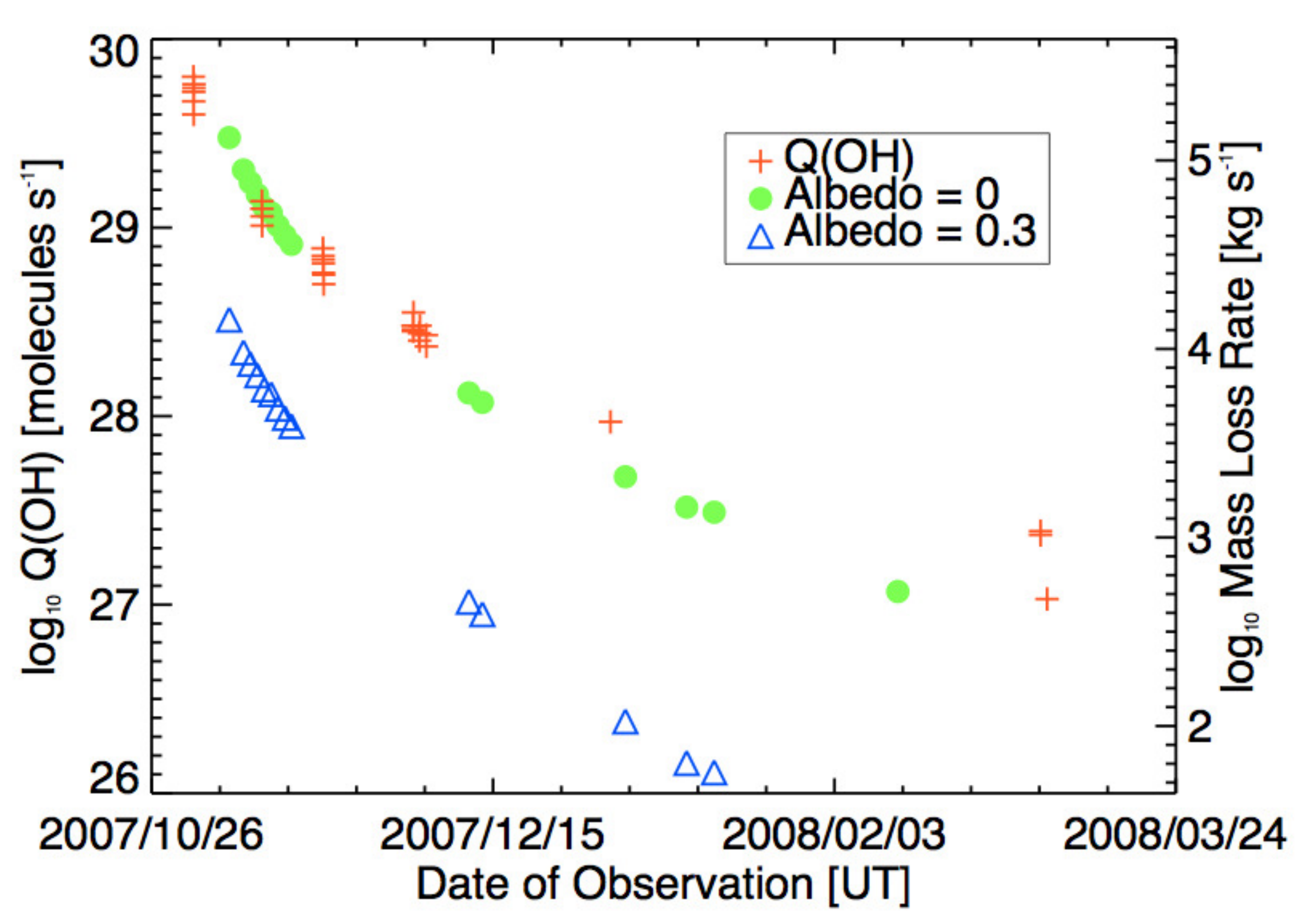}
\caption{A comparison of the derived water production rates in the inner 10$^{4}$ km of the coma (crosses; \citealt{2009AJ....138.1062S}) and the expected mass loss rates for a sublimating ice grain halo composed of grains with albedos of 0 (filled circles) and 0.3 (triangles).  The calculated mass loss rates have been plotted such that the mass loss rate for albedo = 0 on UT 2007 Nov.\ 11.5 coincides vertically with the data point representing the derived water production rate on UT 2007 Nov.\ 11.2.  The decrease in the mass loss rate for an ice grain halo with an albedo of 0 appears consistent with the decrease in water production rates, while the mass loss rate for a halo with higher-albedo grains would decrease more steeply.}
\label{fig:schcomp}
\end{figure}

Finally, we compare the expected mass loss rate from a sublimating ice grain halo to the observed production rates of OH molecules using aperture photometry from this work and production rates taken from \cite{2009AJ....138.1062S}.  The production rates were derived from narrowband photometry of 17P/Holmes' inner coma between UT 2007 Nov.\ 1 and 2008 Mar.\ 5.  Since the published production rates were corrected for 10$^{4}$ km radius apertures, we also use 10$^{4}$ km radius apertures.  We convert the apparent magnitudes to geometric scattering cross-sections using Equation~\ref{crosssectioneq} and assumptions detailed in Section~\ref{sec:fading}.  To estimate the mass loss rate expected from a sublimating ice grain halo, we multiply the cross-section of material (presumed to be 100\% ice) by the density of the ice (assumed to be 1000 kg m$^{-3}$) and the sublimation rate (m s$^{-1}$) for ice grains at the corresponding heliocentric distances of 17P/Holmes.  Figure~\ref{fig:schcomp} compares the expected mass loss rates for ice grain halos composed of grains with albedos of 0 and 0.3 to the derived production rates from \cite{2009AJ....138.1062S}.  We find that the mass loss rate for an ice halo composed of grains with albedo = 0.3 drops off more quickly than the derived water production rate.  The changing mass loss rate for an ice grain halo with grains of albedo = 0 matches the trend in the water production rate well but the albedo conflicts with our previous determination that the extent of the halo is best modeled by sublimating ice grains with albedos of $\sim$ 0.2 - 0.4.  We acknowledge that the model is underconstrained given that the ice grains may not be isothermal, would have a range of sizes and albedos, and those that determine the observed radius of the ice grain halo may not be the grains that supply the sublimated gas.  We conclude by simply saying that the decrease in water production rates agrees loosely with the expected decrease in mass loss rates for an ice grain halo consisting of low-albedo ice grains.

\section{Summary}






We have used aperture photometry to examine the inner coma of 17P/Holmes between UT 2007 Nov.\ 6 and 2008 Feb.\ 12 following its massive outburst in 2007 Oct.\  Our results include:

\begin{enumerate}

\item{The nucleus of 17P/Holmes remained active for at least several months following its spectacular outburst in 2007 Oct., although at a level very low compared to the peak outburst rates.  The inner coma decreased in brightness by a factor of $\sim$ 50 during our observations, but an examination of aperture-crossing times shows that fresh material must have been continually supplied from active regions on the nucleus.}

\item{A second much smaller outburst occurred on UT 2007 Nov.\ 12, ejecting approximately 8 $\times$ 10$^{5}$ kg of material that had traveled up to 5000 km by the time of our observations.  We propose that a fragment broke away from the surface and rapidly broke apart, releasing a significant amount of dust into the coma.}

\item{The surface brightness profile of 17P/Holmes had a measured slope between -0.2 and -0.3.  This is less steep than expected for a steady-state coma consistent with the fragmentation of icy grain aggregates.}

\item{A halo consisting of freshly released ice grains persisted around the nucleus, extending to radii of several thousand km.  This halo was likely responsible for the anomalously high water production rates observed by \cite{2009AJ....138.1062S} several months after the outburst, although the bulk of gas and dust production could originate from distinctly different populations of grains.}

\item{The radius of the ice grain halo is compatible with an origin in isothermal micron-sized grains having a Bond albedo of $\sim$ 0.35.  The ice grains may have been contaminated with small amounts of refractory material.}

\end{enumerate}

Based on observations obtained with MegaPrime/MegaCam, a joint project of CFHT and CEA/DAPNIA, at the Canada-France-Hawaii Telescope (CFHT) which is operated by the National Research Council (NRC) of Canada, the Institut National des Science de l'Univers of the Centre National de la Recherche Scientifique (CNRS) of France, and the University of Hawaii.  This work is based in part on data products produced at the TERAPIX data center located at the Institut d'Astrophysique de Paris.  We appreciate grant support from the NASA Outer Planets Research Program to DJ.  RS acknowledges support from the NASA Postdoctoral Fellowship Program.  Part of the research was carried out at the Jet Propulsion Laboratory, California Institute of Technology, under a contract with the National Aeronautics and Space Administration.

\clearpage

%

\clearpage

\begin{deluxetable}{lccccc}
\tabletypesize{\small}
\tablecaption{Journal of Observations \label{imaging}}
\tablewidth{0pt}
\tablehead{
\colhead{UT Date} &  \colhead{Exposure time [s]} & \colhead{r$_{H}$ [AU] \tablenotemark{1}} & \colhead{$\Delta$ [AU] \tablenotemark{2}} & \colhead{$\alpha$ [deg] \tablenotemark{3}} & \colhead{Scale [km/pixel]}}
\startdata
2007 Nov 6.4 & 5 & 2.49 & 1.62 & 13.7 &218\\ 
2007 Nov 8.5 & 5 & 2.50 & 1.62 & 13.3 & 218\\ 
2007 Nov 9.5 & 5 & 2.50 & 1.62 & 13.1 & 218\\ 
2007 Nov 10.5 & 5 & 2.50 & 1.62 & 12.9 & 218\\ 
2007 Nov 11.5 & 50 & 2.51 & 1.62 & 12.7 & 218\\ 
2007 Nov 12.5 & 5 & 2.51 & 1.62 & 12.5 & 219\\ 
2007 Nov 13.5 & 50 & 2.52 & 1.63 & 12.3 & 219\\ 
2007 Nov 14.5 & 50 & 2.52 & 1.63 & 12.2 & 219\\ 
2007 Nov 15.5 & 50 & 2.52 & 1.63 & 12.0 & 219\\ 
2007 Dec 11.5 & 70 & 2.63 & 1.77 & 12.5 & 238\\ 
2007 Dec 13.4 & 140 & 2.63 & 1.78 & 12.8 & 239\\ 
2008 Jan 3.4 & 200 & 2.73 & 2.01 & 16.5 & 270\\  
2008 Jan 12.3 & 200 & 2.77 & 2.13 & 17.8 & 287\\ 
2008 Jan 16.4 & 200 & 2.78 & 2.19 & 18.3 & 295\\ 
2008 Feb 12.3 & 50 & 2.90 & 2.62 & 19.8 & 352\\ 
\enddata
\tablenotetext{1}{Heliocentric distance}
\tablenotetext{2}{Geocentric distance}
\tablenotetext{3}{Phase angle}
\label{table:obs}
\end{deluxetable}

\clearpage

\begin{deluxetable}{cccccc}
\tabletypesize{\small}
\tablecaption{Characteristics of Coma Material}
\tablewidth{0pt}
\tablehead{
\colhead{UT Date} & \colhead{$m$ \tablenotemark{1}} & \colhead{$H$ \tablenotemark{2}} & \colhead{$\sigma$  \tablenotemark{3}} & \colhead{Mass} & \colhead{Mass Loss Rate}\\
& \colhead{} & \colhead{} & \colhead{[$10^{8}$ m$^{2}$]} & \colhead{[$10^{5}$ kg]} & \colhead{[kg s$^{-1}$]}}
\startdata
2007 Nov.\ 6.4 & 11.7 & 8.1 & 166.0 & 221.4 & 4870 \\
2007 Nov.\ 8.5 & 12.3 & 8.7 & 95.9 & 127.8 & 2810 \\
2007 Nov.\ 9.5 & 12.5 & 8.9 & 79.7 & 106.3 & 2340 \\
2007 Nov.\ 10.5 & 12.7 & 9.1 & 68.2 & 90.9 & 2000 \\
2007 Nov.\ 11.5 & 12.8 & 9.2 & 59.1 & 78.8 & 1730 \\
2007 Nov.\ 12.5 & 12.6 & 9.0 & 71.1 & 94.8 & 2090 \\
2007 Nov.\ 13.5 & 13.0 & 9.4 & 51.5 & 68.6 & 1510 \\
2007 Nov.\ 14.5 & 13.2 & 9.6 & 43.3 & 57.7 & 1270 \\ 
2007 Nov.\ 15.5 & 13.3 & 9.7 & 39.3 & 52.4 & 1150 \\
2007 Dec.\ 11.5 & 15.0 & 11.1 & 10.4 & 13.9 & 310 \\
2007 Dec.\ 13.4 & 15.2 & 11.2 & 9.3 & 12.4 & 270 \\
2008 Jan.\ 3.4 & 16.1 & 11.7 & 6.4 & 8.5 & 190 \\
2008 Jan.\ 12.3 & 16.5 & 11.9 & 5.1 & 6.7 & 150 \\
2008 Jan.\ 16.4 & 16.6 & 11.9 & 5.2 & 6.9 & 150 \\
2008 Feb.\ 12.3 & 17.6 & 12.4 & 3.1 & 4.2 & 90 \\
\enddata
\tablenotetext{1}{Apparent magnitude within 2500 km projected radius aperture}
\tablenotetext{2}{Absolute magnitude within 2500 km projected radius aperture}
\tablenotetext{3}{Cross-section of material within 2500 km projected radius aperture}
\label{table:char}
\end{deluxetable}

\clearpage

\begin{deluxetable}{ccc}
\tabletypesize{\small}
\tablecaption{Surface Brightness Profile Characteristics}
\tablewidth{0pt}
\tablehead{
\colhead{Date (UT)} & \colhead{Surface Brightness Profile Gradient} & \colhead{Ice Grain Halo Radius [km]}}
\startdata
2007 Nov.\ 6.4 & -0.25 & 4870\\ 
2007 Nov.\ 8.5 & -0.26 & 3680\\ 
2007 Nov.\ 9.5 & -0.26 & 3310\\
2007 Nov.\ 10.5 & -0.25 & 3120\\
2007 Nov.\ 11.5 & -0.25 & 3080\\
2007 Nov.\ 12.5 & -0.25 & 4590\\
2007 Nov.\ 13.5 & -0.28 & 3410\\
2007 Nov.\ 14.5 & -0.26 &3300\\
2007 Nov.\ 15.5 & -0.26 &2940\\
2007 Dec.\ 11.5 & -0.26 &3250\\
2007 Dec.\ 13.4 & -0.25 &3730\\
2008 Jan.\ 3.4 & -0.25 &4160\\
2008 Jan.\ 12.3 & -0.25 &4240\\
2008 Jan.\ 16.4 & -0.24 &4330\\
2008 Feb.\ 12.3 & -0.18 &4500\\
\enddata
\label{table:mtable}
\end{deluxetable}



\clearpage

\clearpage

\clearpage

\clearpage


\clearpage

\clearpage

\clearpage

\clearpage

\clearpage

\clearpage

\clearpage

\end{document}